%% file: rrms.tex
\documentclass[apj,twocolumn]{emulateapj}
\usepackage{amsmath}
\usepackage{apjfonts}

\input{Definitions.tex}

\newcommand{\epm}{e^+e^-}   
\newcommand{\TskkeV}{\left(\frac{T_s}{10\keV}\right)}  
\newcommand{\units}[1]{\,\rm{(#1)}}

\begin{document}
\title{Relativistic Radiation Mediated Shocks}
\author{Ran Budnik\altaffilmark{1}, Boaz Katz\altaffilmark{1}, Amir Sagiv and Eli Waxman\altaffilmark{1}}
\begin{abstract}
The structure of relativistic radiation mediated shocks (RRMS) 
propagating into a cold electron-proton plasma is calculated and 
analyzed. A qualitative discussion of the physics of relativistic and 
non relativistic shocks, including order of magnitude estimates for the 
relevant temperature and length scales, is presented. Detailed numerical 
solutions are derived for shock Lorentz factors $\Gamma_u$ in the range 
$6\le\Gamma_u\le30$, using a novel iteration technique solving the 
hydrodynamics and radiation transport equations (the protons, electrons 
and positrons are argued to be coupled by collective plasma processes 
and are treated as a fluid). 
The shock transition (deceleration) region, where the Lorentz factor $ \Gamma $ drops from $ \Gamma_u $ to $ \sim 1 $, is characterized by high plasma temperatures $ T\sim \Gamma m_ec^2 $ and highly anisotropic radiation, with characteristic shock-frame energy of upstream and downstream going photons of a few~$\times\, m_ec^2$ and $\sim \Gamma^2 m_ec^2$, respectively.
Photon scattering is dominated by e$^\pm$ 
pairs, with pair to proton density ratio reaching $\approx10^2\Gamma_u$. 
The width of the deceleration region, in terms of Thomson optical 
depths for upstream going photons, is large, $\Delta\tau\sim\Gamma_u^2$ 
($\Delta\tau\sim1$ neglecting the contribution of pairs) due to Klein 
Nishina suppression of the scattering cross section. A high energy 
photon component, narrowly beamed in the downstream direction, with a 
nearly flat power-law like spectrum, $\nu I_\nu\propto\nu^0$, and an 
energy cutoff at $ \sim \Gamma_u^2 m_ec^2 $ carries a fair fraction of the 
energy flux at the end of the deceleration region. An approximate 
analytic model of RRMS, reproducing the main features of the numerical 
results, is provided.

\end{abstract}
\keywords{~~~shock waves --- radiation mechanisms: nonthermal --- gamma-rays: bursts}
\altaffiltext{1}{Physics Faculty, Weizmann Institute, Rehovot 76100, Israel; ranny.budnik@weizmann.ac.il; boaz.katz@weizmann.ac.il; eli.waxman@weizmann.ac.il}

\section{Introduction}
\label{sec:intro}

Radiation mediated shocks (RMSs) are shocks in which the downstream (DS) energy density is dominated by radiation rather than by particle thermal energy, and in which the fast upstream (US) plasma approaching the shock is decelerated by scattering of photons, generated in the DS and propagating into the US, by the fast US electrons. RMS are expected to occur in a variety of astrophysical flows. The shock waves propagating through, and expelling, the envelopes of massive stars undergoing core collapse supernova explosions, are non relativistic (NR) RMS \citep{Weaver76}. Relativistic RMS (RRMS) may play an important role in, e.g., gamma-ray bursts, trans-relativistic suprenovae, and pulsar accretion flows.
\\ \noindent [1] {\it Gamma Ray Bursts (GRBs).} Within the framework of the collapsar model of GRBs \citep[e.g.][]{Woosley1993}, a highly relativistic jet driven by the collapsed core of a massive star penetrates through the stellar envelope. The shock that decelerates the jet is expected to be a highly relativistic RMS \citep{MezWax2001,Aloy00}.
\\ \noindent [2] {\it Trans Relativistic SNe.} Several recent SN events, that were identified in very early stages of the explosion, have been shown to deposit a significant fraction, $\sim1$\%, of the explosion energy in mildly relativistic,
$\gamma\beta\gtrsim1$, ejecta \citep[][and references therein]{Soderberg06_rel_jet_aj}. The existence of mildly relativistic ejecta components suggests that a mildly relativistic RMS shock traversed the outer envelope of the progenitor. 
\\ \noindent [3] {\it Pulsar accretion flows.} Accretion onto the polar cap of a pulsar is expected to produce a mildly relativistic RMS which is approximately stationary in the neutron star frame \citep[][and references therein]{Burnard91,Becker88}.

NR RMS were studied in detail in \citep{Weaver76}, describing photon propagation using the diffusion approximation and describing the radiation field using two parameters, photon effective temperature and density. These approximations hold for slow shocks, $v/c<0.2$, for which relativistic effects are negligible and the Thomson optical depth of the shock deceleration region is large, $\sim c/v$ (ensuring that the radiation field is nearly isotropic and that the photons are in Compton equilibrium). The NR approximations do not hold for faster shocks. For such shocks, relativistic effects (such as pair production and relativistic corrections to the cross sections of radiative processes) are important, and the radiation field becomes highly anisotropic. 

 A simplified solution for the structure of RRMS, neglecting pair production, photon production and relativistic corrections, was derived by \citet{Levinson08}. This solution may be applicable only in cases where the US plasma holds a significant photon density, which keeps the plasma at low temperatures throughout the shock, much lower than those obtained in a self-consistent solution where the photon density vanishes at US infinity. In a preceding paper \citep{Katz09}
 we derived a simple approximate analytic model for the structure of radiation mediated shocks. This model accurately reproduces the numerical results of \citet{Weaver76} for $v/c\lesssim0.2$, and provides an approximate description of the shock structure at larger velocities, $v/c\rightarrow1$. We confirmed that at shock velocities $v/c\gtrsim 0.1$ the shock transition region is far from thermal equilibrium, with electrons and photons (and positrons) in Compton (pair) equilibrium at temperatures $T_s$ significantly exceeding the far downstream temperature. We have found that $T_s\gtrsim 10\keV$ is reached at shock velocities $v/c\approx 0.2$, and that at higher velocities, $v/c\gtrsim0.6$, the plasma is dominated in the transition region by e$^\pm$ pairs and $60\keV\lesssim T_s \lesssim 200\keV$. We have suggested that the spectrum of radiation emitted during the breaking out of supernova shocks from the stellar envelopes of Blue Super Giants and Wolf-Rayet stars, which reach $v/c>0.1$ for reasonable stellar parameters, may include a hard component with photon energies reaching tens or even hundreds of $\keV$. This may account for the X-ray outburst associated with SN2008D \citep{Soderberg08}, and possibly for other SN-associated outbursts with spectra not extending beyond few $100\keV$ [e.g. XRF060218/SN2006aj \citep{Campana06}]. 

In this paper we derive exact numerical solutions for the steady state structure of RRMS, propagating into a cold upstream plasma of protons and electrons, for shock Lorentz factors $\le 30$ and upstream proper densities $\ll 10^{25}\cm^{-3}$. The solutions are obtained using a novel iteration method for self-consistently solving the energy, momentum and particle conservation equations along with the equation of radiation transport. We assume that the electrons, positrons and protons may be described as a fluid, that the (plasma rest frame) energy distribution of positrons and electrons is thermal, and that the protons are cold. The validity of these assumptions is discussed in detail in \sref{sec:RRMS-phys}. The Radiation mechanisms that are taken into account include Compton scattering, pair production and annihilation and Bremsstrahlung emission and absorption. Other radiation mechanisms, e.g. double Compton scattering, are shown to have a minor effect on our results.

The paper is organized as follows. In section \sref{sec:phys} we
review the physics of RMS, analyze qualitatively the shock
structure, and  motivate  the main assumptions. In section
\sref{sec:formulation} we write down the conservation and transport
equations that are numerically solved, in physical and dimensionless
forms. In section \sref{sec:num_meth} we present the numerical
iteration scheme used to obtain the solutions and apply it to
several test cases. In section \sref{sec:numerical} we present the
numerical solutions of the shock structure and spectrum. In section
\sref{sec:analytic} we give a simple analytic description of the
structure of the shock, which reproduces the main results of the
numerical calculations. In section \sref{sec:nonrel} we present, for
completeness, a preliminary detailed numerical solution of a non
relativistic RMS, and compare it with previously known results.  
In \sref{sec:discussion_RMS} we summarize the main results and discuss their implications.

Throughout this paper the subscripts $u$ and $d$ are used to denote US and DS values respectively. The term "shock frame" refers to the frame at which the shock is at rest (and in which the flow is stationary), and the term "rest frame" refers to the local rest frame of the plasma, i.e. the frame at which the plasma is (locally) at rest. $n$ stands for number density of a species of particles, and if not mentioned otherwise refers to protons. A summary of the notations repeatedly used in this paper appears in appendix \sref{app:notations}.

\section{The physics of RMS}
\label{sec:phys}

In this section we discuss the physics of RMS. In \sref{sec:intro-rms} we define RMS, write down the global requirements that must be satisfied by a physical system in order to allow the formation of RMS, and derive the asymptotic DS conditions. We then focus on NR RMS in \sref{sec:NR-RMS}, writing down the assumptions under which our analysis is carried out, describing the physical mechanisms at play, and providing order of magnitude estimates for the shock width and temperature. Most of the results of \sref{sec:NR-RMS} may be found in earlier papers \citep{Zel'dovich66,Weaver76,Katz09}. The physics of RRMS is discussed in \sref{sec:RRMS-phys}. We highlight the main differences between the relativistic and the NR cases, and describe the assumptions under which the analysis of subsequent sections is carried out.

\subsection{Introduction to RMS}\label{sec:intro-rms}

\subsubsection{Radiation domination}
\label{sec:Definition}

Consider a steady state shock traveling with velocity $ c\beta_u $ through an infinitely thick, cold plasma of protons and electrons, with US rest frame density $ n_u $.
The thermal and radiation pressures  in the asymptotic far DS, which are determined by conservation laws and thermal equilibrium, are given by $  2n_dT_d $ and $ a_{BB}T_d^4/3 $ respectively, where $ n_d $ and $ T_d $ are the far DS proton density and temperature, and  $a_{BB}=\pi^2/15(\hbar c)^3$ is the Stefan-Boltzmann energy density coefficient. The radiation pressure grows much faster than the thermal pressure as a function of  $ \beta_u $, and at high enough $ \beta_u $ the DS pressure is dominated by the radiation. The condition for radiation domination is
\begin{equation}
\frac{a_{BB}T_d^4}{3} \gg  2n_dT_d,
\end{equation}
corresponding to 
\begin{equation}
T_d\gg\left(\frac{6n_d}{a_{BB}} \right)^{1/3}\approx 0.2\left(\frac{n_d}{10^{20}\cm^{-3}}\right)^{1/3}\keV 
\end{equation}
and
\begin{equation}\label{eq:beta_domination}
\beta_u \gg \left( \frac{n_u}{a_{BB}}\right)^{1/6}(m_pc^2)^{-1/2}
\sim 3\times 10^{-4} \left(\frac{n_u}{10^{20}\cm^{-3}}\right)^{1/6}.
\end{equation}
To obtain Eq. \eqref{eq:beta_domination}, note that at low shock velocities, where the radiation pressure is negligible, $ T_d\sim\varepsilon $ and  $ n_d\approx 4n_u $, where $ \varepsilon\approx\beta_u^2m_pc^2/2$ is the kinetic energy per proton in the US.

\subsubsection{Global requirements from a system through which a RMS propagates}
\label{sec:cond-RMS}

In order to sustain a quasi steady state RMS, the system which the shock traverses has to be larger than the shock width. The width of the deceleration, $ L_{dec} $, is $ \beta_u^{-1} $ Thomson optical depths for NR shocks (see \sref{sec:NR-RMS}) and, as we show in this paper, is $ \sim 1 $ Thompson optical depths for relativistic shocks. Hence, systems which RMS traverse much satisfy
\begin{equation}
 L\gg L_{\text{dec}}=\left( \sigma_Tn\beta\right)^{-1},
\end{equation}
Where $L$ is the size of the system, $n$ is the proton density and $\beta c$ is the shock velocity.
The minimum total energy and mass of such systems are 
\begin{equation}\label{eq:Erequirement}
E\sim \frac {\beta^2 }{2}m_pc^2nL^3 > \frac {m_pc^2}{2\sigma_T^3\beta n^2} \approx
3\times 10^{29} n_{20}^{-2}\beta^{-1} \text{erg}
\end{equation}
and 
\begin{equation}
M>L^3nm_p=\frac{m_p}{\sigma_T^3n^2\beta^3}\approx
7\times 10^{50} \left(\frac{n}{\cm^{-3}} \right)^{-2}\left(\frac{\beta}{0.2} \right) ^{-3}\gr.
\end{equation}
respectively.
For example, Eq. \eqref{eq:Erequirement} implies that at ISM typical densities, $n\ll 10^4$, a solar mass rest energy can not drive a RMS. In such cases, the shock would be mediated by other mechanism e.g. collective plasma processes.

\subsubsection{Far DS conditions}
In the far DS, which is in thermal equilibrium, the conditions are completely determined by conservation of energy, momentum and particle fluxes,
\begin{align}\label{eq:udconservation}
&n_d\Gamma_d\bt_d=n_u\Gamma_u\bt_u,\cr
&4\Gamma_d^2\beta_d p_{\gamma,d}=\Gamma_u\beta_u (\Gamma_u-\Gamma_d) n_u m_pc^2,\cr
&  (4\Gamma_d^2\beta_d^2 +1) p_{\gamma,d} =\Gamma_u\beta_u(\Gamma_u\beta_u-\Gamma_d\beta_d) n_u m_pc^2,
\end{align}
where $ p_{\gamma,d}=1/3 a_{BB}T_d^4$ is the far DS radiation pressure, and where the plasma pressure in the DS was neglected.
Eqs. \eqref{eq:udconservation} can be solved for $ \beta_d $ and $ T_d $.   In the NR and ultra relativistic limits the solution reduces to the expressions
\begin{align}\label{eq:Td_NR}
 T_d & \approx \left(\frac{21 n_u \beta_u^2m_pc^2 }{8a_{BB}}\right)^{1/4}\approx 0.41 n_{u,15}^{1/4}\beta_u^{1/2} \keV ,\cr
 \beta_d & \approx \beta_u/7 ,
\end{align}
and
\begin{align}\label{eq:Td_rel}
T_{d} & \approx \left(\frac{2\Gamma_u^2n_um_pc^2}{a_{BB}} \right)^{1/4} \approx0.385\Gamma_{u}^{1/2}n_{u,15}^{1/4}\,\keV,\cr
 \beta_d & \approx 1/3,
\end{align}
respectively, where $n_{u}=10^{15}n_{u,15}\cm^{-3}$. Note that the condition for a radiation dominated DS, Eq. \eqref{eq:beta_domination}, can be obtained by comparing $ T_d $ with $ \varepsilon $.

\subsection{Non relativistic RMS}\label{sec:NR-RMS}
We next focus on NR RMS. By non relativistic shocks we refer to shocks in which neither the protons nor the electrons move with relativistic bulk or thermal velocities throughout the shock. In particular, this implies that the temperature is always much smaller than $m_e c^2$. 

\subsubsection{Assumptions}\label{sec:assum_nr}
The discussion below of NR RMS is valid under the following assumptions \citep{Weaver76}.
\begin{itemize}
\item The electron fluid and the ion fluid move together with the same velocity. This is justified by the presence of collective plasma modes. In the simplest case of protons and electrons, an electrostatic field is sufficient to couple the fluids.
\item The pressure is dominated by radiation throughout the shock transition. This is justified at the end of this sub-section.
\item For typical photons, the optical depth is dominated by Compton scattering. This, combined with the low velocity of the flow implies that the diffusion equation can be used to approximate the spatial transport of the radiation.
\item The Compton $ y $ parameter is much larger than 1 throughout the flow, implying that the energy density is dominated by a component having a Wien spectrum, and that the electron energy spectrum is close to a Maxwellian, with approximately the same temperature. The radiation is well described by two parameters, the temperature and the density of photons  $ n_{\gamma,\text{eff}} $ in the Wein-like component. Note, that a large  Compton $ y $ parameter is sufficient to ensure that the electrons are strongly coupled to the radiation since the radiation dominates the thermal energy density.
\item The main source of photon production is thermal bremsstrahlung.
\end{itemize}

\subsubsection{The shock transition width}
\label{sec:rms-width}

\begin{figure}\label{fig:NR_scheme}
\centering
\includegraphics[scale=0.4]{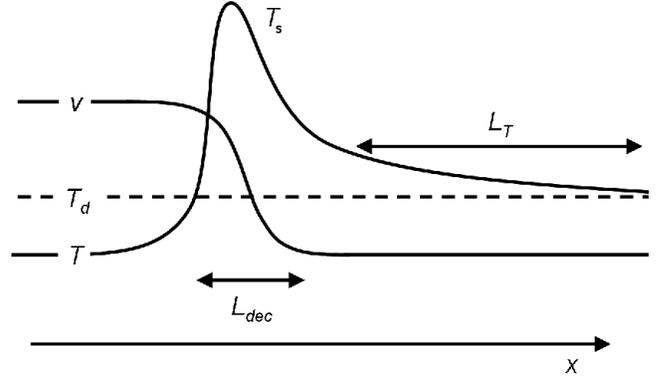}\caption{A schematic description of the structure of a fast NR RMS, in which the radiation departs from thermal equilibrium.}
\end{figure}

Physical quantities approach their far DS equilibrium values on length scales, which may vary by  orders of magnitude for different quantities. In particular, as explained below, the transition width of the velocity is determined by Compton scattering and occurs on length scales, which may be much smaller than the temperature transition width, which is determined by photon production.
A schematic cartoon of the velocity and temperature profiles of NR RMS is shown in fig.~\ref{fig:NR_scheme}.
\paragraph{Velocity transition}
For NR RMS the width $L_{dec}$ of the velocity transition region (see fig. \ref{fig:NR_scheme}) is comparable to the distance $ L_{diff}\sim (\beta_u n_e \sigma_T)^{-1} $ over which a photon can diffuse against the flow before being advected with the flow.
To see that the velocity transition width can not be larger,
note that once a proton reaches a point in the shock where the energy density is dominated by photons, it experiences an effective force
\begin{equation}\label{eq:protondrag}
 \bt_u\frac{d\bt}{dx}m_p c^2\sim \sig_T\bt_u e_\gamma \sim \sig_T n_u\bt_u^3m_pc^2,
\end{equation}
implying a deceleration length of
\begin{equation}\label{eq:ShockWidth}
L_{\text{dec}}\equiv\bt_u \left(\frac{d\bt}{dx}\right)^{-1}\sim \inv{\sig_Tn_u\bt_u}.
\end{equation}
The drag estimated in equation \eqref{eq:protondrag} is unavoidable due to the fact that
once the photons dominate the pressure, they cannot drift with the protons, as this will imply a radiation energy flux greater than the total energy flux.

\paragraph{Thermalization length}
The region of the shock profile over which the temperature changes before it reaches $T_d$ can be extended to distances that are much larger than $L_{diff}$.
To see this, consider the length scale that is required to generate the density of photons of energy $\sim T_d$ in the DS, determined by thermal equilibrium, $n_{\gamma,\text{eq}}\approx p_{\gamma,d}/T_d$,
\begin{equation}\label{eq:LTdefinition}
L_{T}\sim \bt c\frac{n_{\gamma,\text{eq}}}{Q_{\gamma,\rm eff}},
\end{equation}
where $Q_{\gamma,\rm eff}$ is the \textit{effective} generation rate of photons of energy $3T_d$. We use here the term "\textit{effective} generation rate" due to the following important point. Photons that are produced at energies $\ll T_d$ may still be counted as contributing to the production of photons at $T_d$, since they may be upscattered by inverse-Compton collisions with the hot electrons to energy $\sim T_d$ on a time scale shorter than that of the passage of the flow through the thermalization length, $L_T/\beta_d c$.
The Bremsstrahlung effective photon generation rate is  given by
\begin{equation}\label{eq:QgammaBr}
Q_{\gamma,\text{eff}}=\al_e n_pn_e\sig_T c\sqrt{\frac{m_ec^2}{T}}\Lm_{\text{eff}}g_{\text{eff}},
\end{equation}
where $g_{\text{eff}}$ is the Gaunt factor, $\Lm_{\text{eff}}\sim\log[T/(h\nu_{\min})]$  and $ \nu_{\min} $ is the lowest frequency of photons emitted by the plasma which may be upscattered to $ 3T_d $ prior to being absorbed (absorption is dominated by Bremsstrahlung self absorption for far DS values).

The resulting thermalization length is
\begin{equation}\label{eq:LT}
L_T\bt_un_u\sig_T \sim \inv{100\al_e\Lm_{\text{eff}}g_{\text{eff}}}\frac{\vep^2}{\sqrt{m_e c^2 T_d}m_pc^2}.
\end{equation}
This implies that for high shock velocities,
\begin{equation}\label{eq:btNeq}
\bt_u>0.07 n_{15}^{1/30}(\Lm_{\text{eff}}g_{\text{eff}})^{4/15},
\end{equation}
the length required to produce the downstream photon density is much larger than the deceleration scale. For lower shock velocities, thermal equilibrium is approximately maintained throughout the shock.

\subsubsection{Description of the shock structure}
\label{sec:NRShockStructure}

An analytic expression for the velocity, density and pressure profile can be found under the diffusion approximation \citep[e.g.][]{Weaver76}. In particular, the velocity $ \beta c $ at a give position $ x $ along the shock satisfies
\begin{equation}\label{eq:analyticSolution}
x=\frac{1}{21\sig_Tn_u\bt_u}\ln\left[\frac{(\bt_u-\bt)^7}{(7\bt-\bt_u)\bt_u^6}\right].
\end{equation}

The shape of the temperature profile is largely determined by the photon production in one diffusion length into the DS (the first $\beta_d^{-1}$ optical depths of the downstream region, henceforth the immediate DS). In this region the photons mediating the shock are produced. If a photon density of $ \sim a_{BB}T_d^3/3 $ is produced ($L_T\lesssim L_{diff} $), the flow will stay close to thermal equilibrium, and the temperature profile, which can be extracted directly from the analytic pressure profile, essentially follows the velocity profile. 
Otherwise, when $L_T\gg L_{diff} $, the velocity transition of the shock ends without reaching the far DS equilibrium temperature. The radiation pressure $n_{\gamma,\text{eff}} T$ reaches its DS value as soon as the velocity is close to the DS velocity. Down stream of this region, the density of photons $n_{\gamma,\text{eff}}$ grows with distance as more and more photons are being generated and advected with the flow and saturates at the equilibrium  black body photon density $\approx a_{BB}T_d^3/3 $ . Accordingly, $T$ is decreasing throughout the downstream. In this case, we can broadly divide the shock structure into four separate regions.
\begin{enumerate}
\item Near upstream: A few diffusion lengths, $(\bt_s\sig_T n_u)^{-1}$, upstream of the deceleration region. In this region, characterized by velocities that are close to the upstream velocity, $\bt\approx \bt_u$, and temperatures  $T\gg T_u$, the temperature changes from $T_u$ to $\sim T_{s}$. It ends when the fractional velocity decrease becomes significant.
\item Deceleration region: A $(\bt_s\sig_T n_u)^{-1}$ wide region where the velocity changes from $\bt_{u}$ to $\bt_{d}$ and the temperature is roughly constant, $T\simeq T_s$.
\item Immediate downstream:  Roughly a diffusion length, $(\bt_s\sig_T n_u)^{-1}$, downstream of the deceleration region. In this region, characterized by velocities close to the downstream velocity, $\bt\approx\bt_d$, and temperature $T\sim T_s$, the photons that stop the incoming plasma are generated. Upstream of this region $\bt>\bt_d$ and the photon generation rate is negligible. Photons that are generated downstream of this region are not able to propagate up to the transition region. To estimate the temperature value in the immediate DS, $ T_s $, the number of photons produced in the immediate DS by Bremsstrahlung and up-scattered by inverse Compton should be equated to the number of photons required to carry the pressure at that point.
 The production rate,  given by Eq. \eqref{eq:QgammaBr}, combined with diffusion and conservation laws,  leads to the following estimate of the immediate DS temperature for NR RMS \citep{Katz09}
\begin{align}\label{eq:btOfT}
\bt_u=\frac{7}{\sqrt{3}}\left(\inv{2}\al_e \Lm_{\text{eff}}g_{\text{eff}}\right)^{1/4}\left(\frac{m_e}{m_p}\right)^{1/4}\left(\frac{T_s}{m_ec^2}\right)^{1/8} \cr
\approx 0.2\Lm_{\text{eff},1}^{1/4}\left(\frac{g_{\text{eff}}}{2}\right)^{1/4}\TskkeV^{1/8},
\end{align}
where $ \Lm_{\text{eff},1}=10 \Lm_{\text{eff}} $. This result is  in agreement with the numerical results of \citet{Weaver76}.

\item Intermediate downstream: The region in the downstream where most of the far downstream photons are generated and $T$ changes from $T_s$ to $T_d$. This region has a width $L_T$ given by Eq. \eqref{eq:LT}, much grater than $(\bt_s\sig_T n_u)^{-1}$. Thus, diffusion within this region can be neglected. The temperature profile is expected to follow $T\propto x^{-2}$. To see this, note that the photon density at a distance $x$ from the shock is proportional to the integral of the photon generation, $n_{\gamma,\text{eff}}\propto T^{-1/2}x$. Since the photon pressure equals the downstream pressure, we have $n_{\gamma,\text{eff}}\propto T^{-1}$ and $T\propto x^{-2}$ (this is valid for a constant value of $\Lm_{\text{eff}}g_{\text{eff}}$ and is somewhat shallower in reality). Using this dependence of the temperature on distance, the thermalization length can shown to be related to the deceleration length by $L_T\bt_sn_u\sig_T\sim \Lm_{\text{eff}}g_{\text{eff}}|_dT_d^{-1/2}[(\Lm_{\text{eff}}g_{\text{eff}})|_sT_s^{-1/2}]^{-1}$, in agreement with equations \eqref{eq:LT} and \eqref{eq:btOfT}.
\end{enumerate}

\subsubsection{Scaling of the profile with density}\label{subsec:low_density}
The velocity, density and pressure profiles of RMS (as a function of optical depth $\tau=\sig_T x n_u$) are independent of the upstream density. The scaling of the temperature depends on whether or not thermal equilibrium is sustained. In case it is (if $L_T\ll L_{diff} $), the temperature scales with density as $T\propto n_u^{1/4}$. Alternatively, when $L_T\gg L_{diff}$, in the shock regions where the temperature is much higher than its equilibrium value, the temperature profile does not scale with density, $T\propto n_u^0$, see for example eq.~(\ref{eq:btOfT}). To see this, note that in these regions Bremsstrahlung absorption is negligible, while Compton scattering and Bremsstrahlung emission are both two body processes that scale similarly with $n_u$. The conservation and radiation transfer equations are invariant under the scaling of the radiation intensity, densities and length scales across the shock by $ n_u^1$ ,$n_u^1 $ and $n_u^{-1}$ respectively. This scaling is shown explicitly later, in section \sref{sec:equations}. Bremsstrahlung absorption may still be important at low frequencies and affect the structure through logarithmic corrections to the effective photon production rate. In the far downstream, where $T$ approaches its equilibrium value, absorption will no longer be negligible.

\subsubsection{From radiation domination to radiation mediation}\label{sec:dom_to_med_NR}
For shocks satisfying Eq. \eqref{eq:beta_domination}, the far DS pressure is dominated by radiation. It is not a priory trivial that the pressure is dominated by radiation in the velocity transition region of such shocks since photons that are generated in the DS are able to diffuse upstream over a finite distance only. We next illustrate that under a wide range of conditions, the radiation does indeed dominate the pressure in the velocity transition region.

Consider a hypothetical shock, having a DS energy density dominated by radiation, in which the velocity transition is mediated by some mechanism other than radiation. In the absence of radiation, the temperature immediately behind the velocity transition  would be $ T\sim \varepsilon=0.5\bt_u^2 m_pc^2 $. Photons generated in this region can diffuse upstream to a characteristic distance of $ L_{diff} > (\beta n_e\sigma_T)^{-1} $. Under these assumptions, the energy in photons that are produced by Bremsstrahlung is much larger than the available thermal energy,
\begin{equation}\label{eq:dom_to_med_cond_NR}
\frac {e_{\gamma}}{e_{th}}\approx \frac {Q_{Br}TL_{diff}}{\beta c n_p \varepsilon } > \frac{\alpha_e  }{\beta^2} \sqrt{\frac {m_ec^2}{\varepsilon}}\approx 8\alpha_e \frac{m_p}{m_e}\left(\frac{m_ec^2}{\varepsilon} \right)^{3/2} \gg 1,
\end{equation}
where $ Q_{Br} $ is the photon production rate  by Bremsstrahlung at energy $ \sim T $, and where we used $\beta =\beta_u/4 $, appropriate for NR shocks which are not radiation mediated and $\vep\sim T\ll m_ec^2$. Note, that by definition, the condition \eqref{eq:dom_to_med_cond_NR} is roughly equivalent to demanding that the temperature $T_s$ in \eqref{eq:btOfT} be smaller than $\varepsilon$. 

This implies that a shock with a radiation dominated DS and negligible radiation in the velocity transition region cannot exist if the transition region is smaller than $L_{diff}$. Once the pressure is dominated by photons, they will also mediate the shock.

\subsection{Relativistic RMS}\label{sec:RRMS-phys}

\subsubsection{Assumptions}
\label{sec:RMS-assum}

Throughout this paper we make the following assumptions for RRMS:

\begin{enumerate}
\item The electrons, positrons and ions move as a single fluid with the same velocity. This is motivated below by the presence of collective plasma instabilities. 
\item The electron and positron velocity distributions in the rest frame are approximately thermal.  This assumption is justified by the intense radiation field interacting with the electrons and positrons, which quickly eliminates large deviations from the mean velocity.
\item The ions have a negligible contribution to the pressure.
\item The radiation mechanisms dominating the shock are Compton scattering, bremsstrahlung emission and absorption and two photon pair production and pair annihilation.
\end{enumerate}

{
The assumption of a single plasma velocity is motivated by the fact that the plasma time ($ t_{pl} $) is much shorter than the mean time between Compton scatterings ($ t_c $) of an electron, allowing for collective plasma processes to isotropize the velocities of the particles. Indeed, the ratio of these timescales,
\begin{equation}\label{eq:t_pl}
    \frac{t_{pl}}{t_{scat}}=\frac{n_{\gamma}\sigma_{T}c}{\omega_{pl}}=
    \frac{n_{e}\frac{n_{\gamma}}{n_{e}}\sigma_{T}c}{\sqrt{\frac{4\pi     
    n_{e}e^{2}}{m_{e}}}}\approx10^{-9}n_{e,19}^{1/2}\frac{n_{\gamma}}{n_{e}},
\end{equation}
where $n_{e}=10^{19}n_{e,19}\cm^{-3}$, is much lower than unity given that $n_{\gamma}/n_{e}$ is not
very large, see \sref{sec:immDS_subson}.
}

The second assumption we make regarding the plasma - the existence of an effective temperature, is somewhat more subtle. In principle, the electrons and positrons can have a general distribution function.
However, most of the shock is characterized by a strong dominance of radiation energy density over particle thermal energy density. This leads to the electrons and positrons being "held" in momentum space by the radiation, since each scattering changes the energy of the electron considerably, if it departs significantly from the average photon energy. The only way to maintain a very non-thermal electron spectrum is by having a radiation spectrum which is not dominated by a typical photon energy, e.g. a power law. Our numerical results show that the radiation energy density is dominated by photons of limited energy range in the rest frame of the plasma, and that when a high energy photon tail appears, the photons populating this tail have a very low cross section for interaction with electrons or other photons. This supports the assumption that the energy distribution of electrons and positrons may be characterized by some typical "thermal" energy, greatly simplifying the calculations.

A note is in place here regarding Coulomb collisions.
The effective cross section for Coulomb collisions of electrons on protons is $ \sigma\sim e^4/\varepsilon_k^2 $, where $ \varepsilon_k $  is the electron kinetic energy. When the energy of the electron is of the order of $ m_ec^2 $, the cross section is similar to $ \sigma_T $, the Thomson cross section. This implies that Coulomb collisions in RRMS play a marginal role in equilibrating the motion of particles in the plasma, as the photon density inside the shock is typically of the order of the electron density. Unlike plasma instabilities, this process can not account for the equilibration of the distribution function of the particles. At low energies, i.e. NR RMS, Coulomb collisions may become dominant (see Weaver 1976) due to a much larger effective cross section.


\subsubsection{Velocity and temperature transition regions' widths}
\paragraph{Velocity transition}

The line of arguments presented in \sref{sec:NR-RMS} for estimating the velocity transition width can not be directly extended to relativistic shocks since KN corrections to the Compton scattering cross section depend on the a priori unknown photon frequency and plasma temperature, which vary throughout the transition region. Note, that as expalined above, pair production and relativistic corrections to the cross sections become important already at non relativistic upstream energies $\varepsilon=\beta_u^2m_pc^2/2\sim100$~MeV, since the temperature of the plasma within the deceleration region reaches a considerable fraction of $m_{e}c^{2}$ for this value of $\varepsilon$. The production of pairs also changes the simple estimate, since it changes both the scatterers' number density and the shock optical depth. Finally, an additional complication is introduced by the strong dependence of the scattering mean free path on the photon's direction of propagation, expected due to the relativistic velocity of the plasma.

\paragraph{Thermalization length}
The thermalization length can be estimated in a way similar to the NR case, since the width of the temperature transition is much larger than the deceleration width [see Eq. \eqref{eq:btNeq}], and the scale is set by the lowest temperature, i.e. $ T_d $, which is non relativistic. The thermalization then takes place over
\begin{equation}
\sim  \inv{100\al_e\Lm_{\text{eff}}g_{\text{eff}}}\frac{\vep^2}{\sqrt{m_e c^2 T_d}m_pc^2}
\end{equation} 
Thomson optical depths. Since $ \varepsilon=(\Gamma_u-1)m_pc^2 $ and $ T_d $ is many orders of magnitude smaller, this width is always very large in terms of Thomson optical depths. 
 
\subsubsection{Immediate DS}\label{sec:immDS_subson}

We next give a rough estimate of the average temperature in the first few optical depths of the immediate DS of RRMS \citep{Katz09}. The assumption we use is that the electron-positron pairs and the radiation are in Compton Pair Equilibrium (CPE). This assumption is valid since the velocity is $ \lesssim c/3 $, and since the $ y $ parameter arising from mildly relativistic temperatures is large, as shown below. The numerical calculations are not based on this assumption, and its self consistency is discussed in \sref{sec:simplified-im-ds}. Following the NR RMS analysis, $T_s$ is estimated by equating the number density of photons produced by Bremsstrahlung and by inverse Compton emission of thermal pairs with the number density of photons needed to carry the energy flux at the end of the deceleration region.

Assuming that the number density of pairs is much larger than that of protons, and neglecting Double Compton emission, the ratio of photon to electron-positron number densities may be written as
\begin{equation}\label{eq:MainEquation_rel}
\frac{n_{\gamma,\text{eff}}}{n_l}=\inv{3}\al_e\Lm_{\text{eff}} \bar g_{\text{eff,rel}}(\hat T)\bt_d^{-2},
\end{equation}
where $ n_{\gamma,\text{eff}} $ is the density of photons in the Wein-like component (see \sref{sec:assum_nr})  and  the free-free emission is written in the form
\begin{equation}\label{eq:QffRel}
Q_{\gamma,\text{eff}}=\al_e \sig_T c n_l^2 \Lm_{\text{eff}}\bar g_{\text{eff,rel}}(\hat T).
\end{equation}
Here $\bar g_{\text{eff,rel}}$ is the total Gaunt factor [defined by Eq.
\eqref{eq:QffRel}] including all lepton-lepton Bremsstrahlung emission. For $10<\Lm_{\text{eff}}<20$ and
$60\keV<T<m_ec^2$, the approximation
\begin{equation}\label{eq:fBrApp}
\bar g_{\text{eff,rel}}\approx\Lm_{\text{eff}}/2
\end{equation}
agrees with the results of \citet{Svensson84} to an accuracy of better than $25\%$. At these high temperatures, the Compton $y$ parameter is large and
radiative Compton emission is negligible. Substituting Eq.~\eqref{eq:fBrApp} in
Eq.~\eqref{eq:MainEquation_rel} we find
\begin{equation}\label{eq:PhotonGenerationRel}
\frac{n_{\gamma,\text{eff}}}{n_l}\approx 2.5 \left(\frac{\Lm_{\text{eff}}}{15}\right)^2\left(3\beta_d \right)^{-2}.
\end{equation}
In the regime $200\keV<T<m_ec^2$, pair production equilibrium is approximately given by
\begin{equation}\label{eq:EquilibriumHighT}
n_\gamma/n_l\approx 0.5 m_ec^2/T.
\end{equation}
Comparing equations \eqref{eq:PhotonGenerationRel} and \eqref{eq:EquilibriumHighT}, we see that if ${T\gtrsim 200\keV}$
there would be too many photons generated per lepton. Much lower temperatures lead to insufficient photon production, as can be deduced from the NR case. We conclude that for
relativistic shocks, 
\begin{equation}\label{eq:Ts_rel}
    T_s \sim 200\keV.
\end{equation}
The weak dependence of the immediate DS temperature on parameters is due to the rapid increase of pair density with $T$ at $ T\sim m_ec^2 $. 

\paragraph{Subsonic region}  The US flow is "super-sonic", in the sense that the plasma velocity, $\beta_u c$, is larger than the plasma speed of sound, $\beta_{ss}c$. The production of a large number of pairs in the immediate DS, $n_+/n_p\gg1$, and the heating of the plasma at this region to relativistic temperatures, $T_s \sim 0.4 m_ec^2$, implies a "sub-sonic" flow, $\beta_{ss}>\beta$, in the immediate DS. The large number of pairs implies that the average plasma particle mass is close to $m_e$, for which the temperature, $T_s \sim 0.4 m_ec^2$, gives a speed of sound which is close to it's highly relativistic value, $\beta_{ss}=1/\sqrt 3$ (see appendix \sref{sec:sound_speed} for a detailed calculation of the speed of sound). $\beta_{ss}=1/\sqrt 3$ is larger than the plasma velocity in the immediate DS, which is close to its far DS value, $ \beta_d \le 1/3 $. 

Note, that we are referring here to the plasma speed of sound neglecting the (dominant) contribution of the radiation to the pressure. This speed of sound describes the propagation of (small) disturbances in the plasma on length (time) scales which are short compared to the mean free path (time) for electron-photon collisions. As explained in \S~\ref{sec:RMS-assum}, see eq.~(\ref{eq:t_pl}), collective plasma modes are expected to lead to a fluid like behavior of the plasma on length and time scales much shorter than the electron-photon collision mean free path.

In the far DS, the flow becomes super-sonic again, $\beta_{ss}<\beta_d$. This implies that for relativistic shocks the flow crosses two sonic points, accompanied by singularities of the differential conservation equations [Eqs. \eqref{eq:en_flux}, \eqref{eq:mom_flux}]:
\begin{itemize}
\item At the first sonic point, the flow changes from supersonic to subsonic. This is a hydrodynamically unstable point { which results in a hydrodynamic shock.
A steady state hydrodynamic flow can not smoothly cross a sonic point going from supersonic to subsonic velocities because downstream of the sonic point, upstream going characteristics converge to the sonic point \citep[e.g.][]{Zel'dovich66}, infinitely steepening a continuous profile at that point and resulting in a shock. 
We show below that while most of the deceleration of the plasma is continuous, a (sub-)shock across which the velocity jump is small, $\delta(\Gamma\beta)\sim0.1$ (see fig.~\ref{fig:struct_g_DS.eps}), is indeed required to exist at the end of the deceleration region. This sub-shock must be mediated by the same processes that are assumed to isotropize the particles' velocities in the fluid rest frame  on a scale much shorter than the radiation mean free path [e.g. plasma instabilities, see eq.~(\ref{eq:t_pl})].}
\item At the second sonic point the flow passes from a subsonic to a supersonic region. This is a stable point which has no special significance, and is simply part of the thermalization tail of the shock. 
\end{itemize}

\subsubsection{Structure}

The structure of RRMS differs from that of NR RMS. The main differences are:
\begin{itemize}
\item The deceleration length is much larger than the naive estimate: The length, measured in Thomson optical depths of $ e^- $ $ e^+ $, grows with the upstream Lorentz factor $ \Gamma_u $ in a manner faster than linear (for NR RMS it is $ \sim \beta_u^{-1} $);
\item Pair production has a significant contribution to the deceleration of the plasma.
\item As explained above, a "hydrodynamic" sub-shock (possibly mediated by plasma instabilities) across which the velocity jump is small, $\delta(\Gamma\beta)\sim0.1$, is required to exist at the end of the deceleration region;
\item The radiation is highly anisotropic, and exhibits a high energy tail with a typical cutoff energy of $ \sim \Gamma_u^2 m_ec^2$.
\end{itemize}

\subsubsection{From Radiation domination to radiation mediation}

 Expanding the reasoning given in \sref{sec:dom_to_med_NR} for NR RMS, we argue here that relativistic shocks which are radiation dominated, i.e. in which the DS energy density is dominated by radiation, must also be radiation mediated. Let us assume the contrary, i.e. that the energy density in the deceleration region is not dominated by radiation and that deceleration is therefore not mediated by radiation. In this case, the plasma reaches a temperature $\sim \Gamma_u m_p c^2$ at the end of the deceleration region, and then gradually thermalizes as it flows further into the DS. Since $T_d<m_ec^2$ (see eq.~\ref{eq:Td_rel}), let us consider the point in the downstream where the temperature reaches $T\sim10$~MeV. Since the velocity at this point already reached its DS value, $\beta\sim \beta_d \le 1/3$, photon transport is well described in this region by the diffusion approximation, with diffusion length $ L_{diff}\approx 3 (n_e\sigma_T)^{-1} $. An electron crossing this diffusion length produces a large number of $ \sim 10 \MeV $ photons,
\begin{align}
\frac{n_{\gamma,10}}{n_e}\approx \frac{L_{diff}}{n_e \beta c}Q_{\gamma,eff}\approx 
\frac{ \alpha_e \bar g_{ff,rel} \Lambda_{eff}}{\beta^2} \sim \cr 10 \frac{\bar g_{ff,rel}(10\MeV)\Lambda_{eff}}{100} (3\beta_d)^{-2}
\end{align}
(using conservative estimates for the Gaunt factor and the logarithmic correction). 
{ This ratio is much larger than its CPE value  $ n_{\gamma,\text{eff}} \sim n_e $, expected at $ T\gg m_ec^2 $.
For such a high ratio of photons to electrons, a photon will produce a pair on another photon on a time scale much shorter than its scattering time scale. Such a deviation from equilibrium on a length scale $ \sim 3 $ scattering optical depths is not self consistent.} We conclude therefore that the temperature can not significantly exceed $ m_ec^2 $ at the point where the deceleration is complete. This implies, in turn, that most of the energy at the end of the deceleration must be carried out by radiation.

\section{RMS equations and boundary conditions}
\label{sec:formulation}
{bf structural change in this section. Intro added.}
In this section we write down the equations of RMS that are numerically solved based on assumptions 1-4 given in \sref{sec:RMS-assum}.
In \sref{sec:equations} we write down the hydrodynamic and radiation transfer equations in physical and dimensionless form and define the variables that are solved for. In \sref{sec:rad_mech_detail} we provide expressions for radiation scattering (Compton), production and absorption (Bremsstrahlung) and pair production and annihilation. A summary of all the equations in dimensionless form is given in \sref{sec:eq_summary}. The boundary conditions are described in \sref{sec:bound_cond}.

\subsection{Hydrodynamic and radiation transfer equations}
\label{sec:equations}
The equations governing the structure of a steady planar shock propagating along the $z$ direction are
\begin{equation}
\frac{d}{dz_{sh}}T_{sh}^{0z}=0\label{eq:en_flux},\end{equation}
\begin{equation}
\frac{d}{dz_{sh}}T_{sh}^{zz}=0\label{eq:mom_flux},\end{equation}
\begin{equation}
n_{p}=n_{p,u}\frac{\Gamma_{u}\beta_{u}}{\Gamma\beta},\label{eq:particle_cons}\end{equation}
\begin{equation}
\frac{d(\Gamma\beta n_{+})}{dz_{sh}}=\frac{Q_{+}}{c},\label{eq:pair_num}\end{equation}
\begin{equation}
\mu_{sh}\frac{dI_{\nu_{sh}}(\mu_{sh})}{dz_{sh}}= \eta_{sh}(\mu_{sh},\nu_{sh})
 -I_{\nu_{sh}}(\mu_{sh})\chi_{sh}(\mu_{sh},\nu_{sh})\label{eq:transfer}.
\end{equation}
The first 3 eqs. describe the conservation of energy, momentum and proton number. The forth eq. describes the production and annihilation of positrons, and the fifth eq. describe the transport of photons. $z_{sh}$ is the shock frame distance along the shock propagation direction, $\beta c$ is the plasma velocity in the shock frame,
$\Gamma=1/\sqrt{1-\beta^2}$ is the corresponding Lorentz factor, $n_{p}$ is the proper proton density, $n_{+}$ is the proper  positron density, $I_{\nu_{sh}}(\mu_{sh})$ is the shock frame specific intensity at (shock frame) frequency $\nu_{sh}$ and direction $\mu_{sh}=\cos\theta_{sh}$ ($\theta_{sh}$ is the azimuthal angle with respect to $z$), and $T_{sh}^{\alpha\beta}$ is the shock frame energy-momentum tensor.
 $Q_{+}=\partial n_{+}/\partial t$
is the net positron production rate (production minus annihilation), which is frame independent.
 $\eta$ and   $\chi$ are the emissivity and absorption coefficients, respectively, and are
functions of the plasma parameters and of the local radiation field described in \sref{sec:rad_mech_detail}. We use $\{\nu,I_\nu,\mu,\eta,\chi\}$ to denote quantities measured in the plasma rest frame, and add a subscript "{\it sh}" to denote values of these quantities measured in the shock frame.

\begin{figure}\center
\includegraphics[scale=0.7]{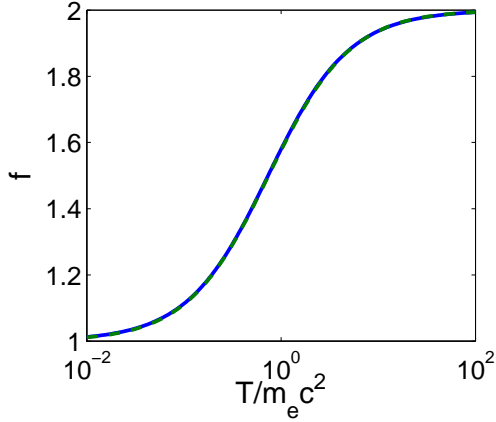}\caption{A comparison of the exact value of $ f(T ) $ (solid line), calculated numerically for a Maxwellian distribution, and the approximation given by Eq.~\eqref{eq:f_eos} (dashed line).}
\label{fig:f32}
\end{figure}

The energy and momentum are carried by the plasma and the radiation, 
\begin{equation}
T^{\alpha\beta}_{sh}=T^{\alpha\beta}_{sh,pl}+T^{\alpha\beta}_{sh,rad},
\end{equation}
where the subscripts $ pl $ and $ rad $ refer to the plasma and radiation contributions respectively. 
The radiation part of $T_{sh}^{\alpha\beta}$, $ T^{0z}_{sh,rad}=F_{rad,sh} $ and  $ T^{zz}_{sh,rad}=P_{rad,sh} $, is given by 
\begin{equation}\label{F_def}
F_{rad}=\int d\Omega \mu d\nu I_{\nu}(\mu),
\end{equation}
\begin{equation}\label{P_def}
P_{rad}=c^{-1}\int d\Omega \mu^2 d\nu I_{\nu}(\mu)
\end{equation}
(see appendix D for rules of transformation between rest frame and shock frame measured quantities). As mentioned in the introduction, we assume that the protons, electrons and positrons may be described as a fluid of single velocity $c\beta(z_{sh})$, that the energy distribution of the electrons and positrons is thermal, with temperature $T(z_{sh})$, and that the protons are cold. Under these assumptions,
\begin{equation}
T^{0z}_{pl,sh}=\Gamma^2\beta\left(e_{pl}+P_{pl} \right) ,
\end{equation}
and
\begin{equation}
T^{zz}_{pl,sh}=P_{pl}+\Gamma^2\beta^2\left(e_{pl}+P_{pl} \right) ,
\end{equation}
where the proper energy density $ e_{pl} $ and pressure $ P_{pl} $ are given by
\begin{equation}\label{eq:epl}
e_{pl}=n_pm_pc^2+\left(n_e+n_+ \right)m_ec^2+\frac 32f\left(T \right) \left(n_e+n_+ \right)T ,
\end{equation}
and
\begin{equation}\label{eq:ppl}
P_{pl} =\left(n_e+n_+ \right)T 
\end{equation}
(note, that we neglected the thermal pressure of the cold protons). $ f(T) $ is dimensionless and is approximated by the following interpolation between the NR ($ f=1 $) and relativistic ($ f=2 $) values,
\begin{equation}\label{eq:f_eos}
f(T)=\frac 12 \tanh\left( \frac{\ln(T/m_ec^2)+0.3}{1.93}\right) +\frac 32.
\end{equation}
This approximation describes the equation of state of Maxwell-Boltzmann distributed plasmas to an accuracy better than $\sim 2 \times 10^{-3}$ for all temperatures, as shown in fig. \ref{fig:f32}.

\subsubsection{Dimensionless equations}

We define the following dimensionless quantities:
\begin{align}\label{eq:dimless_def}
 \hat{T} & =  \frac{T}{m_{e}c^{2}},\cr
 \hat{\nu} & =  \frac{h\nu}{m_{e}c^{2}},\cr
 x_+ & = n_+/n_p,\cr
 \hat{z}_{sh} & =  \Gamma_{u}n_{u}\sigma_{T}z_{sh},\cr
 d\tau_{*} & =  \Gamma(1+\beta)(n_{e}+n_{+})\sigma_{T}dz_{sh},\cr
 \hat I & =  \frac{I}{\Gamma_{u}^{2}\beta_{u}n_{u}(m_{p}/m_{e})hc}.
\end{align}
With these definitions, and using the explicit forms of $T_{sh}^{\alpha\beta}$ derived above, the energy and momentum conservation equations take the form
\begin{align}\label{eq:scaled_en}
& \frac{\Gamma}{\Gamma_{u}}\left\{ 1+\left(1+2x_{+}\right)\frac{m_{e}}{m_{p}}\left[1+\hat{T}\left(1+\frac{3}{2}f(\hat T)\right)\right]\right\}
 +\cr & + 2\pi\hat{F}_{rad,sh}= 1+\frac{m_{e}}{m_{p}},
\end{align}
\begin{align}\label{eq:scaled_mom}
& \frac{\Gamma\beta}{\Gamma_{u}\beta_{u}}\left\{ 1+\left(1+2x_{+}\right)\frac{m_{e}}{m_{p}}\left[1+\hat{T}\left(\frac{1}{\left(\Gamma\beta\right)^{2}}+1+\frac{3}{2}f(\hat T)\right)\right]\right\}
 +\cr & +\frac{1}{\beta_{u}}2\pi\hat{P}_{rad,sh}= 1+\frac{m_{e}}{m_{p}},
\end{align}
where
\begin{equation}
\hat{F}_{rad,sh}=\frac{F_{rad,sh}}{2\pi\Gamma_{u}^{2}\beta_{u}n_{u}m_{p}c^{3}},
\end{equation}
\begin{equation}
\hat{P}_{rad,sh}=\frac{cP_{rad,sh}}{2\pi\Gamma_{u}^{2}\beta_{u}n_{u}m_{p}c^{3}},
\end{equation}
are the scaled energy and momentum fluxes of the radiation field. 

The transfer equation, eq. \eqref{eq:transfer}, takes the form 
\begin{align}\label{eq:transfer_scaled}
\mu_{sh}\frac{d\hat I_{\nu_{sh}}(\mu_{sh})}{d\tau_{*}}= \hat\eta_{sh}(\mu_{sh},\hat \nu_{sh})
 -\hat I_{\hat \nu_{sh}}(\mu_{sh})\hat\chi_{sh}(\mu_{sh},\hat\nu_{sh}).
\end{align}
The emissivity and absorption coefficients are the sum of the contributions due to the various processes considered
\begin{equation}
\hat \eta_{tot} (\mu ,\hat\nu)=\sum \hat\eta_{proc}(\mu ,\hat\nu),
\end{equation}
\begin{equation}
\hat \chi_{tot} (\mu ,\hat\nu)=\sum \hat \chi_{proc}(\mu ,\hat\nu).
\end{equation}
The transformation relations for the scaled emissivity and absorption are
\begin{equation}
\hat \eta=\frac{\eta}{\Gamma(1+\beta)\sigma_T(n_e+n_+)}\frac{m_e}{m_p\Gamma_u^2\beta_un_uhc},
\end{equation}
\begin{equation}
\hat \chi=\frac{\chi}{\Gamma(1+\beta)\sigma_T(n_e+n_+)}.
\end{equation}

Finally, the equation describing the evolution of pair density may be written as
\begin{equation}\label{eq:pair_scaled}
\frac{dx_+}{d\tau_*}=\hat Q_+,
\end{equation}
where the scaled rate of pair production is
\begin{equation}
\hat Q_+=\frac{Q_+}{\Gamma^2\beta(1+\beta)n_p(n_e+n_+)\sigma_T c}.
\end{equation}

We describe next the various radiative processes included. 
\subsection{Radiation mechanisms}\label{sec:rad_mech_detail}
The radiative processes we take into account are Compton scattering, Bremsstrahlung emission and absorption and two photon pair production and annihilation. Other processes, which we neglect, do not modify the results significantly. The leading corrections are due to double Compton scattering ($ \gamma+e \rightarrow 2\gamma+e $), three photon pair annihilation
($ e^+e^-\rightarrow 3\gamma $) and pair production on nuclei. Other processes, such as muon and pion pair production and synchrotron emission, are less significant.

\subsubsection{Compton scattering}\label{sec:compton_eqs}

The contribution of Compton scattering to $\eta$ and $\chi$ is 
\begin{equation}
\eta_{s}(\mu ,\nu)=(n_e+n_+)\int d\Omega' d\nu'  \frac{d\sigma_s}{d\nu' d\Omega'} \left( \nu' , \Omega'  \rightarrow  \nu , \Omega  \right) I_{\nu'}(\Omega'),
\end{equation}
\begin{align}
&\chi_{s} (\mu ,\nu)=(n_e+n_+)\sigma_c(\nu,T) \times
\end{align}
where the total cross section,
\begin{equation}
\sigma_c ({\nu},{T})=  \int d\Omega' d\nu' \frac{d\sigma_s}{d\nu' d\Omega'} \left( \nu , \Omega  \rightarrow  \nu' , \Omega'  \right),
\end{equation}
may be written as
\begin{align}\label{eq:tot_scat_cs}
\sigma_c (\hat{\nu},\hat{T})= & \int d\Omega' d\nu' \frac{d\sigma_s}{d\nu' d\Omega'} \left( \nu , \Omega  \rightarrow  \nu' , \Omega'  \right) \cr
= & \sigma_T\frac 34 \Bigg[ \frac{1+\zeta}{\zeta^3}\left\lbrace \frac{2\zeta(1+\zeta)}{1+2\zeta}-\ln (1+2\zeta)\right\rbrace \cr & +\frac{\ln(1+2\zeta)}{2\zeta}-\frac{1+3\zeta}{(1+2\zeta)^2} \Bigg].
\end{align}
Here, $ \zeta \equiv \hat{\nu}(1+2\hat{T})  $ [see e.g. \citet{RL79}].
The normalized emissivity and absorption are
\begin{align}\label{eq:compton_em_scaled}
 \left[ \Gamma(1-\beta\mu_{sh})\right]^2\hat \eta _{s,sh}(\hat \nu_{sh},\Omega_{sh}) = \hat \eta_s(\hat \nu,\Omega)= \frac{1}{\Gamma(1+\beta)}  \cr
\times\int d\Omega' d\hat \nu'  \frac{d\tilde \sigma_s}{d\hat \nu' d\Omega'} \left(\hat \nu' , \Omega'  \rightarrow  \hat\nu , \Omega  \right)\hat I_{\hat \nu'}(\Omega') ,
\end{align}
\begin{equation}\label{eq:compton_ab_scaled}
 \left[ \Gamma(1-\beta\mu_{sh})\right]^{-1}\hat \chi_{s,sh}(\hat \nu_{sh})= \hat \chi_{s}(\hat \nu)
=\frac{1}{\Gamma(1+\beta)} \tilde \sigma_c(\hat \nu,\hat T),
\end{equation}
where $ \tilde \sigma \equiv \sigma/\sigma_T$ and the transformations between the shock frame and plasma rest frame values of $ \nu $, $ \mu $ are given in appendix \sref{sec:trtans-def}.

Since using the exact form of the differential cross section for Compton scattering greatly increases the computational resources demands, we use instead an approximation described in appendix \sref{app:compton}. In particular, we assume isotropic scattering in the rest frame of the plasma, i.e. $ d\sigma_s \left( \nu , \Omega  \rightarrow  \nu' , \Omega'  \right) $ independent of $ \Omega' $.

\subsubsection{Pair production and annihilation}

\paragraph{Pair annihilation}
The photon emission arising from annihilation of pairs has the form
\begin{equation}\label{eq:eta_annihilation}
\eta_\nu= \frac {1}{4\pi}\dot n_\nu h\nu=\frac{h\nu n_en_+\sigma_Tc f_\pm(\nu,T)r_\pm(T)}{4\pi},
\end{equation}
where $ r_\pm $ is a dimensionless function of $ T $ accounting for the rate of annihilation and $ f_\pm $ is the spectral distribution of the photons, where
\begin{equation}
\int f_\pm(\nu,T)d\nu =1.
\end{equation}
The approximation we use for $ f_\pm $ is based on the analysis of \cite{Zdziarski80}, who fits an analytic function to the results of Monte Carlo calculations. For the annihilation rate we use, based on \citet{Svensson1982a},
\begin{equation}\label{eq:pair_annihilation_rate}
r_\pm(\hat T) = \frac{3}{4}
\left[ 1 + \frac{2\hat{T}^2}{\ln \left(2 \eta_E
\hat{T} + 1.3 \right)} \right]^{-1}\;,
\end{equation}
where $ \eta_E=e^{-\gamma_E}\approx0.5616 $, and $ \gamma_E\approx0.5772 $ is Euler's constant.
The normalized emissivity is given, based on Eq. \eqref{eq:eta_annihilation}, by
\begin{eqnarray}\label{eq:pair_ann_em_scaled}
 \hat \eta_\pm(\hat \nu,\Omega) &=&\left[ \Gamma(1-\beta\mu_{sh})\right]^2\hat \eta _{\pm,sh}(\hat \nu_{sh},\Omega_{sh})  \nonumber \\ &=&
 \frac{ (x_+ +1)x_+ \hat \nu f_\pm(\hat \nu,\hat T)r_\pm(T)}{4\pi(2x_+ +1)\Gamma_u\Gamma^2 \beta(1+\beta)}\frac{m_e}{m_p}.
\end{eqnarray}
The annihilations rate in Eq. \eqref{eq:pair_num} is simply
\begin{equation}
\dot Q=-\frac 12 n_en_+\sigma_Tcr_\pm(T),
\end{equation}
and the scaled contribution to Eq. \eqref{eq:pair_scaled} is
\begin{equation}\label{eq:pair_ann_q_scaled}
\hat Q_+=-\frac{ x_+(x_++1)r_\pm(T)}{2\Gamma^2\beta(1+\beta)(1+2x_+)}.
\end{equation}

\paragraph{Pair production}
The two photon pair production contribution to the absorption in the transfer equation is
\begin{align}
\chi_{\nu,\gamma\gamma}(\mu)=\int\sigma_{\gamma\gamma}(\nu,\nu',\mu,\Omega') \times \cr \frac{I_{\nu'}(\Omega')}{ch\nu'}(1-\cos\theta_{1})
\Theta[\nu\nu'(1-\cos\theta_{1})-2\nu_{p}^{2}]d\Omega'd\nu',
\end{align}
where $ \theta_{1} $ is the angle between $ \mu $ and $ \Omega' $.
The scaled absorption can be written as
\begin{align}\label{eq:pair_prod_abs_scaled}
\hat\chi_{\hat\nu,\gamma\gamma}(\mu)=\frac{\Gamma_{u}\beta (m_{p}/m_{e})}{(1+\beta)(2x_+ +1)} \int\tilde \sigma_{\gamma\gamma}(\hat \nu,\hat \nu',\mu,\Omega')\times \cr \frac{\hat I_{\nu'}(\Omega')}{\hat \nu'}
(1-\cos\theta_{1})\Theta[\hat \nu \hat \nu'(1-\cos\theta_{1})-2]d\Omega'd\hat\nu'
\end{align}
($ \hat \chi $ should be calculated at the same frame for which $ \hat I' $ is given).
For the cross section we use  [e.g. \citet{padma1}]
\begin{align}
& \sigma_{\gamma\gamma}(s)=\frac 38 \frac {\sigma_T}{s} \times \cr & \left[\left( 2+\frac 2s -\frac{1}{s^2}\right)\cosh ^{-1}s^{1/2}-\left(1+\frac 1s \right)\left(1-\frac 1s \right)^{1/2}    \right] ,
\end{align}
where
\begin{equation}
s=\frac 12 h\nu h\nu' (1-\mu \mu')
\end{equation}
is the center of momentum energy squared.
To shorten the computing time we integrate over $ \phi' $ assuming that $ \sigma_{\gamma\gamma} $ changes slowly with $ \phi' $ and that $ \Theta[\nu\nu'(1-\cos\theta_{1})-2\nu_{p}^{2}] $ has the same value for most $ \phi' $ values, obtaining approximately
\[
<1-\cos(\theta_{1})>_{\phi}=1-\mu\mu'.\]
To find the positron production rate $ Q_{+} $ we use the rate of photon loss to this process,
\begin{equation}
 Q_{+}=-\frac 12 \dot{n}_\gamma =\frac 12 \int \frac{I_\nu(\mu)}{h\nu}\chi_{\nu,\gamma\gamma}(\mu)d\nu d\Omega .
\end{equation}
The scaling of the production rate follows,
\begin{equation}\label{eq:pair_prod_rate_scaled}
\hat Q_+=\frac {\Gamma_{u}m_{p}}{2m_{e}} \int \frac{\hat I_{\hat\nu}(\mu)}{\hat\nu}\hat \chi_{\hat \nu,\gamma\gamma}(\mu) d \hat\nu d\Omega.
\end{equation}

\subsubsection{Bremsstrahlung}

Bremsstrahlung emission includes contributions from $e^-p$ and
$e^+p$\, encounters, as well as from $e^-e^-, e^+e^+$ and $e^-e^+$
encounters, which become important sources of photon production at
high temperatures.
The emission can be expressed by \citep{Svensson1982a}
\begin{equation}\label{eq:em_brems}
\dot n_{\gamma,ff}(\Omega,\nu) = \frac{1}{\pi^2}
\sqrt{\frac{2}{\pi}} \alpha_{e} \sigma_T m_e^{1/2} c^2 n_i^2
\frac{e^{-h\nu/T}}{\sqrt{T} \nu} \lambda_{ff}\;,
\end{equation}
where $\alpha_{e}$ is the fine structure constant, and
\begin{equation}
\lambda_{ff}(x_+, T) = (1+x_+)\lambda_{ep} + \left[x_+^2+(1+x_+)^2\right]\lambda_{ee} +
x_+(1+x_+)\lambda_{+-}
\end{equation}
is a numerical factor accounting for the presence of
electron-positron pairs and for relativistic corrections at high
temperature.
We use a prescription for bremsstrahlung emission based on \citet{skibo95} (note that there is an errata correction to this paper), which gives a general fit for the Gaunt factor as a function of temperature, positron density and the emitted frequency.
The transformation between the different notations is $ \lambda_{ff}=\frac{\pi}{2\sqrt 3}g_{s} $, where $ g_{s} $ is the Gaunt factor as given in \citet{skibo95}.

The emissivity resulting from Eq. \eqref{eq:em_brems} is
\begin{align}
& \eta_{ff,\nu}(\mu)= h\nu \dot n_{\gamma,ff}(\Omega,\nu)= \cr
& \frac{h}{\pi^2}
\sqrt{\frac{2}{\pi}} \alpha_{e} \sigma_T m_e^{1/2} c^2 n_i^2
\frac{e^{-h\nu/T}}{\sqrt{T} } \lambda_{ff}.
\end{align}
The normalized emissivity then reads
\begin{align}\label{eq:ff_em_scaled}
& \left[ \Gamma(1-\beta\mu_{sh})\right]^2\hat \eta _{ff,sh}(\hat \nu_{sh},\Omega_{sh}) = \hat \eta_{ff}(\hat \nu,\Omega)= &  \cr
& \frac{\alpha_{e}m_e/m_p}{\pi^2\Gamma_u\Gamma^2\beta(1+\beta)(1+2x_+)}
\sqrt{\frac{2}{\pi}}
\frac{e^{-\hat \nu/ \hat T}}{\sqrt{\hat T} } \lambda_{ff}.
\end{align}

\paragraph{Minimal $ \nu $} 
Coulomb screening suppresses bremsstrahlung emission at impact
parameters larger than the Debye length $\lambda_D=\sqrt{T/4\pi
e^2 (n_e+n_+)}$, implying a low energy cutoff for bremsstrahlung
emission (Weaver, 1976b)
\begin{equation}\label{eq:epsilon_sc}
\epsilon_{sc} \simeq \frac{\gamma_{e,th}^2
\beta_{e,th}}{\lambda_D} \hbar c\;,
\end{equation}
where $\gamma_{e,th}$ is the Lorentz factor associated with the
random (``thermal'') motion of the electrons, and $\beta_{e,th}$
is the associated velocity (in units of $c$). Setting
$\gamma_{e,th} \simeq 1 + 3 T/m_ec^2$ we get for the non relativistic case ($ T\ll m_ec^2 $)
\begin{equation}\label{eq:numin_nr}
\epsilon_{sc,nr} \simeq 2.87
\times 10^{-6}\; n_{i,15}
^{1/2} (1+2x_+)^{1/2}\; \rm{KeV},
\end{equation}
and for the relativistic case ($ T\gg m_ec^2 $)
\begin{equation}\label{eq:numin_rel}
\epsilon_{sc,rel} \simeq 9.12 \times 10^{-10}\; n_{i,15}^{1/2} (1+2x_+)^{1/2} \left(
\frac{T}{\rm{KeV}} \right)^{3/2}\; \rm{KeV},
\end{equation}
where $ n_i=n_{i,15}10^{15} \cm^{-3} $.

We note that since our calculation explicitly describes upscattering and bremsstrahlung self absorption, there is no need to introduce (as was done, for example, in Weaver 1976) a cutoff to the Bremsstrahlung emission at low frequencies, for which the flow dynamical time scale or the self absorption time scale are shorter than the time required for a low energy photon to be upscattered to $ T $. 

{\it Bremsstrahlung self absorption}.
Using Kirchhoff's law and the calculated value of $ \eta_{\nu,ff} $ in the rest frame of the plasma we have
\begin{equation}
\chi_{\nu,ff}=\frac{\eta_{\nu,ff}}{B_{\nu}(T)}[\mathrm{cm^{-1}]},
\end{equation}
where
\begin{equation}
B_{\nu}(T)=\frac{2h\nu^{3}}{c^{2}}\frac{1}{e^{h\nu/k_{B}T}-1}
\end{equation}
is Plank's spectrum.

The normalized Plank spectrum is
\begin{equation}
\hat{B}_{\hat\nu}=\frac{2m_{e}^{4}c^{3}}{h^{3}m_{p}}\frac{1}{\Gamma_{u}^{2}\beta_{u}n_{u}}\frac{\hat{\nu}^{3}}{e^{\hat{\nu}/\hat{T}}-1},
\end{equation}
and the normalized absorption  is
\begin{align}\label{eq:ff_abs_scaled}
& \left[ \Gamma(1-\beta\mu_{sh})\right]^{-1}\hat \chi_{ff,sh}(\hat \nu_{sh})= \hat \chi_{ff}(\hat \nu)
 =\frac{\hat \eta_{\hat \nu,ff}}{\hat B_{\hat \nu}(\hat T)}   \cr &
 = \frac{  \alpha_{e}h^{3} \lambda^{(ff)} }
{\sqrt 2 \pi^{5/2}m_{e}^{3}c^{3}\Gamma(1+\beta)(1+2x_+)}
\frac{n_i \left(1-e^{-\hat \nu/ \hat T} \right) }{\hat{\nu}^{3}\sqrt{\hat T} }.
\end{align}

\subsubsection{Summary}\label{sec:eq_summary}
To summarize: we use equations \eqref{eq:scaled_en}, \eqref{eq:scaled_mom}, \eqref{eq:transfer_scaled} and \eqref{eq:pair_scaled}, to determine the variables
$\hat T(\tau_{*})$, $\beta(\tau_{*})$, $x_{+}(\tau_{*})$ and $\hat I_{\hat \nu_{sh}}(\mu_{sh})(\tau_{*}).$
The contributions of the radiative processes to the transfer equation [eq. \eqref{eq:transfer_scaled}] are given by eqs. \eqref{eq:compton_em_scaled}, \eqref{eq:compton_ab_scaled}, \eqref{eq:pair_ann_em_scaled}, \eqref{eq:pair_prod_abs_scaled}, \eqref{eq:ff_em_scaled} and \eqref{eq:ff_abs_scaled}. The contributions to $ \hat Q $ in the positron fraction equation [eq.  \eqref{eq:pair_scaled}] are given by eqs. \eqref{eq:pair_ann_q_scaled} and \eqref{eq:pair_prod_rate_scaled}.

\subsection{Boundary conditions}\label{sec:bound_cond}
We obtained solutions of the equations given above over a finite optical depth range around the shock transition, that satisfies the following requirements:
\begin{itemize}
\item The solution includes a subsonic region downstream of a supersonic region with continuous radiation field $I_{\nu_{sh}}(\mu_{sh})$ and positron flux across the sub-shock separating the two regions (see \sref{sec:immDS_subson});
\item The radiation momentum flux in the last  several photon mean free paths away from the shock transition in the US region is negligible compared to the far US  {\it electron} momentum flux;
\item The width of the subsonic region is sufficiently large compared to the photon mean free path, such that the solution is insensitive to the precise boundary conditions that are applied at the DS edge, while remaining short enough as to avoid reaching the second supersonic region which exists DS of the subsonic region.
\end{itemize}

The boundary conditions in the far upstream are $I_{\nu_{sh}}(\mu_{sh}>0,\, z_{sh}=-\infty)=0,$
i.e. no incident radiation at the upstream (In practice we use an effective "reflector" in the US end of the calculation, to avoid numerical fluctuations and shorten the iteration time. It does not affect the shock structure). In addition, the positron number is taken as $ 0 $ at the US boundary. 

The boundary conditions at the far downstream are given by thermal equilibrium. Since the calculation does not reach the far DS, we use a boundary condition in the DS which corresponds to isotropy of the radiation field in the rest frame of the far DS. { This is done by equating the intensity and spectrum of US going radiation at the DS boundary to that of the DS going radiation.  
For numerical reasons, we multiply the reflected radiation by a factor which is close to unity, this has a negligible effect on the shock structure.
In addition, we impose an upper limit on the photon energy of the reflected radiation, typically $ 3m_ec^2 $. The physical reasoning for this upper limit is that high energy photons that cross this point in the DS either scatter and lose most of their energy (as $ \hat T \ll 1 $ at that point and further away), or more likely, produce an $ e^+e^- $ pair that is swept DS with the flow.}

\section{The numerical method}\label{sec:num_meth}
We briefly describe below the numerical method we use for solving the equations.

\subsection{Iteration scheme}\label{sec:iter_scheme}
We start with an initial guess for the shock profile, $S^{0}=\left\{ \hat T^0(\tau_{*}),\beta^0(\tau_{*}),x_{+}^0(\tau_{*}),\hat I^0_{\hat \nu_{sh}}(\mu_{sh},\tau_{*})\right\}$, and modify it iteratively until a solution of the equations is obtained. The iterations are performed as follows:
\begin{enumerate}
\item Compute $\hat \eta(S^{n})$ and $\hat \chi(S^{n})$ using the profile $S^{n}$;
\item Integrate directly the transfer eq., eq.~\eqref{eq:transfer_scaled}, using $\hat \eta(S^{n})$ and $\hat \chi(S^{n})$, to obtain $\hat I_{\hat \nu_{sh}}^{n+1}(\mu_{sh},\tau_{*})$;
\item Use eqs.~\eqref{eq:scaled_en}, \eqref{eq:scaled_mom}
and \eqref{eq:pair_scaled} with the new radiation field, $\hat I_{\hat \nu_{sh}}^{n+1}(\mu_{sh},\tau_{*})$,
to obtain the new profile $S^{n+1}$;
\end{enumerate}
{ Usage of "partial iterations", where $S^{n+1}$ is replaced with a weighted average of $S^n$ and $S^{n+1}$, was required in order to achieve convergence and stability.
}

{
At any given $\tau_*$, the energy and momentum conservation equations, Eqs.~\eqref{eq:scaled_en}, \eqref{eq:scaled_mom}, have a supersonic and a subsonic solution for $\bt$ and $\hat T$ given $x_+$, $\hat I_{\hat \nu_{sh}}(\mu_{sh})$. The position of the sub-shock (see \sref{sec:immDS_subson}) was set to $\tau_*=0$, upstream of which the supersonic solution was chosen and downstream of which, the subsonic solution was chosen.}

{ We significantly reduced the computational time of the calculation, by separating the spatial grid into two regions, and preforming the above iterations on each. The downstream going photons on the downstream boundary of the first region were used as a boundary condition for the second region and vice versa. We preformed macro iterations in which we updated these boundary conditions until a self consistent profile was obtained across the border between the regions.}

\subsection{Discretization}

\begin{figure}\centering
\includegraphics[scale=0.45]{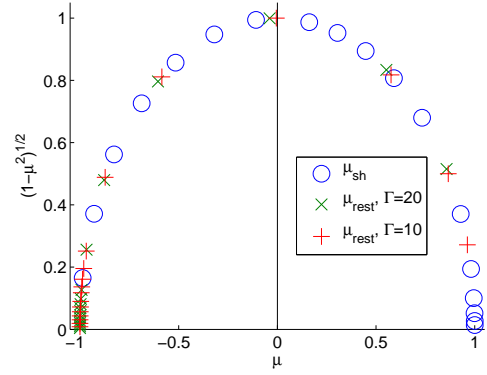}\caption{$ \Gamma_u=20 $, distribution of 18 $  \mu's $ in three frames: shock frame, US frame and $ \Gamma=10 $ frame.}
\label{fig:mu_dist}
\end{figure}

We use a discrete approximation of $ I_{sh,\nu_{sh}}(\mu_{sh}) $, 
\begin{equation}
I_{sh,\nu_{sh}}(\mu_{sh})=\sum \hat I_{sh,ij}f_\sqcap(\nu_{sh},\nu_{sh,i},\nu_{sh,i+1})f_\sqcap(\mu_{sh},\mu_{sh,j},\mu_{sh,j+1}),
\end{equation}
where $f_\sqcap(x,x_1,x_2)=\Theta(x-x_1)\Theta(x_2-x)$ is the top hat function and $\Theta$ is the step function. The distribution of $ \nu_{sh,i} $ is logarithmic in the range $ \nu_{min}$ to $ \nu_{max}$. Typical values are $ h\nu_{max}=10\Gamma_u^2 m_ec^2 $ and $ h\nu_{min}=10^{-8}m_ec^2 $. The distribution of $ \mu_{sh,j} $ is set to account for relativistic beaming of the radiation in the shock frame as well as for a relatively isotropic component in all frames, from US to DS. This is achieved by a logarithmic separation of $ \mu_{sh} $ in the US direction between $ \mu_{sh}=0 $ and $ \mu_{sh}=1 $, with $ 1-\max (\mu_{sh}) < \Gamma_u^{-2}$. The $ \mu_{sh} <0$ directions are chosen as the zeros of a Legendre polynomial, the same as the common Gaussian quadrature.
 A typical division is shown in fig. \ref{fig:mu_dist}. We note that in order to account correctly for the relativistic beaming using Gaussian quadrature, for instance, would require a much larger number of azimuthal directions for high values of $ \Gamma_u $. The convergence of the solutions with respect to the resolution is demonstrated in \sref{sec:NumConv}.

\subsection{Test problems}

The numerical scheme and its implementation were tested thoroughly to ensure the results are valid. The tests verified a correct description of the different radiation mechanisms in steady state problems including, e.g., Compton scattering with pair production, bremsstrahlung emission with self absorption. We present here only  two of the tests, demonstrating the suitability of the numerical scheme for dealing with repeated Compton scattering and pair production and annihilation. The test results are compared with analytic solutions or Monte Carlo simulations and are shown to reproduce them well. 

\subsubsection{Comptonization in a cloud of low and medium optical depths}

 A thin, stationary planar layer of plasma with Thomson optical depth $ \tau_T $ in the $ z $ direction (perpendicular to the symmetry plane) and a given temperature $ T $ is irradiated at one end, $ \tau_*=0 $, by a $ \delta $ function in $ \nu $, directed along the $ z $ axis,
\begin{equation}
I_{\nu}(\mu>0,\tau=0)=I_0\delta(\nu-\nu_0)\delta(\mu-1).
\end{equation}
At $ \tau_*=\tau_T $ a free boundary condition, $ I(\mu<0,\tau_*=\tau_T)=0 $, is applied.
In order to reach the steady state solution for the radiation field, the iteration scheme of the radiative transfer equation is used until the radiation field converges. 

\begin{figure}[ht]
\centering
\includegraphics[scale=0.5]{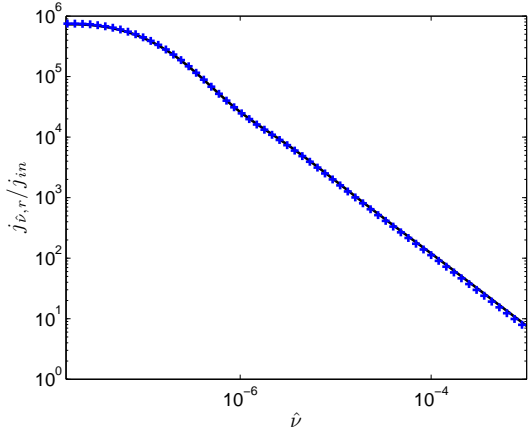}\caption{The specific photon flux leaving a cloud of plasma with $ \hat T=1 $ and width  $ \tau_T=1 $. The radiation entering the cloud has a single frequency $ \hat \nu_0=10^{-8} $. The results of the code are marked with blue pluses and the Monte Carlo results are shown as a black line.}
\label{fig:cloud_med}
\end{figure}

\begin{figure}[ht]
\centering
\includegraphics[scale=0.5]{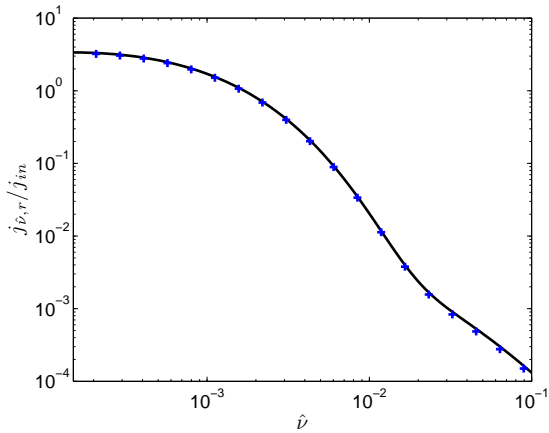}\caption{The specific photon flux leaving a cloud of plasma with $ \hat T=1 $ and width  $ \tau_T=0.01 $. The radiation entering the cloud has a single frequency $ \hat \nu_0=10^{-4} $. The results of the code are marked with blue pluses and the Monte Carlo results are shown as a black line.}
\label{fig:cloud_thin}
\end{figure}

The results of these calculations are compared with an independent Monte Carlo simulation of the setups using the same approximate  Compton kernel, as described in \sref{sec:compton_eqs}.
The specific photon flux escaping through the free boundary at $\tau_*=\tau_{T}$,
\begin{equation}
j_{\hat \nu}=\int_0^1  \frac {I_{\hat\nu}}{\hat \nu}\mu d\mu,
\end{equation}
was calculated for two cases with $ \hat T=1 $, one with $ \tau_T=1 $, $ \hat \nu_0=10^{-8} $ and the other with $ \tau_T=0.01 $, $ \hat \nu_0=10^{-4} $.  The resulting spectra are shown in  Figs.  \ref{fig:cloud_med} and \ref{fig:cloud_thin} for $ \tau_T=1 $ and $ \tau_T=0.01 $ respectively. In each figure the results of the code (blue pluses) and the  Monte Carlo simulation (black lines) are shown. As can be seen, there is an excellent agreement between the two independent methods for calculating the spectrum of escaping photons.


\subsubsection{Pair quasi equilibrium for given T}
This test checks the numerical description of the (integral) pair production and annihilation. We use a setup with a given Wien spectrum of the radiation field,
\begin{equation}
I_{\hat \nu}(\mu)\propto \hat \nu ^2e^{-\hat \nu/\hat T}.
\end{equation}
For a given $\hat T $, we find the equilibrium value of $ x_+=n_+/n_p $ for which the positron production and annihilation rates cancel each other analytically and numerically. A comparison between the two values obtained is given
 in  table \ref{tab:xpos} for different temperatures.

Note that $x_{+}$ does not necessarily grow with $\hat{T}$, since
we use different densities $n$ for convenience. We obtain an accuracy
of a few \% except for very low temperature, where higher resolution
is needed in order to account for the exponential cutoff near $\hat{\nu}=1$.
The resolution used here is $\nu_{n+1}/\nu_{n}=1.4$, $N_{\mu}=12$.
\begin{table}
\centering\begin{tabular}{|c|c|c|}
\hline
$\hat{T}$ & $x_{analytic}$ & $x_{num}$\tabularnewline
\hline
\hline
0.3 & 550 & 425\tabularnewline
\hline
0.5 & 541 & 500\tabularnewline
\hline
0.8 & 421 & 421\tabularnewline
\hline
1.5 & 259 & 266\tabularnewline
\hline
10 & 43 & 43\tabularnewline
\hline
\end{tabular}
\caption{Equilibrium values of $ x_+ $, balancing the pair production and annihilation rates at different temperatures. }
\label{tab:xpos}
\end{table}

\section{Numerical results}\label{sec:numerical}
In this section we present the numerical results, solving  equations
\eqref{eq:en_flux}-\eqref{eq:transfer} self consistently for
different values of the upstream Lorentz factor $ \Gamma_u $.  We
divide the presentation of the results into 2 parts: The structure
(\sref{subsec:struct}) and the radiation spectrum
(\sref{subsec:spectrum}). The structure is the spatial distribution
of integral parameters such as temperature, velocity (or Lorentz
factor), pair density and radiation pressure. The spectrum is the
distribution of radiation intensity at different angles and photon
energies (at given locations across the shock), measured in a
specific reference frame. Two important frames of reference are the
shock frame, in which the solution is a steady state solution, and the local
rest frame of the plasma, which is useful for understanding the
interaction between the radiation and the plasma.

\subsection{Structure}\label{subsec:struct}
\begin{figure*}[ht]
\begin{minipage}[t]{0.5\linewidth}
\centering
\includegraphics[scale=0.5]{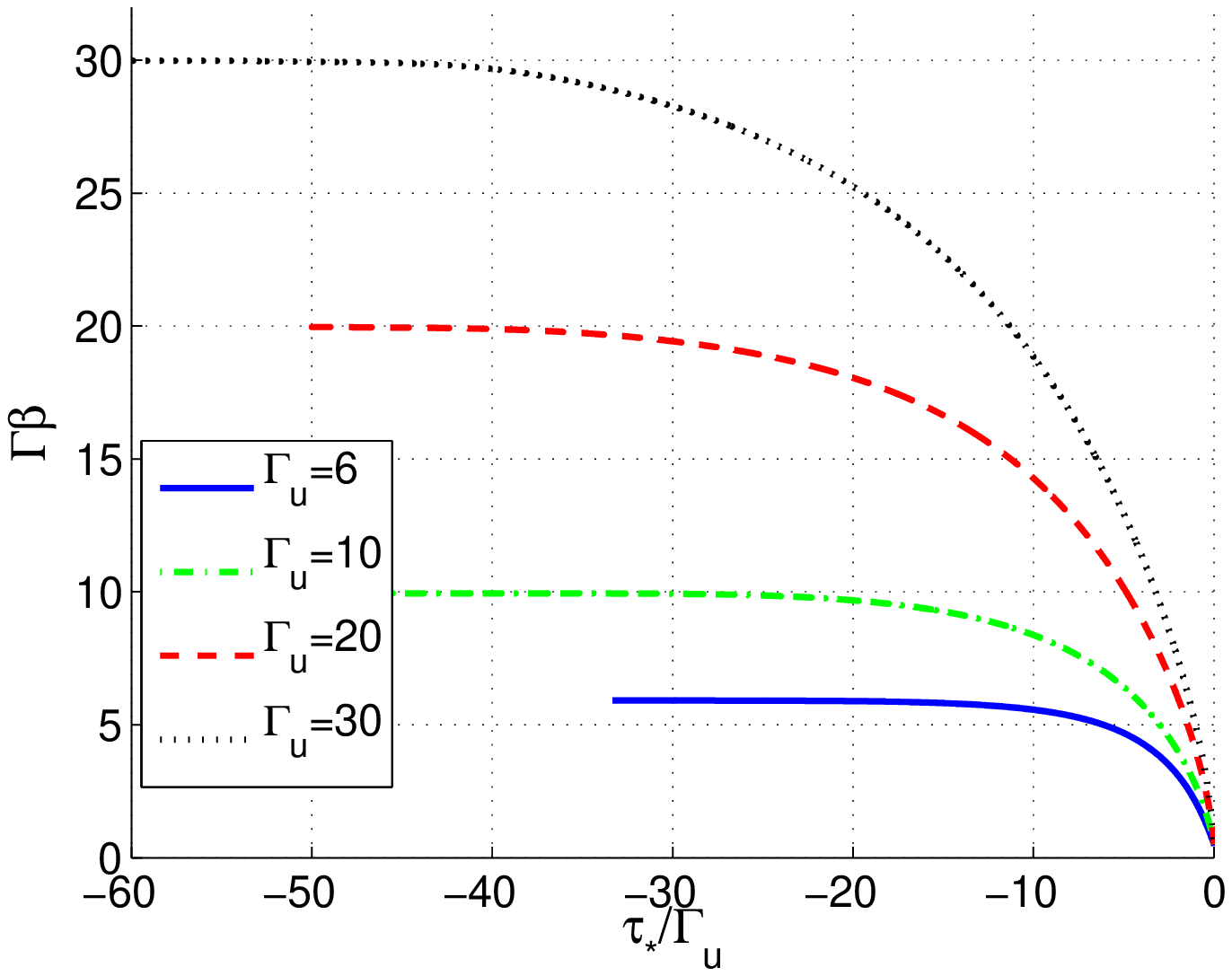}\caption{The relativistic velocity of the flow $ \Gamma \beta $ vs. $ \tau_*/\Gamma_u $ for different values of $ \Gamma_u $, from the US to the subshock ($\tau_*=0$). }
\label{fig:struct_g.eps}
\end{minipage}
\hspace{0.5cm}
\begin{minipage}[t]{0.5\linewidth}
\centering
\includegraphics[scale=0.5]{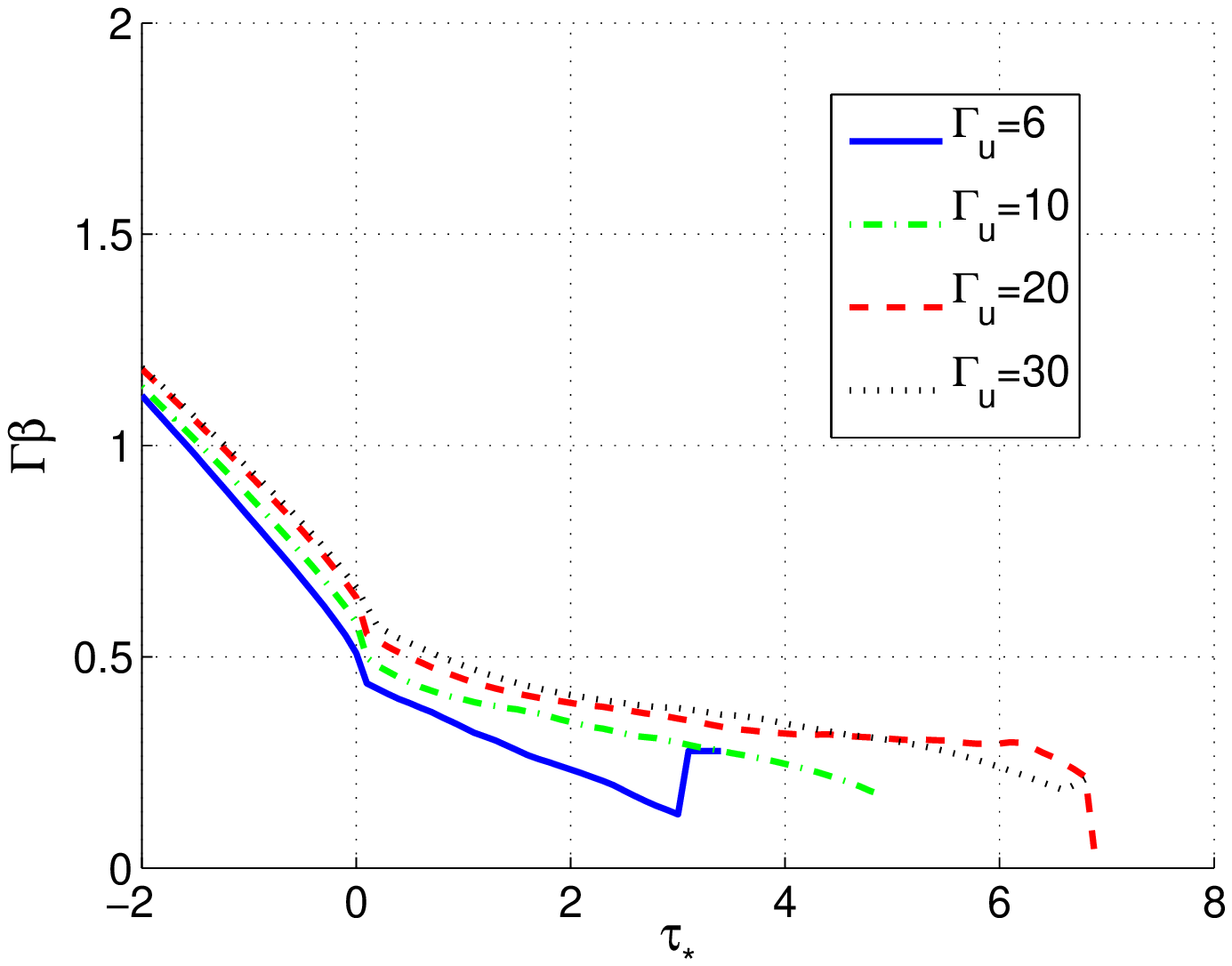}\caption{The relativistic velocity of the flow $ \Gamma \beta $ vs. $ \tau_*$ for different values of $ \Gamma_u $, around the subshock ($\tau_*=0$). Notice that the last mean free path on the right hand side is influenced by the boundary conditions, however the flow near the subshock is not affected by this boundary condition. }
\label{fig:struct_g_DS.eps}
\end{minipage}
\end{figure*}

\begin{figure*}[ht]
\begin{minipage}[t]{0.5\linewidth}
\centering
\includegraphics[scale=0.5]{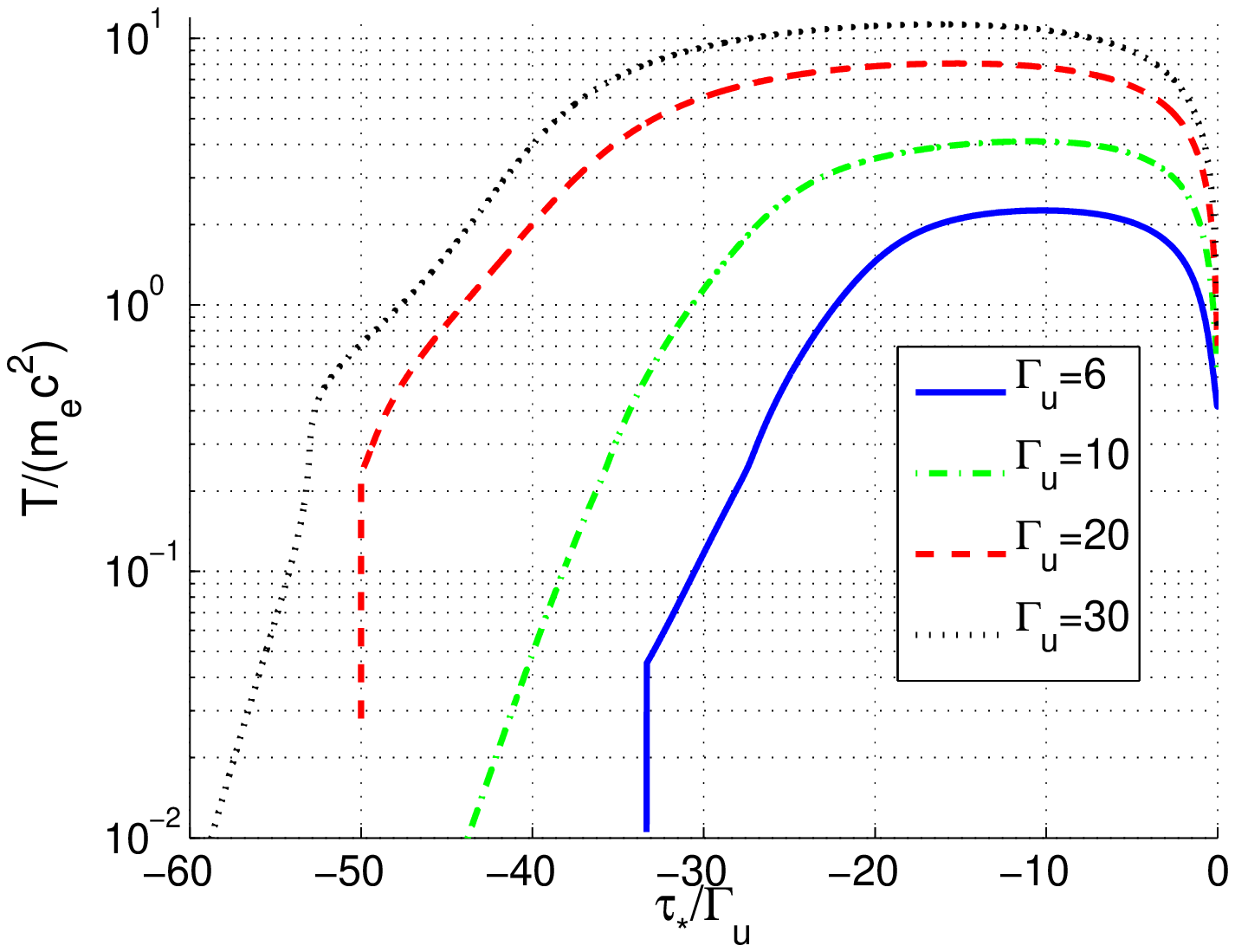}\caption{The normalized temperature $ \hat T $ vs. $ \tau_*/\Gamma_u $ for different values of $ \Gamma_u $, from the US to the subshock. }
\label{fig:struct_T}
\end{minipage}
\hspace{0.5cm}
\begin{minipage}[t]{0.5\linewidth}
\centering
\includegraphics[scale=0.5]{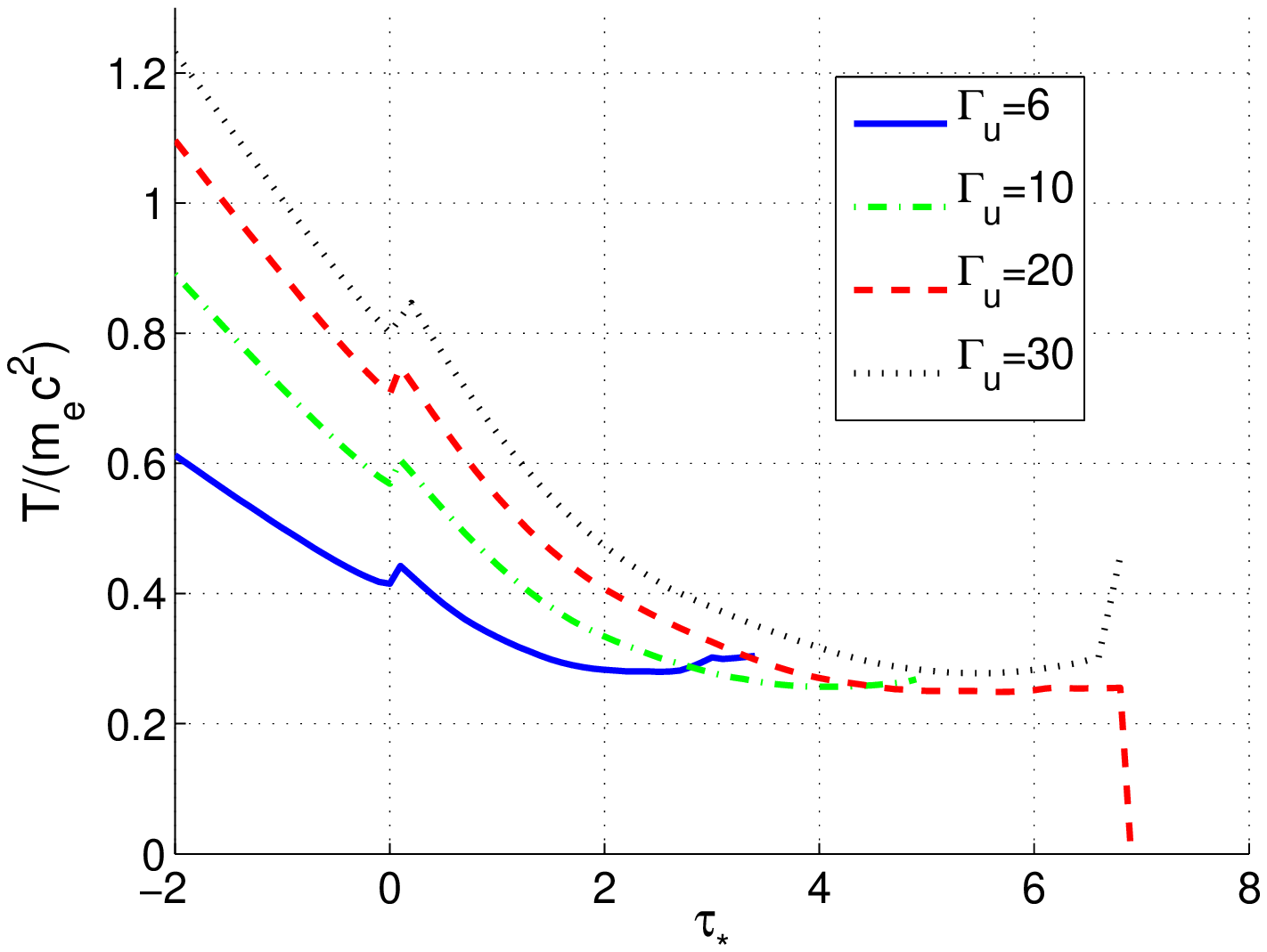}\caption{The normalized temperature $ \hat T $ vs. $ \tau_*$ for different values of $ \Gamma_u $, around the subshock (Notice that the last mean free path on the righ hand side is influenced by the boundary conditions).  }
\label{fig:struct_T_DS}
\end{minipage}
\end{figure*}

\begin{figure*}[ht]
\begin{minipage}[t]{0.5\linewidth}
\centering
\includegraphics[scale=0.5]{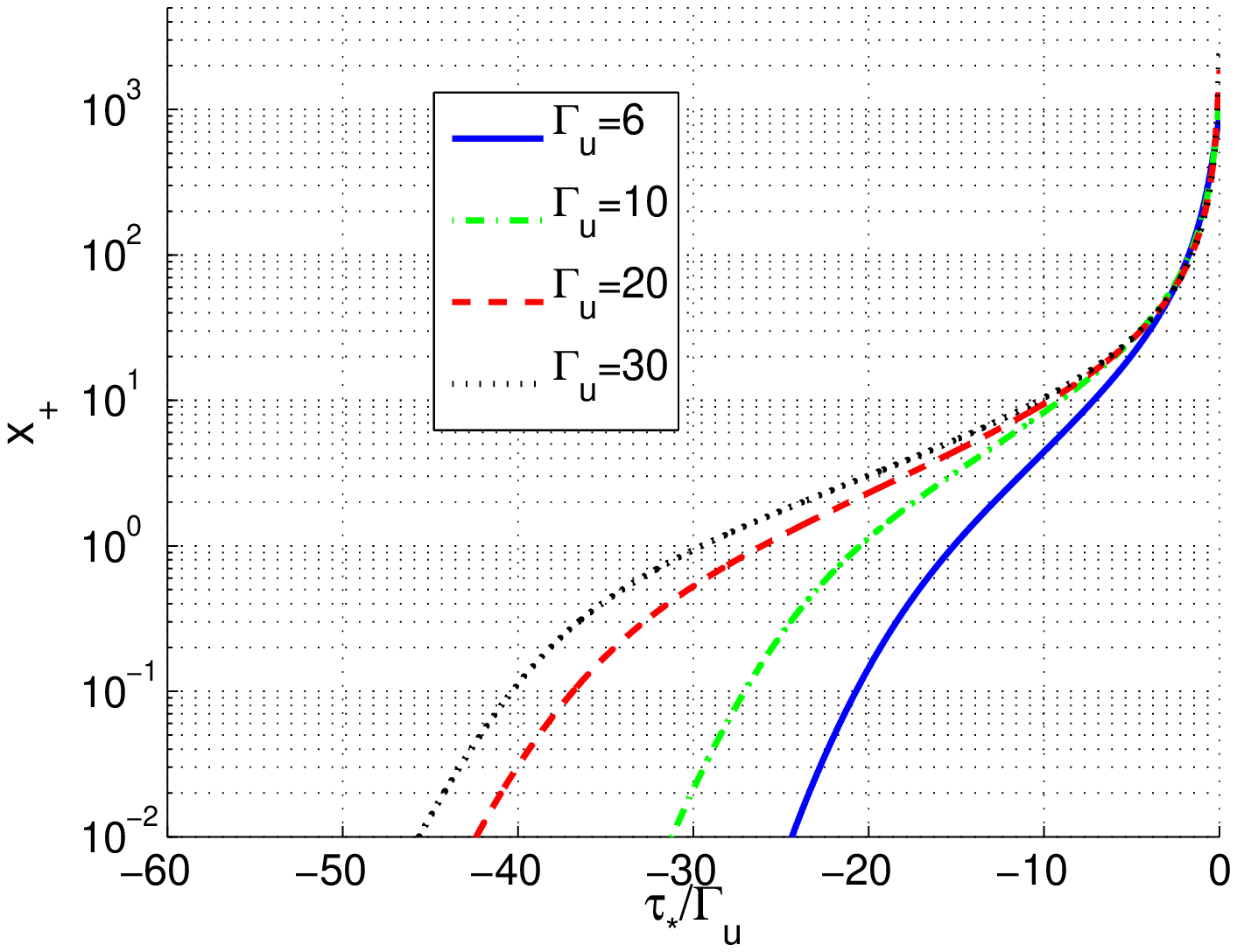}\caption{The positron to proton ratio $ x_+ $ vs. $ \tau_*/\Gamma_u $ for different values of $ \Gamma_u $, from the US to the subshock.}
\label{fig:struct_x}
\end{minipage}
\hspace{0.5cm}
\begin{minipage}[t]{0.5\linewidth}
\centering
\includegraphics[scale=0.5]{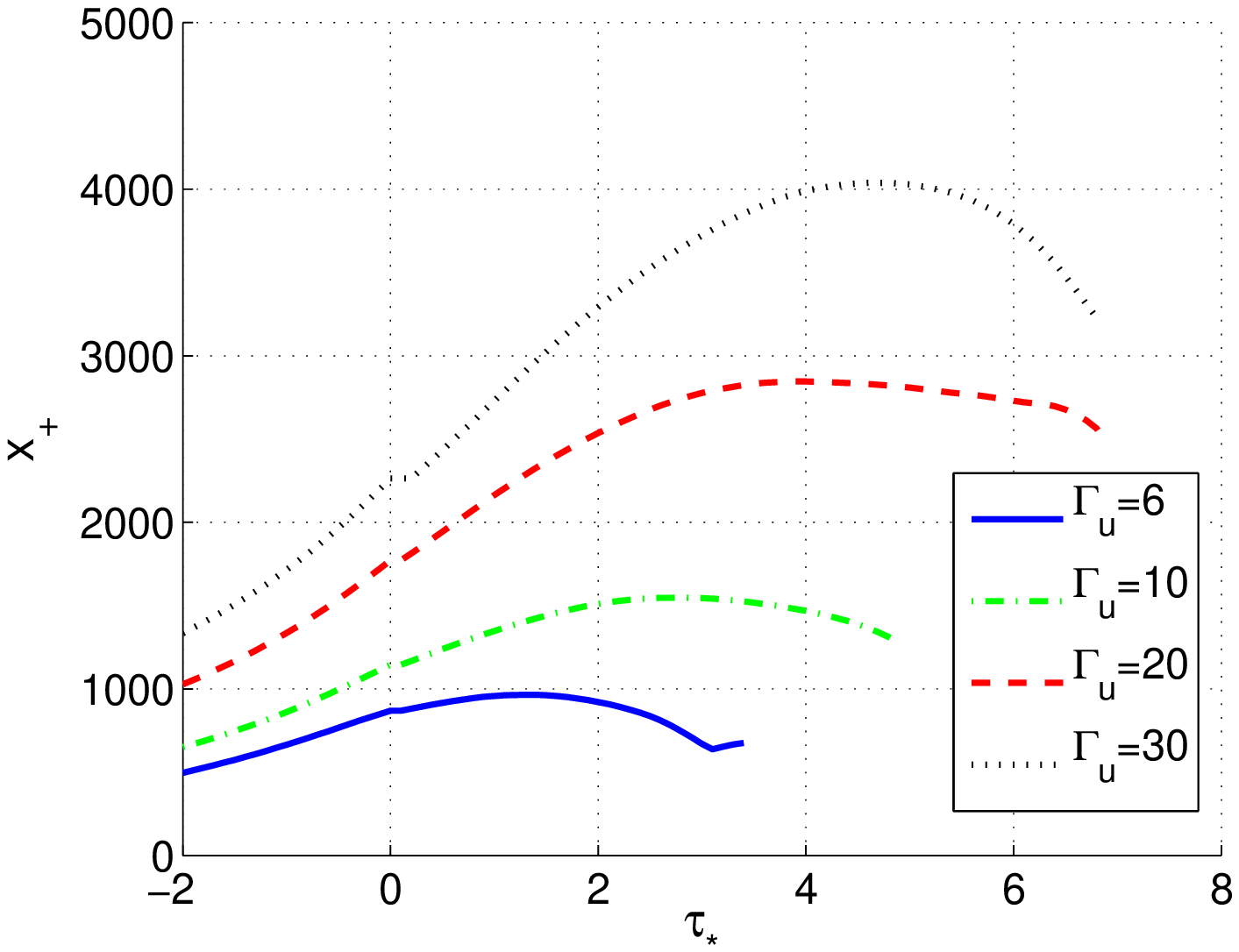}\caption{The positron to proton ratio $ x_+ $ vs. $ \tau_* $ for different values of $ \Gamma_u $, around the subshock.}
\label{fig:struct_x_DS}
\end{minipage}
\end{figure*}

{ The values of $ \Gamma\beta$, $ \hat T $ and $ x_+ $   for $ \Gamma_u=6,~10,~20$ and $30$ are shown in figures \ref{fig:struct_g.eps} to \ref{fig:struct_x_DS} as functions of the Thomson optical depth for upstream going photons $ \tau_* $ [defined in Eq. \eqref{eq:dimless_def}] or $ \tau_*/\Gamma_u $. Figures zoomed on the DS region ($ \tau_* \ge 0 $) are separately  given.} 
The results are calculated for $n_{u}=10^{15}\cm^{-3}$, over regimes where bremsstrahlung absorption is negligible (i.e. they are in the
low density limit, see \sref{subsec:low_density}). 

The shock profiles can be divided
to 4 regions:
\begin{enumerate}
\item Far upstream - The velocity is constant, while the radiation intensity
and positron fraction grow exponentially until they hold a significant
fraction of the energy and momentum of the flow.
\item The velocity transition - Here the flow decelerates considerably, reaching
a velocity close to the downstream velocity. For RRMS this regime
is bound by a subshock.
\item Immediate downstream - In the first $\beta_{d}^{-1}$ optical depths behind the velocity transition the flow approximately stays at constant velocity,
while the plasma and radiation are in CE. A gradual cooling by
bremsstrahlung emission and inverse Compton scattering takes place. This region
produces the radiation that diffuses upstream and decelerates the
incoming plasma.
\item Far downstream - Further than approximately $\beta_{d}^{-1}$ optical depths into
the downstream, from where most photons can not diffuse upstream.
From this point on, a slow thermalization takes place accompanied by
a slow decline in the plasma temperature and photon energies, ending
when the temperature reaches the downstream temperature. The decline in temperature leads first to a decrease in positron number, until the pair density becomes negligible compared to that of the original electrons ($ x_+ < 1 $) at $ T\sim 50\mathrm{keV} $. Then the thermalization continues until
bremsstrahlung absorption takes over and thermal radiation at
equilibrium is established.
\end{enumerate}
We do not solve the equations in the fourth region  since the solution there is straightforward
(the radiation is isotropic and in equilibrium with the plasma). Also, note that since the far downstream is supersonic,
a second sonic point is expected in RRMS. This, however, is a stable
point with no special physical significance.

\begin{figure*}[ht]
\begin{minipage}[t]{0.5\linewidth}
\centering
\includegraphics[scale=0.5]{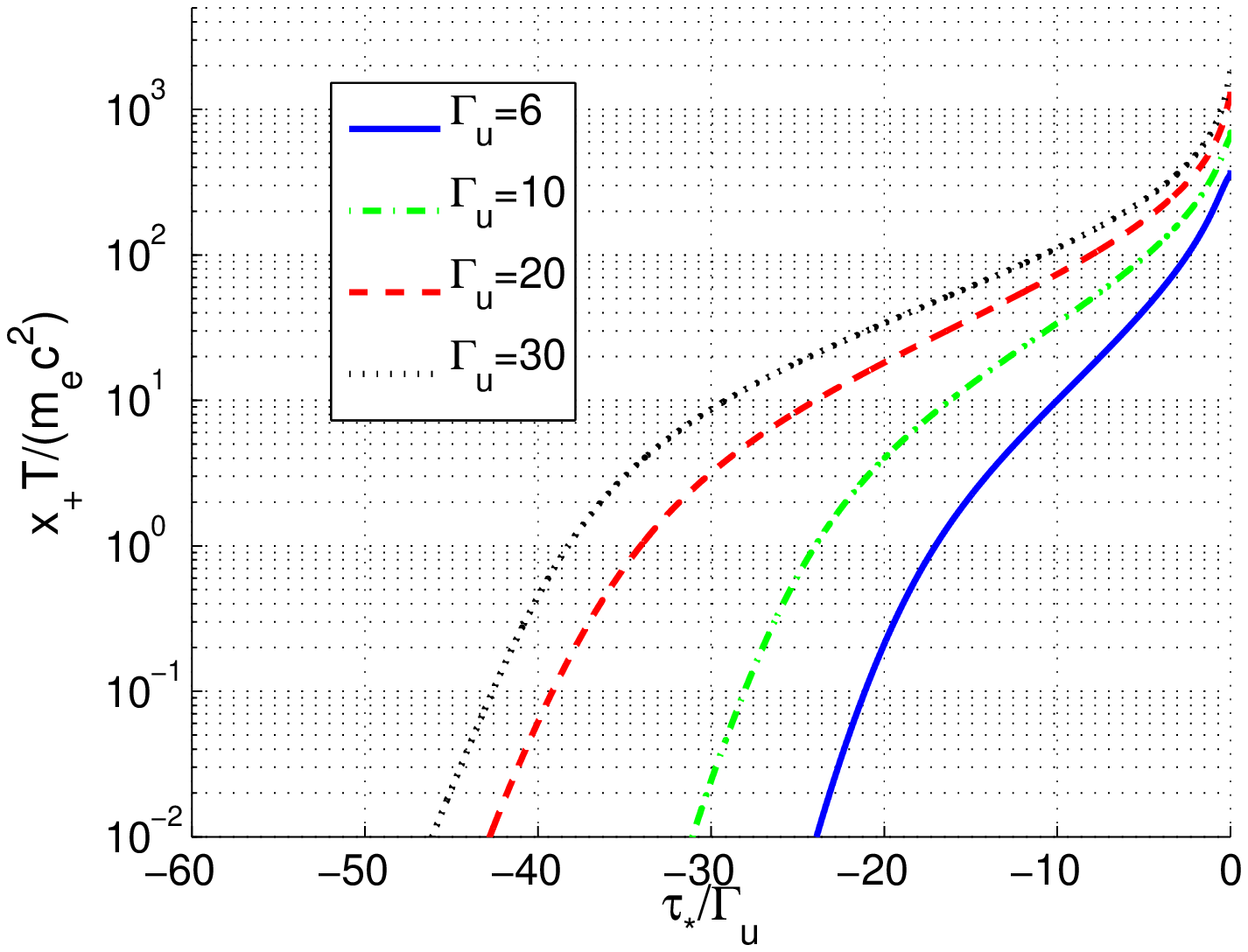}\caption{The rest frame normalized positron pressure $ x_+\hat T $ vs. $ \tau_*/\Gamma_u $ for different values of $ \Gamma_u $, from the US to the subshock. }
\label{fig:struct_xT}
\end{minipage}
\hspace{0.5cm}
\begin{minipage}[t]{0.5\linewidth}
\centering
\includegraphics[scale=0.5]{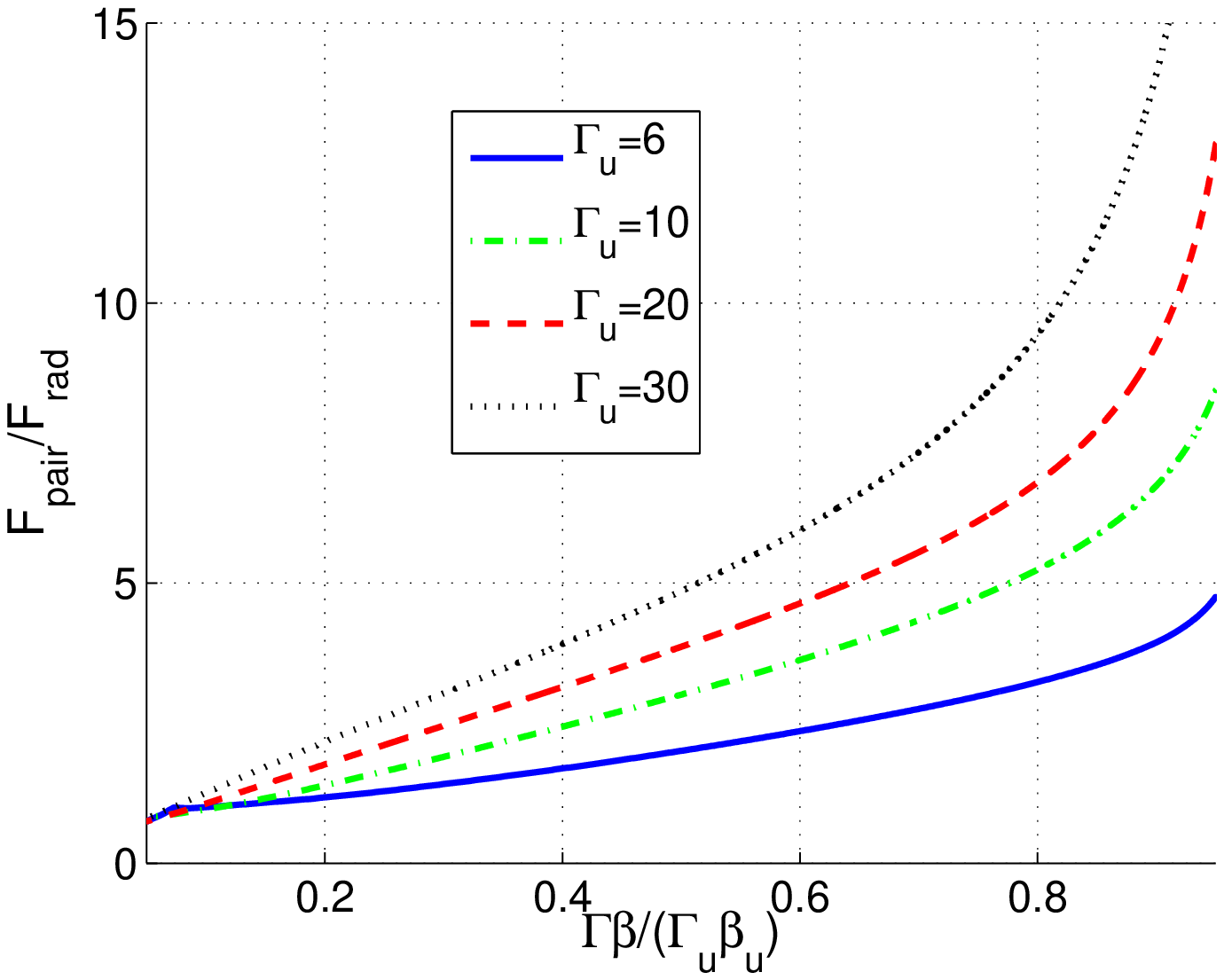}\caption{The ratio of thermal energy flux carried by electrons and positrons to the radiation energy flux $ \hat F_{rad,sh} $ vs. $ \Gamma\beta/(\Gamma_u\beta_u) $ for different values of $ \Gamma_u $.}
\label{fig:struct_Fpos_F}
\end{minipage}
\end{figure*}

\begin{figure*}[ht]
\begin{minipage}[t]{0.5\linewidth}
\centering
\includegraphics[scale=0.7]{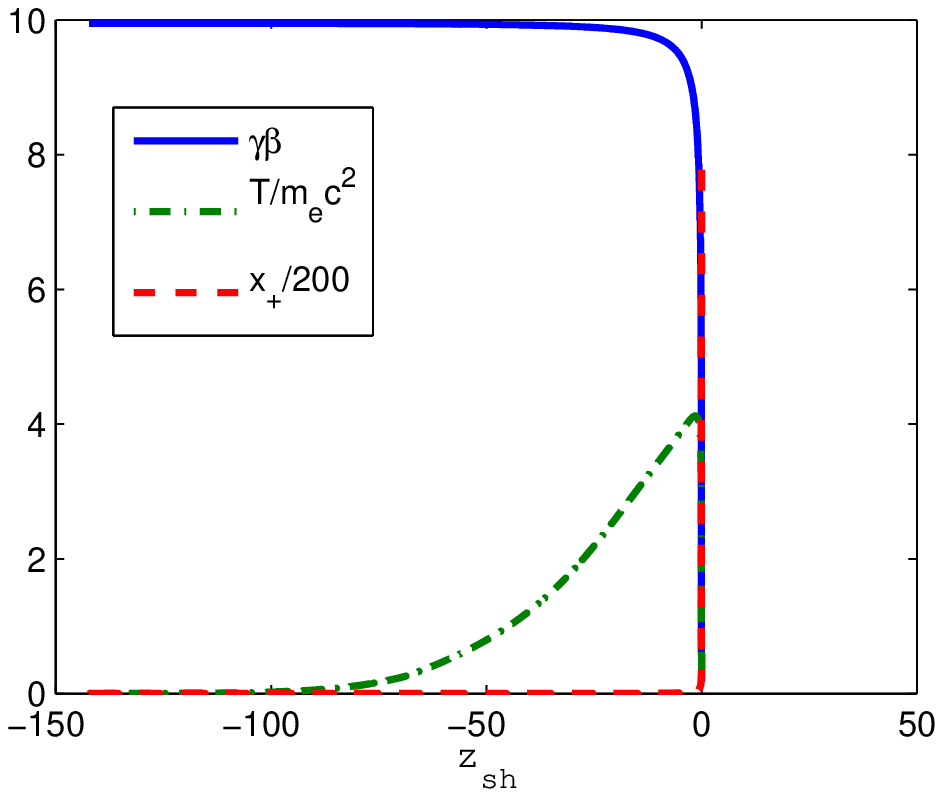}\caption{The relativistic velocity $ \Gamma\beta $, the normalized temperature $ \hat T $ and the positron to proton ratio $ x_+ $ vs. normalized distance $ \hat z_{sh}=\Gamma_un_u\sigma_Tz_{sh}  $ for $ \Gamma_u=10 $. }
\label{fig:struct_vs_z}
\end{minipage}
\hspace{0.5cm}
\begin{minipage}[t]{0.5\linewidth}
\centering
\includegraphics[scale=0.7]{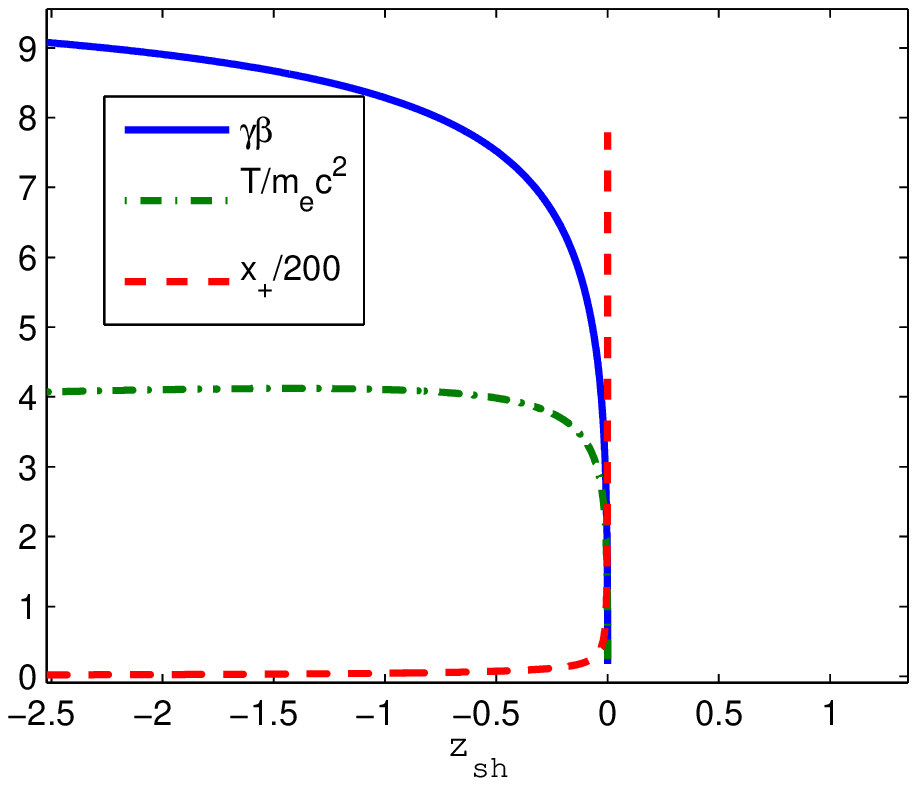}\caption{Same as fig.~\ref{fig:struct_vs_z}, showing only the DS region.}
\label{fig:struct_vs_z_zoom}
\end{minipage}
\end{figure*}

\begin{figure}
\includegraphics[scale=0.5]{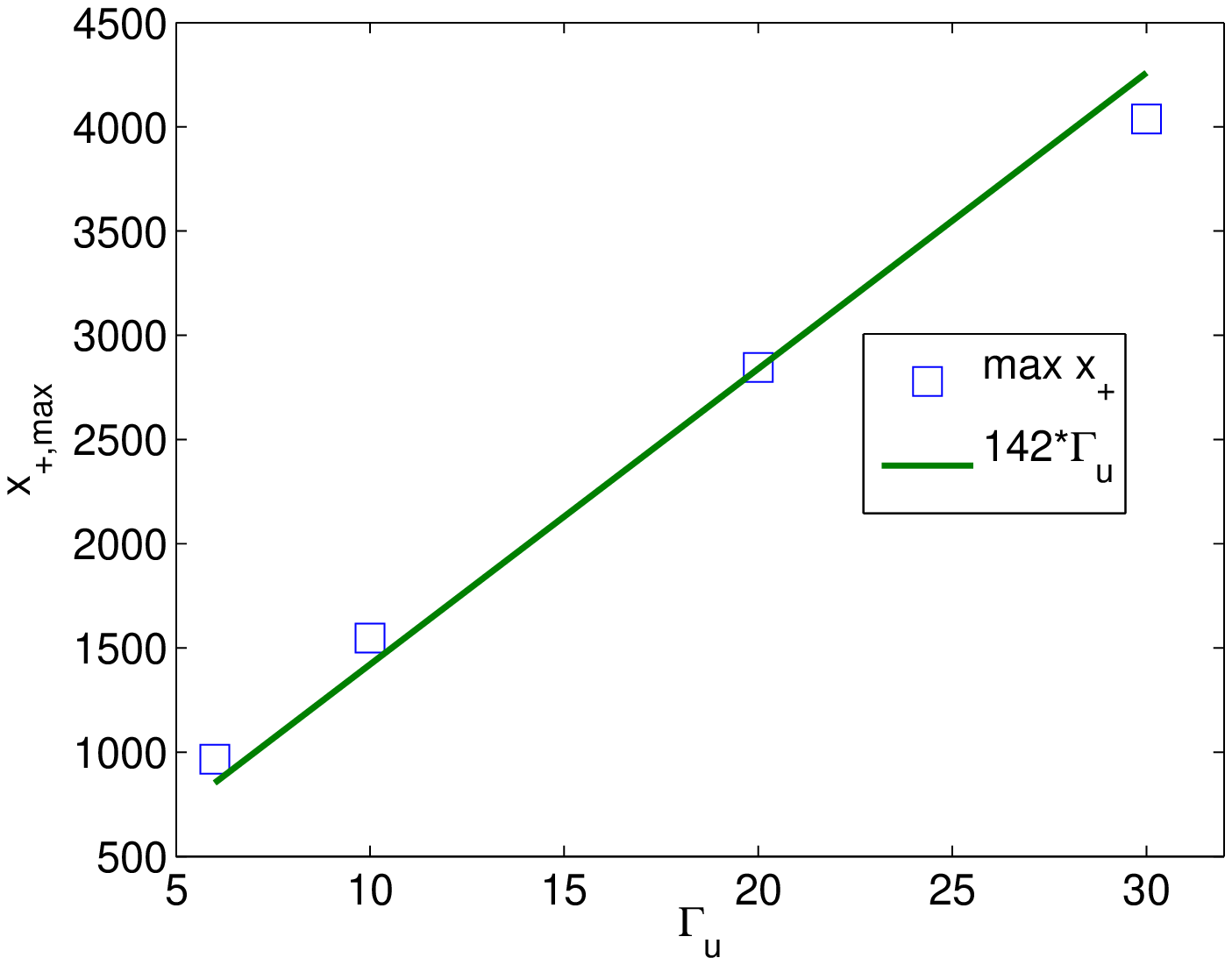}\caption{The maximum value of $ x_+ $ for different $ \Gamma_u $ values. The approximation $ x_{+,\max }=142\Gamma_u $ is accurate to better than 10\% in the range we investigated.}
\label{fig:xmax}
\end{figure}

Figures \ref{fig:struct_g.eps} and \ref{fig:struct_g_DS.eps} show,
for different values of $ \Gamma_u $, the structure of the
relativistic velocity $ \Gamma\beta $ across the shock. It can be
seen that the deceleration length in units of $ \tau_* $ is much larger than unity and grows with
$ \Gamma_u $ in a manner faster than linear. A subshock is
obtained at the sonic point, with a discontinuous deceleration
of $ \delta(\Gamma\beta)\sim 0.1 $. Behind the subshock, the
velocity approaches its far DS value in a few Thomson optical
depths. The last optical depth is affected by the boundary
conditions imposed on the right hand side. This effect will be discussed in \sref{sec:ds_length}

Figures \ref{fig:struct_T} and \ref{fig:struct_T_DS} show, for different values of $ \Gamma_u $, the structure of the temperature $ \hat T $ across the shock. The far US shows an exponential growth of $ \hat T  $ as a function of $ \tau_* $. The temperature then saturates at a maximum which is approximately linear in $ \Gamma_u $, and then decreases towards the subshock. Behind the subshock the temperature jumps, reaching a value of $ \hat T_{jump}\sim 0.5 $, which grows with $ \Gamma_u $, and then cools with a typical distance of a few Thomson optical depths ($ \tau_* $).

Figures \ref{fig:struct_x} and \ref{fig:struct_x_DS} show , for different values of $ \Gamma_u $, the
structure of the positron to proton number ratio, $ x_+ $, across the shock. The growth of $ x_+ $ as a function of $ \tau_* $ when approaching the subshock is super exponential, and its value reaches a maximum a few optical depths behind the subshock. The maximal value is approximately linear in $ \Gamma_u $ (see figure \ref{fig:xmax}).
Figure \ref{fig:struct_xT} shows $ x_+\hat T $ across the shock, which represents the pressure of the positrons and their relative importance in setting the speed of sound in the plasma, compared to the protons. The value of $ x_+\hat T $ goes above a few hundreds at the subshock for $ \Gamma_u\ge 6 $.

Figure \ref{fig:struct_Fpos_F} shows the ratio of thermal energy
flux carried by electrons and positrons to the radiation energy
flux, $ F_{sh} $, vs. $ \Gamma\beta/(\Gamma_u\beta_u) $. The energy
flux (``taken'' from the protons) is dominated by thermal and rest
mass energy flux of the electrons and positrons during most of the
transition rather than by radiation energy flux. The energy is transferred
to the radiation when the flow approaches the DS velocity, and the
two fluxes are comparable around the subshock. Comparing the results
at a fixed point (e.g. $ \Gamma=\Gamma_u/2 $), this ratio grows with
$ \Gamma_u $.

Figures  \ref{fig:struct_vs_z} and \ref{fig:struct_vs_z_zoom} show the relativistic velocity $ \Gamma\beta $, the temperature $ \hat T $ and $ x_+ $ as a function of the scaled distance $ \hat z_{sh}=\Gamma_u n_u \sigma_T z_{sh} $, for $ \Gamma_u=10 $. These figures illustrate that the shock width is comparable to the upstream Thomson mean free path, as $ \hat z_{sh} $ is approximately measured in these units.

\subsection{Spectrum}
\label{subsec:spectrum}

\begin{figure*}[ht]
\begin{minipage}[t]{0.5\linewidth}
\centering
\includegraphics[scale=0.45]{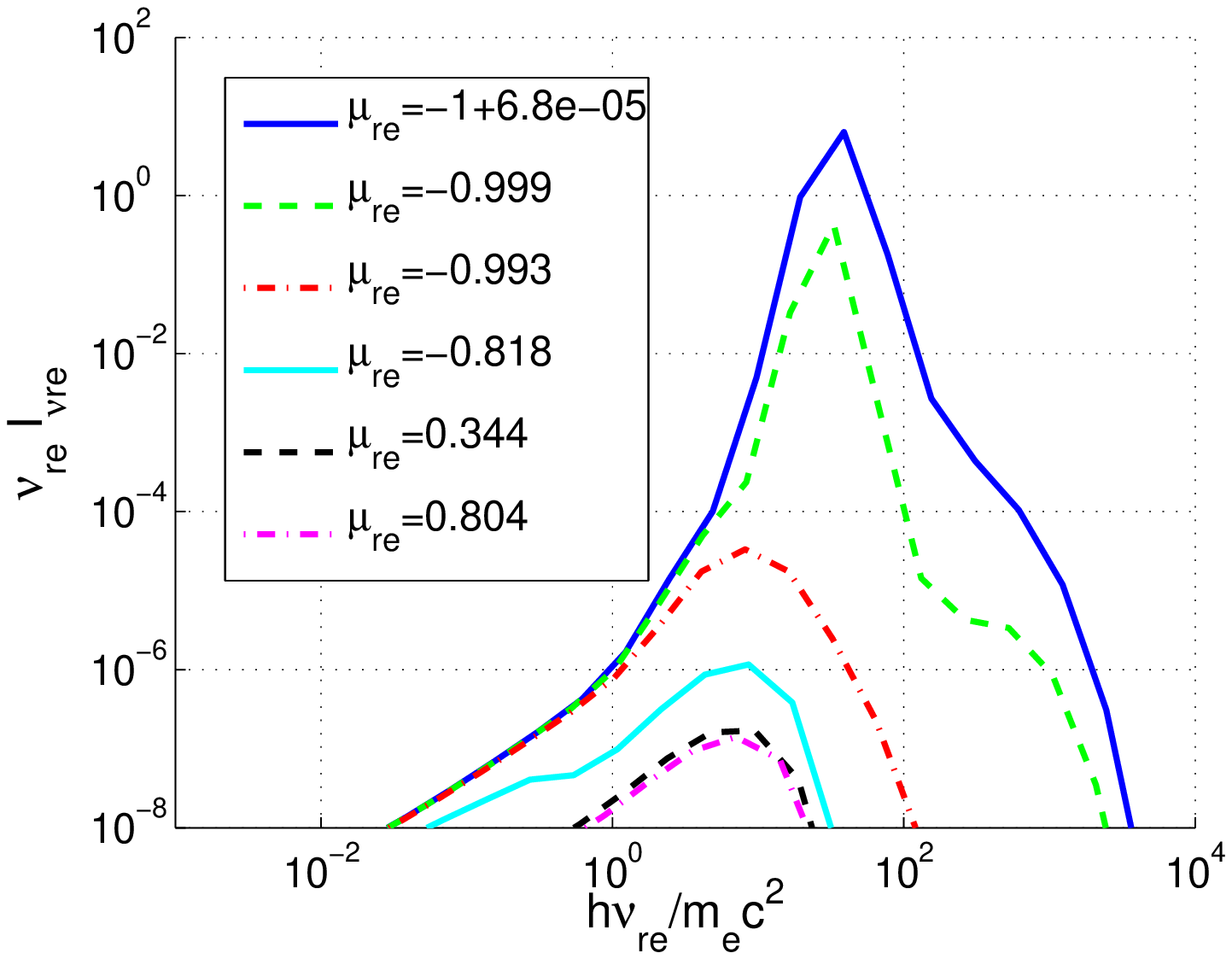}\caption{The plasma rest frame radiation spectrum $ \hat \nu \hat I _{\hat \nu } $ vs. $ \hat{\nu} $, for  $ \Gamma_u=10 $ in the US ($ \Gamma=9.9 $).}
\label{fig:spec_10_US_re}
\end{minipage}
\hspace{0.5cm}
\begin{minipage}[t]{0.5\linewidth}
\centering
\includegraphics[scale=0.45]{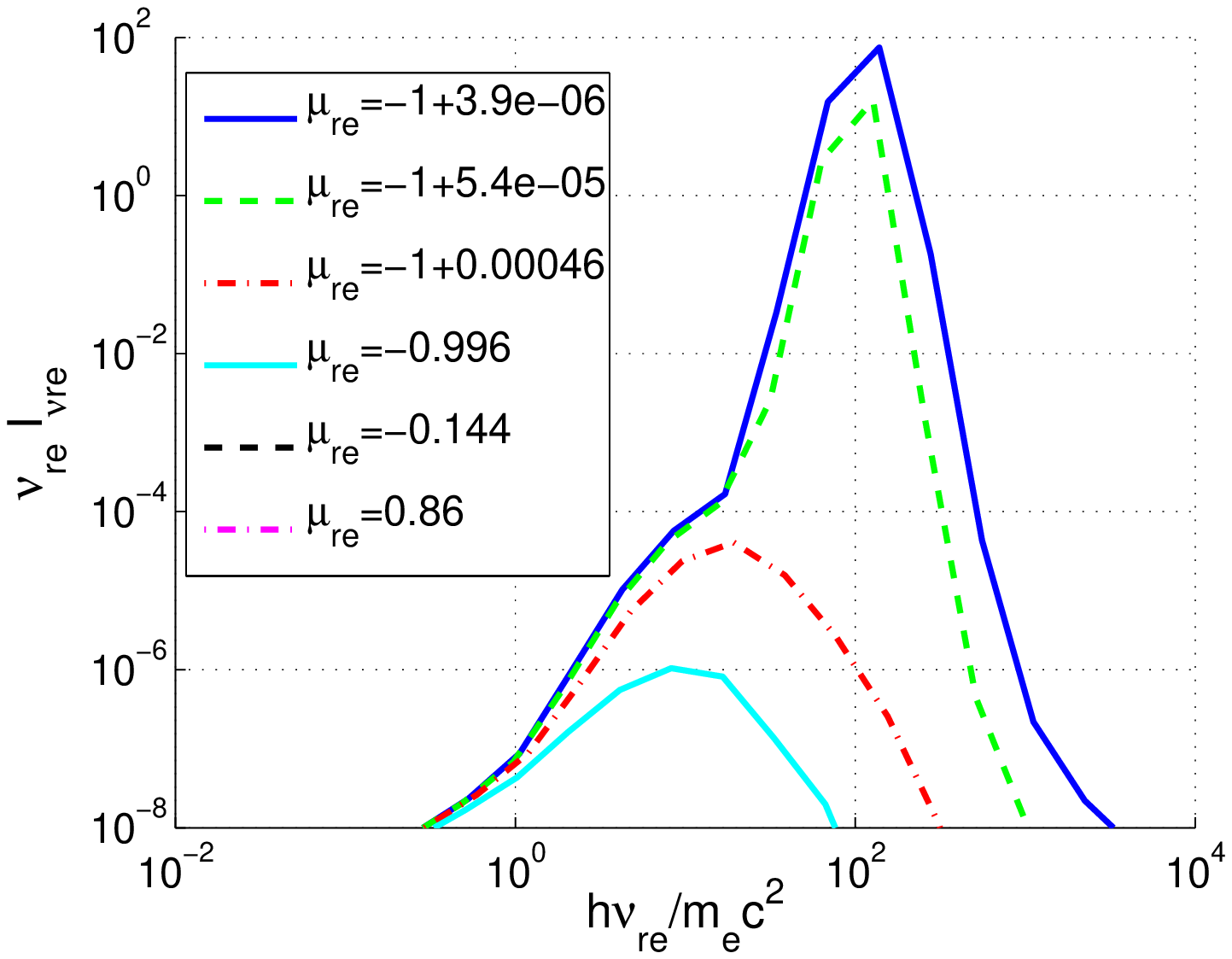}\caption{The plasma rest frame radiation spectrum $ \hat \nu \hat I _{\hat \nu } $ vs. $ \hat{\nu} $, for  $ \Gamma_u=30 $ in the US ($ \Gamma=29.7 $).}
\label{fig:spec_30_US_re}
\end{minipage}
\end{figure*}

\begin{figure*}[ht]
\begin{minipage}[t]{0.5\linewidth}
\centering
\includegraphics[scale=0.45]{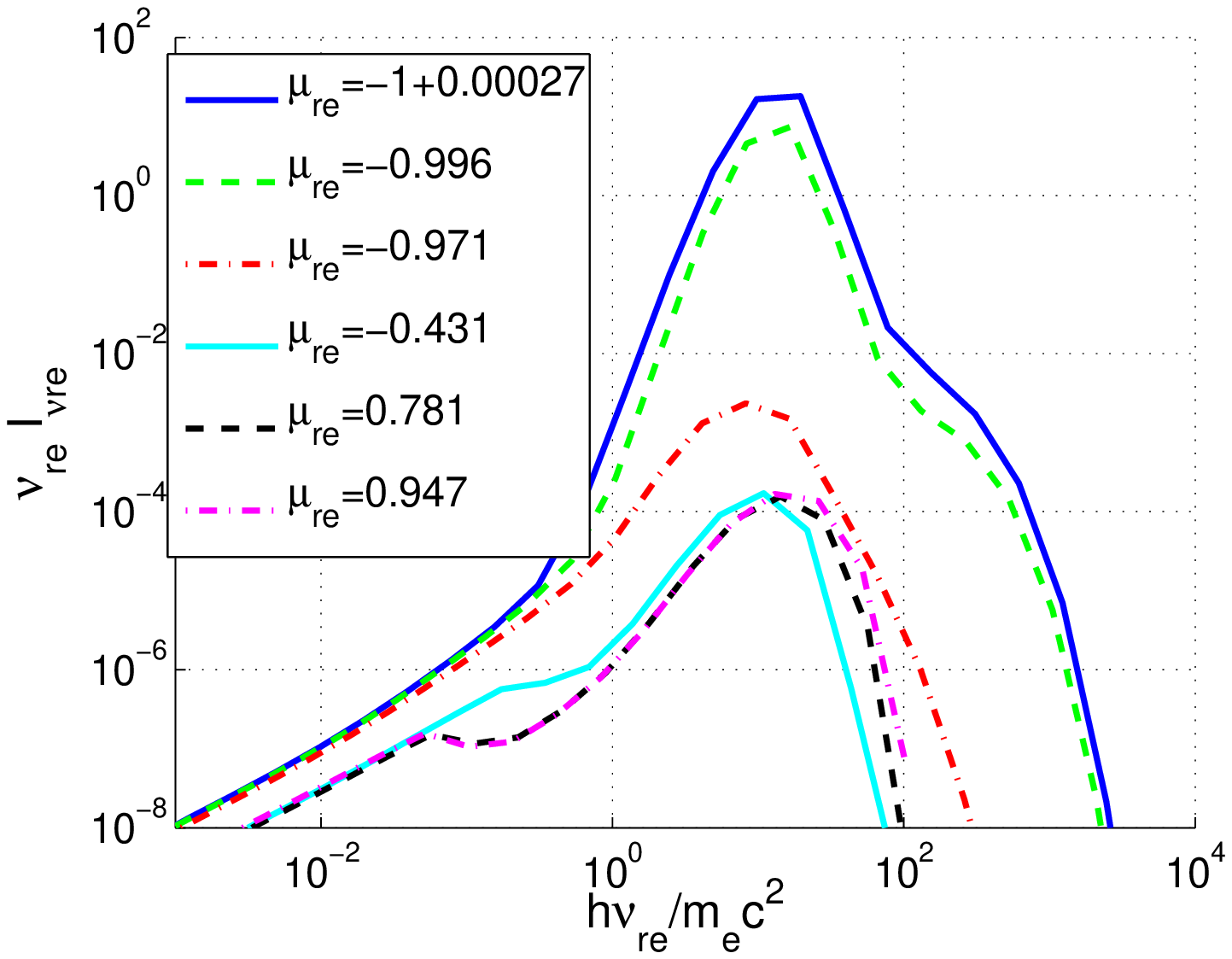}\caption{The plasma rest frame radiation spectrum $ \hat \nu \hat I _{\hat \nu } $ vs. $ \hat{\nu} $, for  $ \Gamma_u=10 $ in the middle of the transition ($ \Gamma=5 $).}
\label{fig:spec_10_MID_re}
\end{minipage}
\hspace{0.5cm}
\begin{minipage}[t]{0.5\linewidth}
\centering
\includegraphics[scale=0.45]{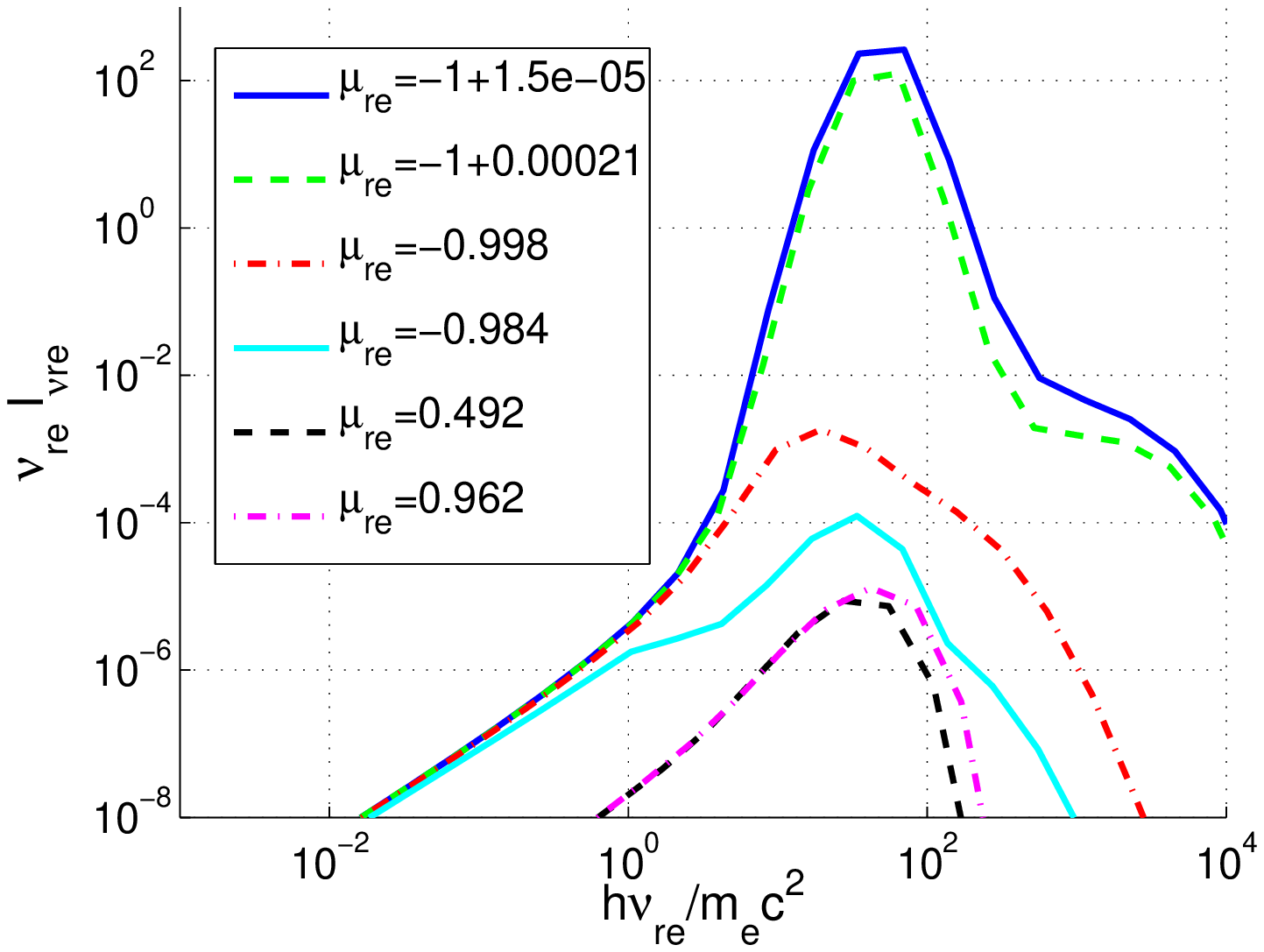}\caption{The plasma rest frame radiation spectrum $ \hat \nu \hat I _{\hat \nu } $ vs. $ \hat{\nu} $, for  $ \Gamma_u=30 $ in the middle of the transition ($ \Gamma=15 $).}
\label{fig:spec_30_MID_re}
\end{minipage}
\end{figure*}

\begin{figure*}[ht]
\begin{minipage}[t]{0.5\linewidth}
\centering
\includegraphics[scale=0.45]{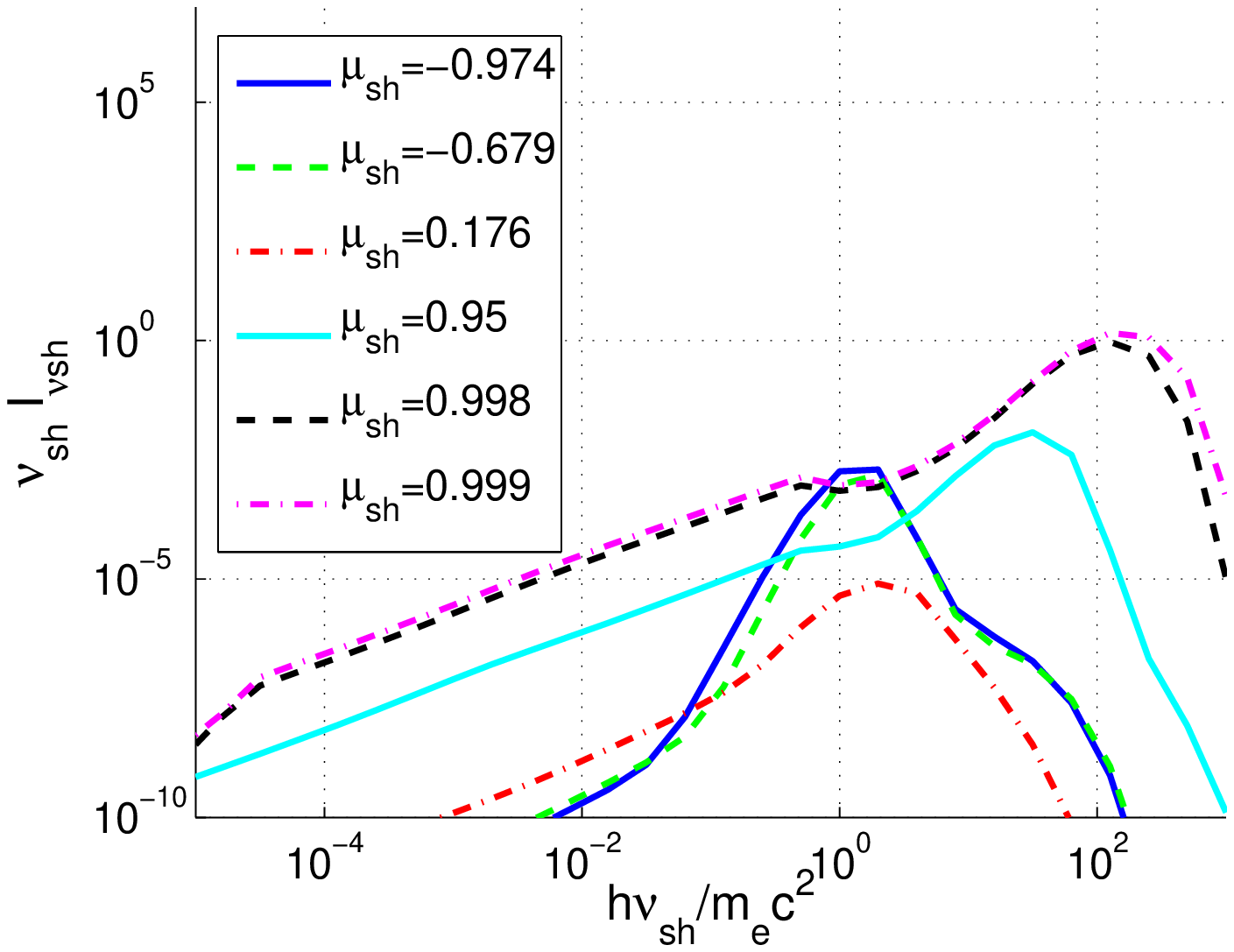}\caption{The shock frame radiation spectrum $ \hat \nu_{sh} \hat I_{sh,\hat \nu_{sh} } $ vs. $ \hat{\nu}_{sh} $, for  $ \Gamma_u=10 $ witihn the transition region ($ \Gamma=5 $).}
\label{fig:spec_10_MID_sh}
\end{minipage}
\hspace{0.5cm}
\begin{minipage}[t]{0.5\linewidth}
\centering
\includegraphics[scale=0.45]{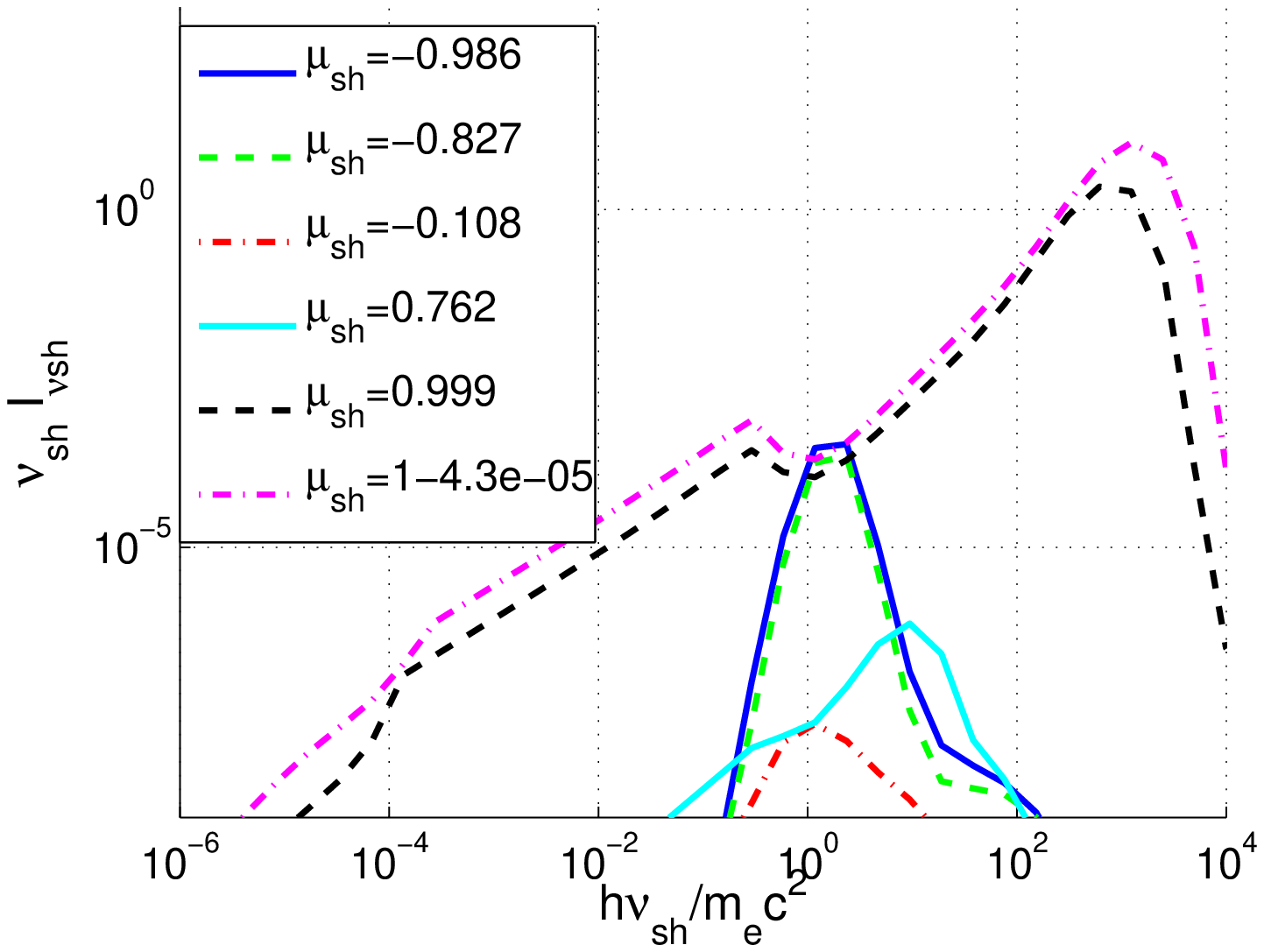}\caption{The shock frame radiation spectrum $ \hat \nu_{sh} \hat I_{sh,\hat \nu_{sh} } $ vs. $ \hat{\nu}_{sh} $, for  $ \Gamma_u=30 $ witihn the transition region ($ \Gamma=15 $).}
\label{fig:spec_30_MID_sh}
\end{minipage}
\end{figure*}

\begin{figure*}[ht]
\begin{minipage}[t]{0.5\linewidth}
\centering
\includegraphics[scale=0.45]{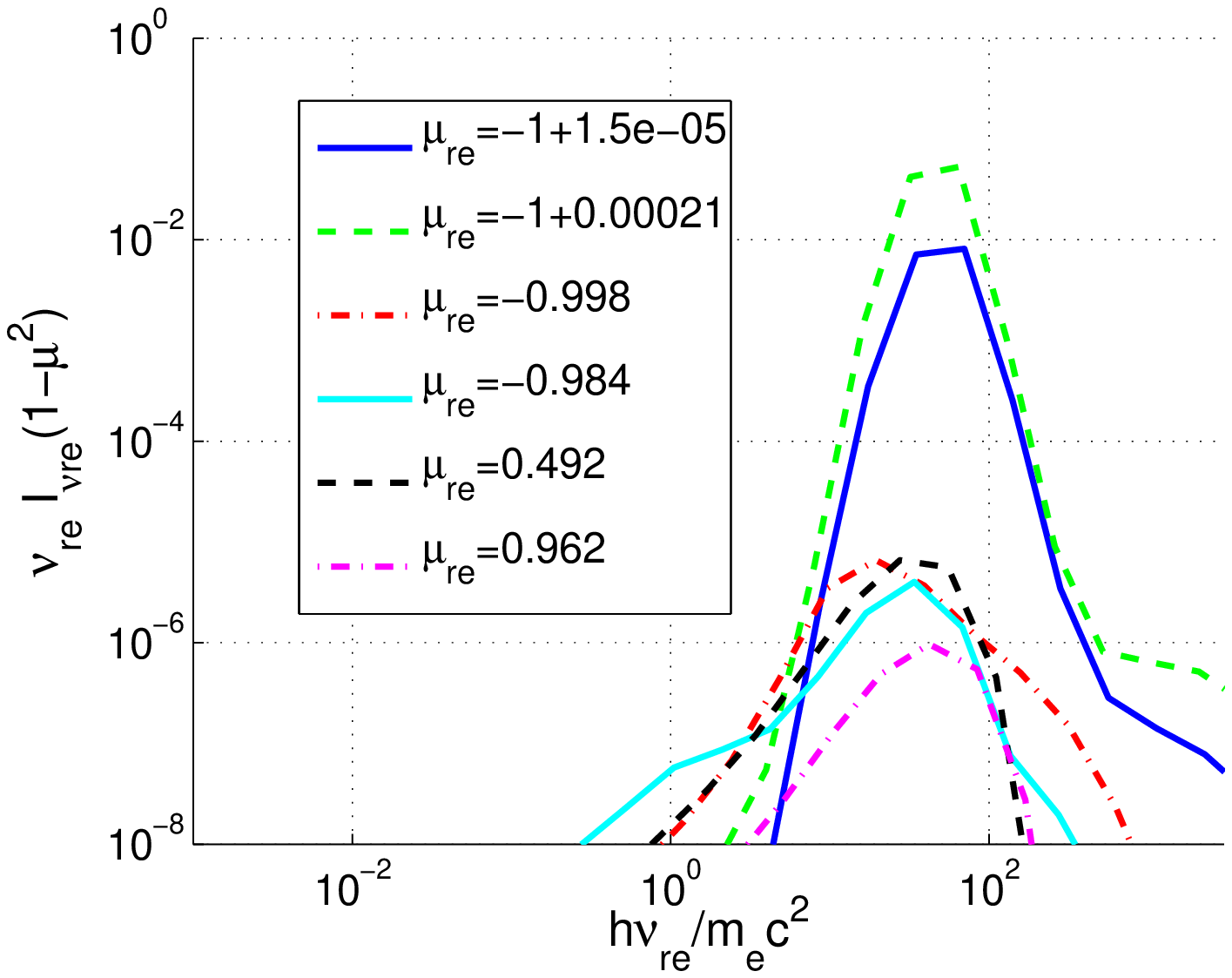}\caption{The solid angle weighted rest frame radiation spectrum, $ \hat \nu_{re} \hat I_{re,\hat \nu_{re} }(1-\mu_{re}^2) $, for  $ \Gamma_u=30 $ witihn the transition region ($ \Gamma=15 $).}
\label{fig:spec_30_mid_rest_mus}
\end{minipage}
\hspace{0.5cm}
\begin{minipage}[t]{0.5\linewidth}
\centering
\includegraphics[scale=0.45]{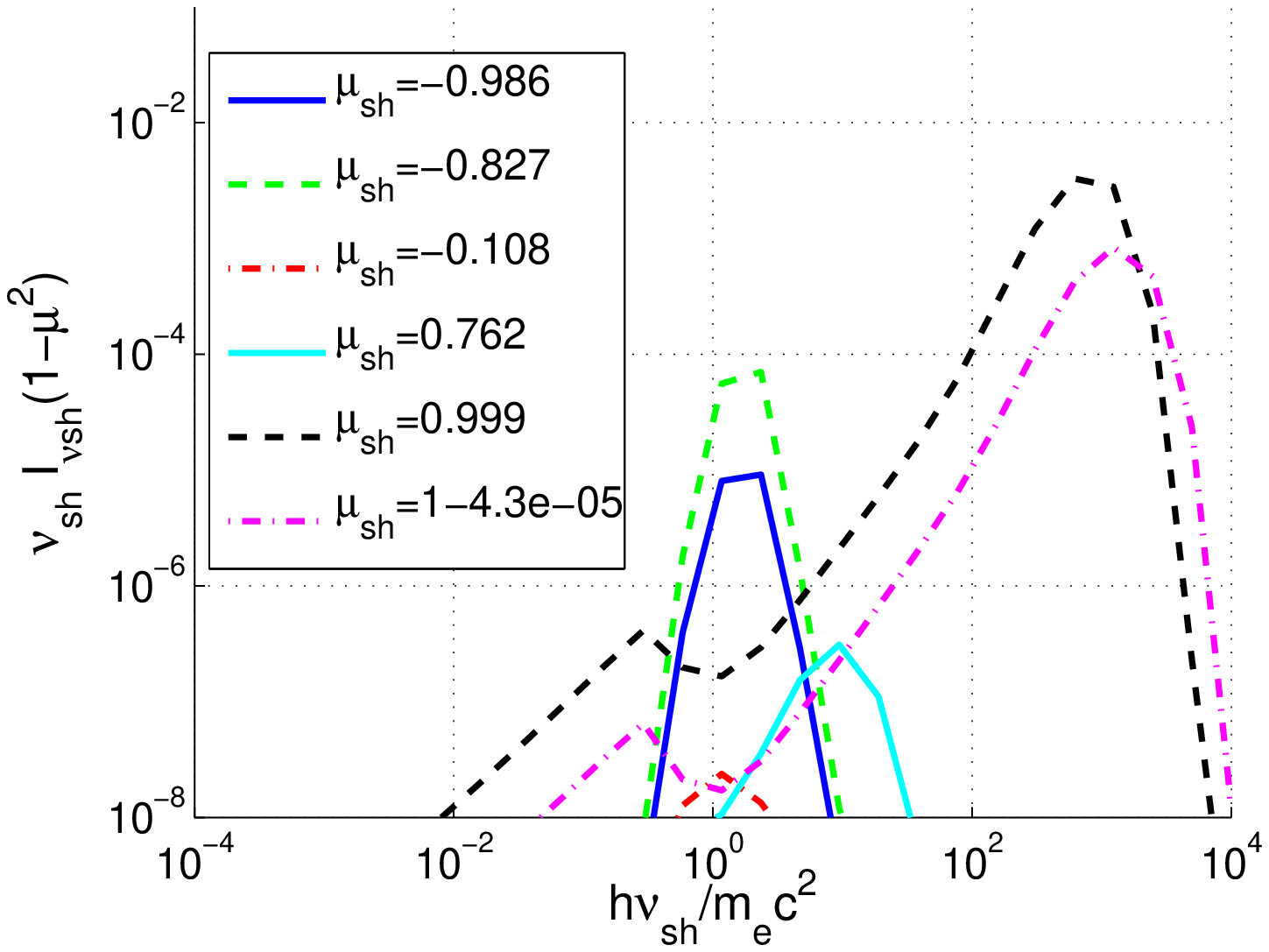}\caption{The solid angle weighted shock frame radiation spectrum, $ \hat \nu_{sh} \hat I_{sh,\hat \nu_{sh} }(1-\mu_{sh}^2) $, for  $ \Gamma_u=30 $ witihn the transition region ($ \Gamma=15 $).}
\label{fig:spec_30_mid_shock_mus}
\end{minipage}
\end{figure*}

\begin{figure*}[ht]
\begin{minipage}[t]{0.5\linewidth}
\centering
\includegraphics[scale=0.45]{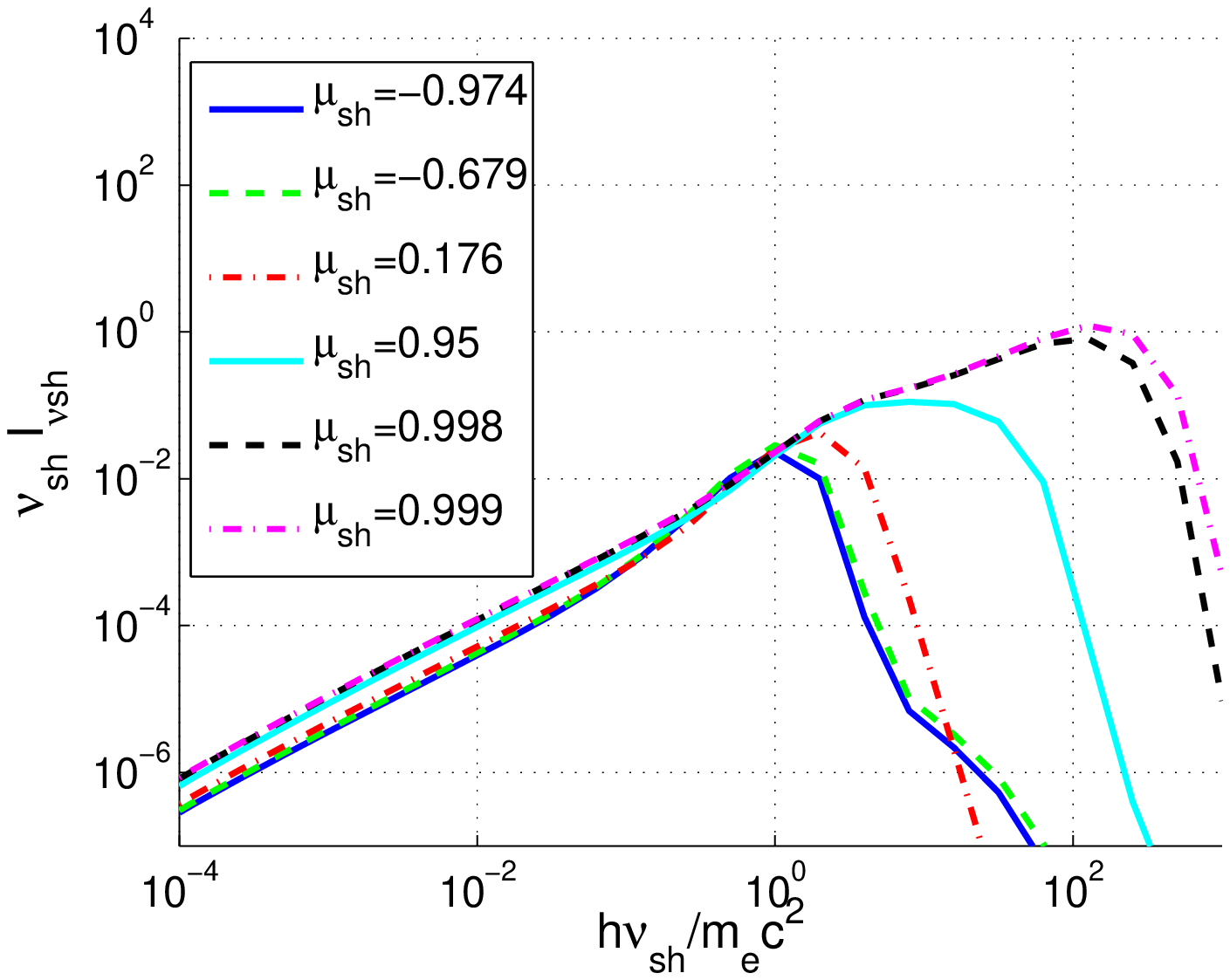}\caption{The shock frame radiation spectrum $ \hat \nu_{sh} \hat I_{sh,\hat \nu_{sh} } $ vs. $ \hat{\nu}_{sh} $, for  $ \Gamma_u=10 $ in the immediate DS ($ \tau_*=1 $).}
\label{fig:spec_10_DS_sh}
\end{minipage}
\hspace{0.5cm}
\begin{minipage}[t]{0.5\linewidth}
\centering
\includegraphics[scale=0.45]{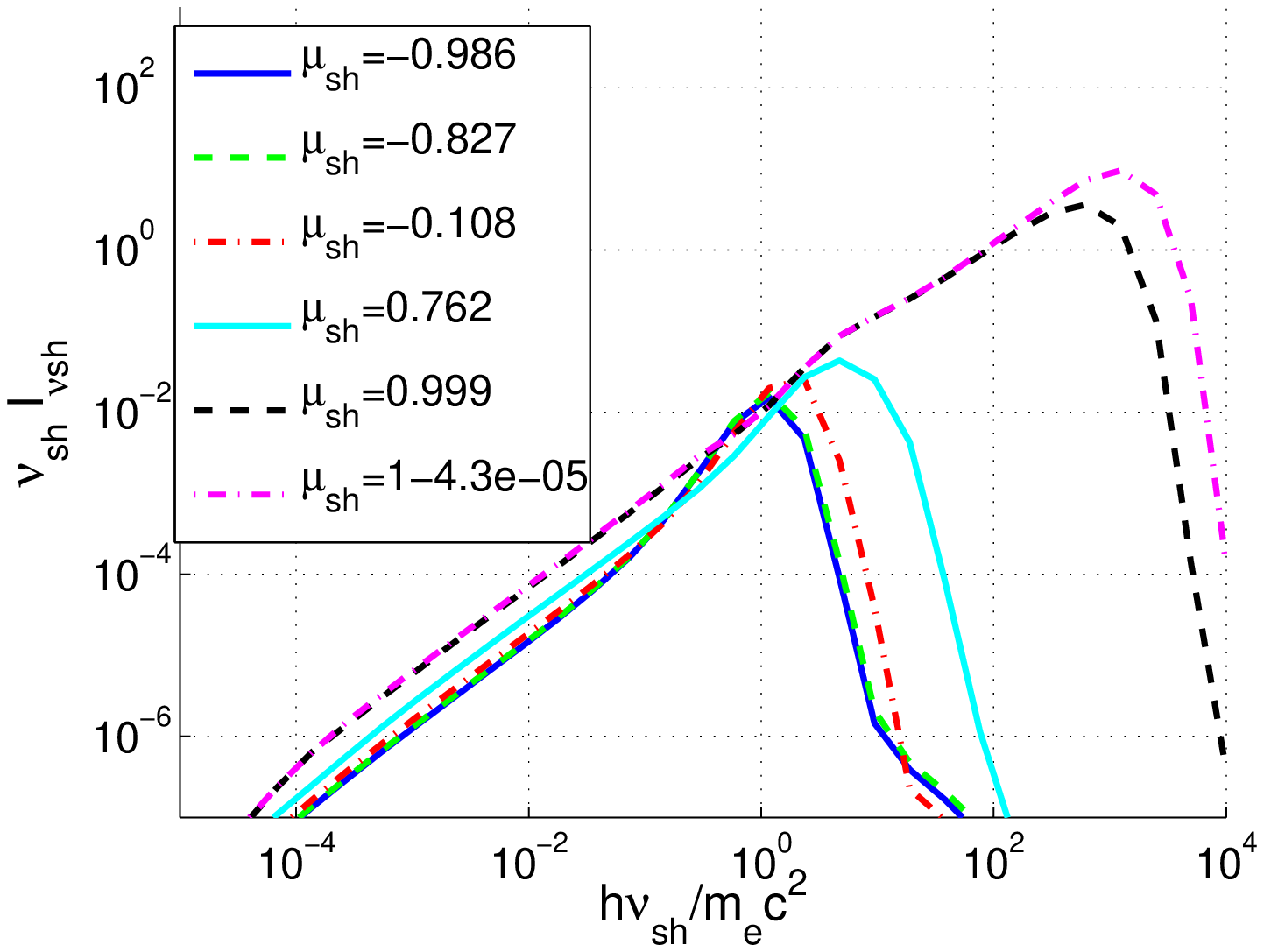}\caption{The shock frame radiation spectrum $ \hat \nu_{sh} \hat I_{sh,\hat \nu_{sh} } $ vs. $ \hat{\nu}_{sh} $, for  $ \Gamma_u=30 $ in the immediate DS ($ \tau_*=1 $).}
\label{fig:spec_30_DS_sh}
\end{minipage}
\end{figure*}

Figures \ref{fig:spec_10_US_re} to \ref{fig:spec_30_DS_sh} show the radiation spectrum at different points along the shock profile for the cases $ \Gamma_u=10 $ and $ \Gamma_u=30 $. { The normalization of the intensity and frequency is given in Eq.  \eqref{eq:dimless_def}.} The points of interest are:
\begin{enumerate}
\item The upstream -  where $ \Gamma=0.99 \Gamma_u $. At this point we show the spectrum in the rest frame of the plasma (Figs. \ref{fig:spec_10_US_re} and \ref{fig:spec_30_US_re} for $ \Gamma_u=10 $ and $ \Gamma_u=30 $, respectively).
\item The transition - where $ \Gamma=\Gamma_u/2 $. At this point we show the spectrum in the rest frame of the plasma (Figs. \ref{fig:spec_10_MID_re} and \ref{fig:spec_30_MID_re} for $ \Gamma_u=10 $ and $ \Gamma_u=30 $, respectively), and in the shock frame (Figs. \ref{fig:spec_10_MID_sh} and \ref{fig:spec_30_MID_sh} for $ \Gamma_u=10 $ and $ \Gamma_u=30 $, respectively).
\item The immediate DS - One Thomson optical depth ($ \tau_* =1$) downstream of the subshock. At this point we show the spectrum in the shock frame (Figs. \ref{fig:spec_10_DS_sh} and \ref{fig:spec_30_DS_sh} for $ \Gamma_u=10 $ and $ \Gamma_u=30 $, respectively).
\end{enumerate}

\begin{figure*}[ht]
\begin{minipage}[t]{0.5\linewidth}
\centering
\includegraphics[scale=0.5]{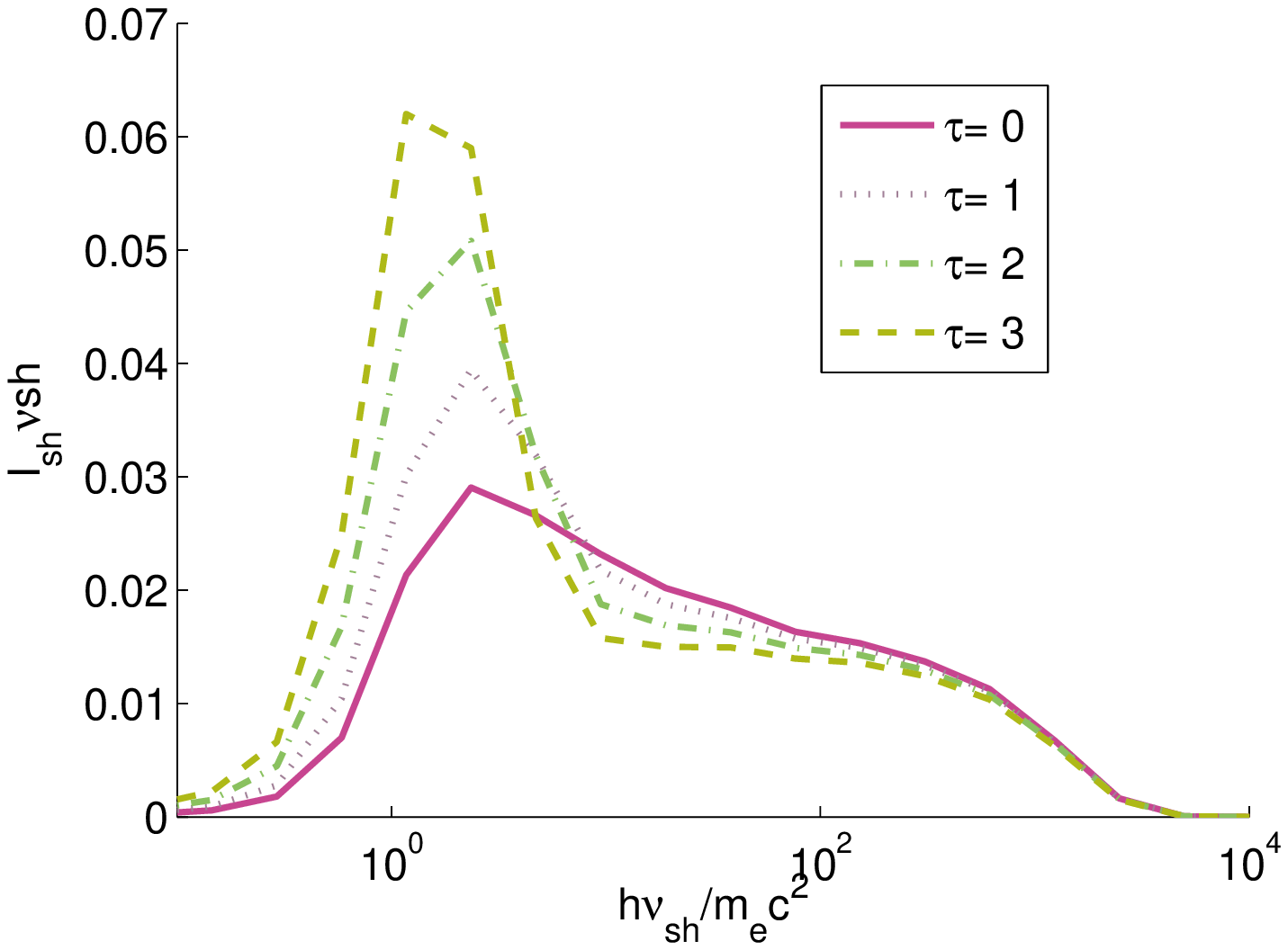}\caption{The  integrated spectrum in the shock frame, $ \hat \nu_{sh} \int \hat I_{sh,\hat \nu_{sh}}(\mu_{sh})d\mu_{sh} $, at different depths inside the immediate DS ($ \tau_*>0$) for $ \Gamma_u=30 $.}
\label{fig:spec_int_30}
\end{minipage}
\hspace{0.5cm}
\begin{minipage}[t]{0.5\linewidth}
\centering
\includegraphics[scale=0.5]{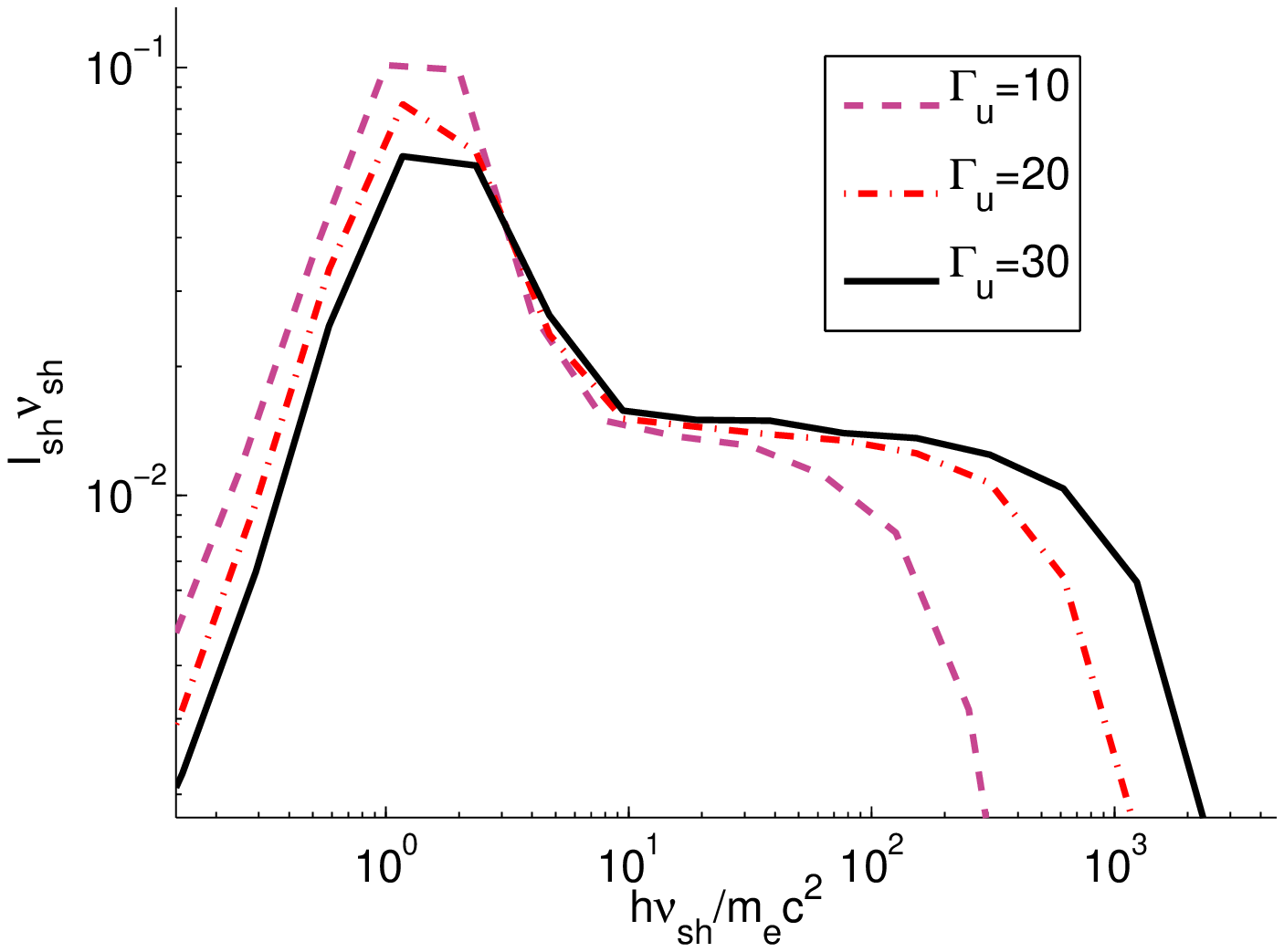}\caption{The  integrated spectrum in the shock frame, $ \hat \nu_{sh} \int \hat I_{sh,\hat \nu_{sh}}(\mu_{sh})d\mu_{sh} $, in the immediate DS ($ \tau_*=3$) for $ \Gamma_u=10 $, $ \Gamma_u=20 $ and $ \Gamma_u=30 $. }
\label{fig:int_spec_ds_10_20_30}
\end{minipage}
\end{figure*}

We now give a short description of the main characteristics of the spectrum at different locations across the shock. An extensive analysis and an analytic description of the results is given in section \sref{sec:analytic}.
\begin{itemize}
\item{\bf Upstream: }
The rest frame spectrum (figs. \ref{fig:spec_10_US_re} and \ref{fig:spec_30_US_re}) is strongly dominated by  a photon component beamed in the US direction, with a typical energy of $ \sim 3\Gamma_u m_ec^2 $, and a much weaker, isotropic component with energy $ \sim \Gamma_u m_ec^2  $. In the shock frame (not shown here), the dominant component is  beamed in the DS direction, with characteristic energy $ \sim \Gamma_u^2 m_ec^2$. There is also a weaker and not strongly beamed  US going component with energy somewhat higher than $ m_ec^2 $.

\item{\bf Transition region: }
The radiation in this region is extremely anisotropic in both the shock frame and the rest frame of the plasma. In the rest frame (figs. \ref{fig:spec_10_MID_re}, \ref{fig:spec_30_MID_re}) the radiation is dominated by a high energy component beamed in the US direction, with a typical energy of $ h\nu\approx \Gamma m_ec^2 $, where $ \Gamma $ is the local Lorentz factor. An isotropic component, which is much weaker in intensity and with typical photon energy similar to the beamed component, also exists. In the shock frame (figs. \ref{fig:spec_10_MID_sh}, \ref{fig:spec_30_MID_sh}) the spectrum is composed of a
dominant narrowly beamed component in the DS direction with typical photon energy $ h\nu \sim \Gamma_u^2 m_ec^2 $, and of a much weaker intensity of US going photons with typical energy of $ h\nu\sim m_ec^2 $.

The spectrum in both frames contains highly beamed components. In order to estimate the amount of energy carried by the beams, we show in figures \ref{fig:spec_30_mid_rest_mus} and \ref{fig:spec_30_mid_shock_mus} the intensity $ I $ multiplied by $ 1-\mu^2 $, which for $ 1-\abs{\mu}\ll 1 $ is proportional to the solid angle. 
In the rest frame, the $ h\nu_{re}\approx \Gamma m_ec^2 $ component dominates the total energy, while in the shock frame the energy carried by the US going $ h\nu_{sh}\approx  m_ec^2 $ photons is comparable to that of DS going $ h\nu_{sh}\approx \Gamma^2 m_ec^2 $ photons. 

\item{\bf Immediate DS: }
Figs. \ref{fig:spec_10_DS_sh} and \ref{fig:spec_30_DS_sh} show that the spectrum is composed of two components: a relatively isotropic component with $ h\nu\sim m_ec^2 $, and a component narrowly beamed into the DS direction with energy reaching  $ h\nu \sim \Gamma_u^2 $.

Figures \ref{fig:spec_int_30} and \ref{fig:int_spec_ds_10_20_30} show the spectrum integrated over $ \mu $ in the immediate DS. The integrated spectrum is dominated by photons of energies $ \sim m_ec^2 $, but includes a significant high energy tail. The high energy component  holds 10\%-20\% of the total energy flux of the radiation and is analyzed in \ref{sec:HE-beam}.
\end{itemize}

\subsection{Compton scattering and pair production optical depths}
\label{sec:results_opt_depth}

\begin{figure*}[ht]
\begin{minipage}[t]{0.5\linewidth}
\centering
\includegraphics[scale=0.5]{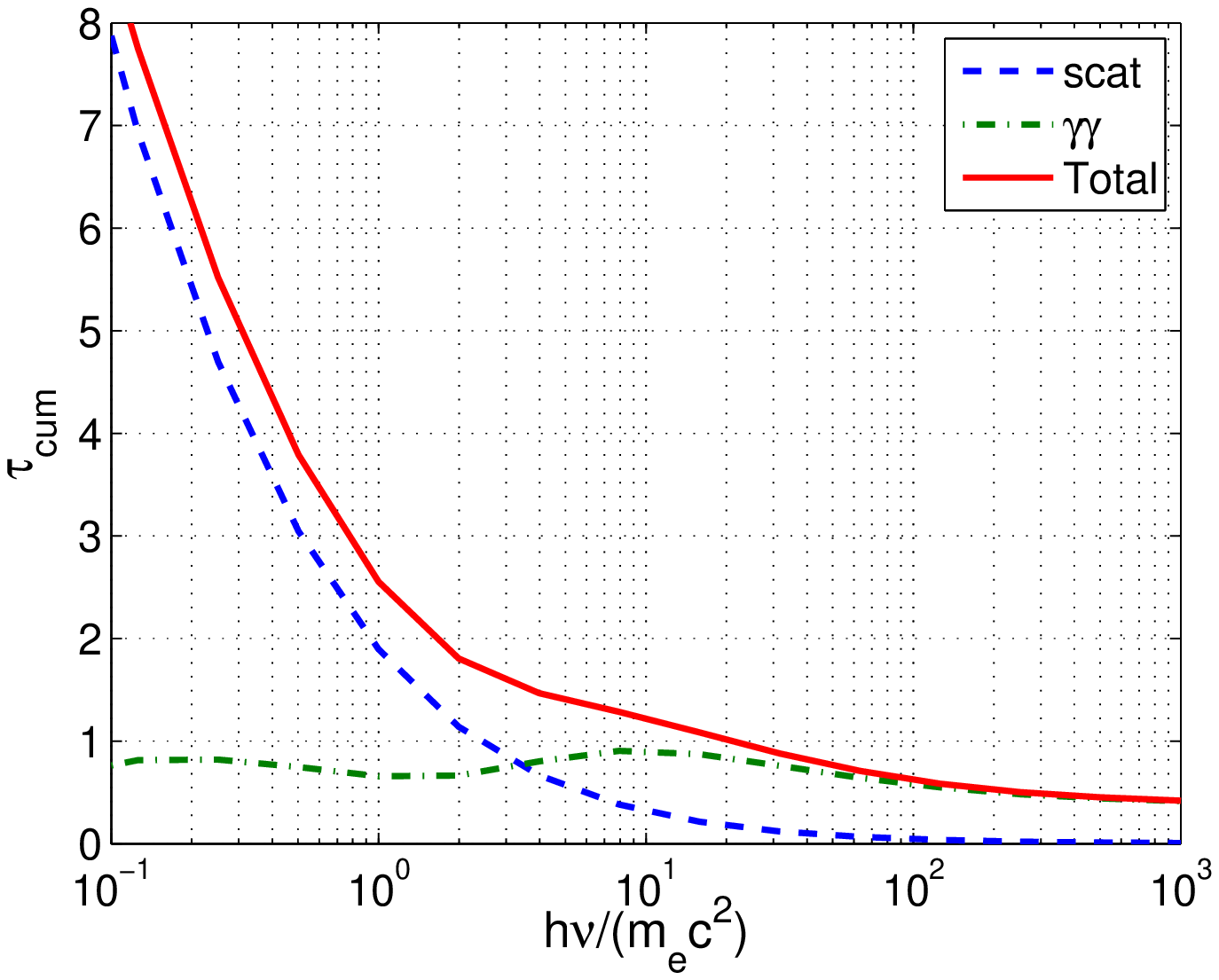}\caption{Cumulative optical depth of US going photons from the subshock to $ \Gamma=\Gamma_u/2 $ vs. shock frame frequency $ \hat \nu_{sh} $, due to Compton scattering and photon-photon pair production, $ \Gamma_u=10 $.  }
\label{fig:tau_us_halfGu_10}
\end{minipage}
\hspace{0.5cm}
\begin{minipage}[t]{0.5\linewidth}
\centering
\includegraphics[scale=0.5]{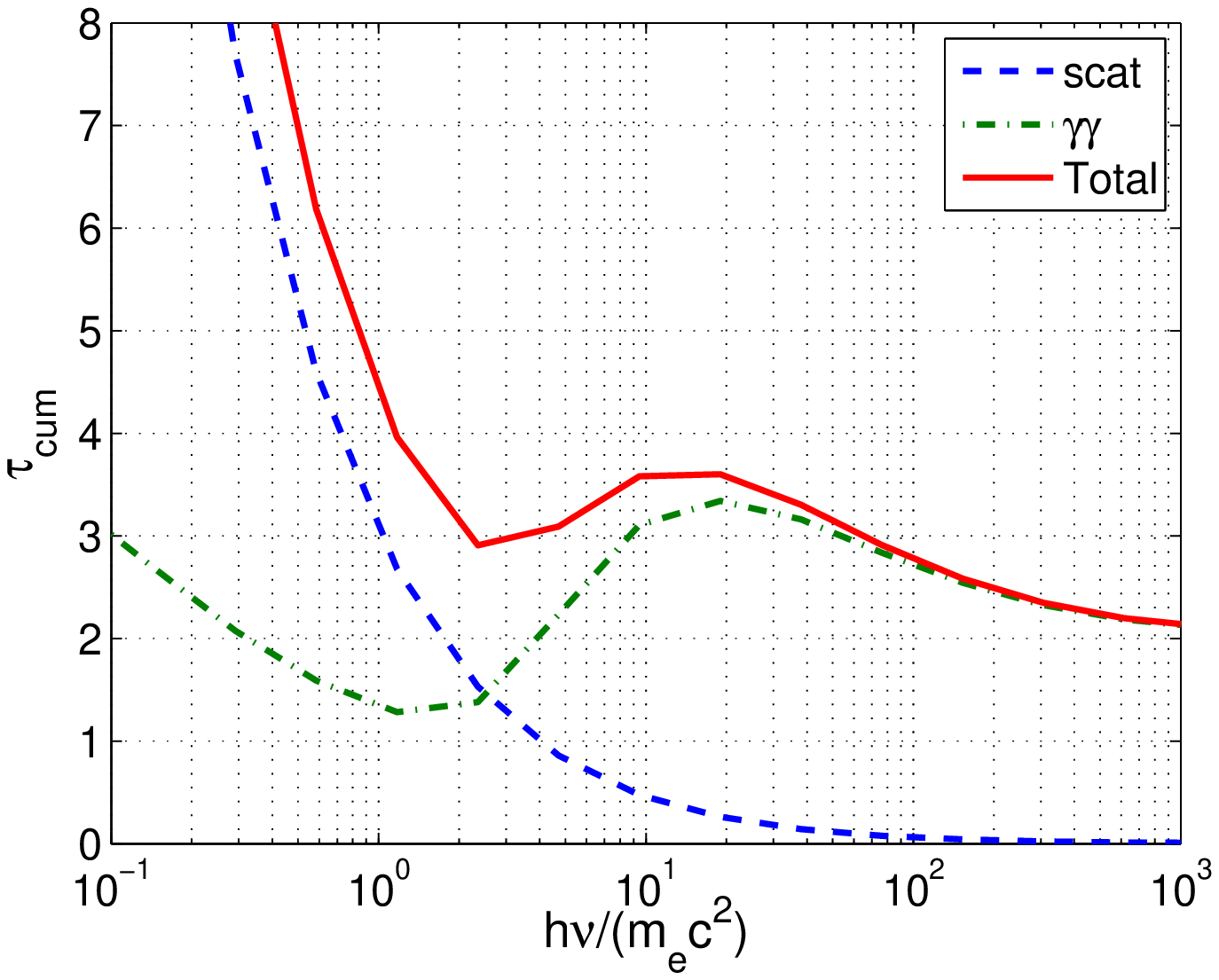}\caption{Cumulative optical depth of US going photons from the subshock to $ \Gamma=\Gamma_u/2 $ vs. shock frame frequency $ \hat \nu_{sh} $, due to Compton scattering and photon-photon pair production, $ \Gamma_u=30 $. }
\label{fig:tau_us_halfGu_30}
\end{minipage}
\end{figure*}

The dominant  mechanisms affecting the radiation in the transition
region are Compton scattering and photon-photon pair production. To
determine the relative importance of the two processes and obtain a handle
on some of the important physical features of the deceleration
mechanism, we examine the optical depth for US going and DS going
photons in the transition region, for the cases $ \Gamma_u=10 $ and
$\Gamma_u=30$. Figures \ref{fig:tau_us_halfGu_10} and
\ref{fig:tau_us_halfGu_30} show the cumulative optical depths for US
going photons leaving the subshock and reaching the point where $
\Gamma=\Gamma_u/2 $ as a function of shock frame frequency. It is
clear that many of the photons with $ \hat \nu_{sh}\gtrsim 1 $  will
make it from the immediate DS to the middle of the transition, while
low energy photons $ \hat \nu_{sh}\ll 1$ will be scattered on the
way.

\begin{figure*}[ht]
\begin{minipage}[t]{0.5\linewidth}
\centering
\includegraphics[scale=0.5]{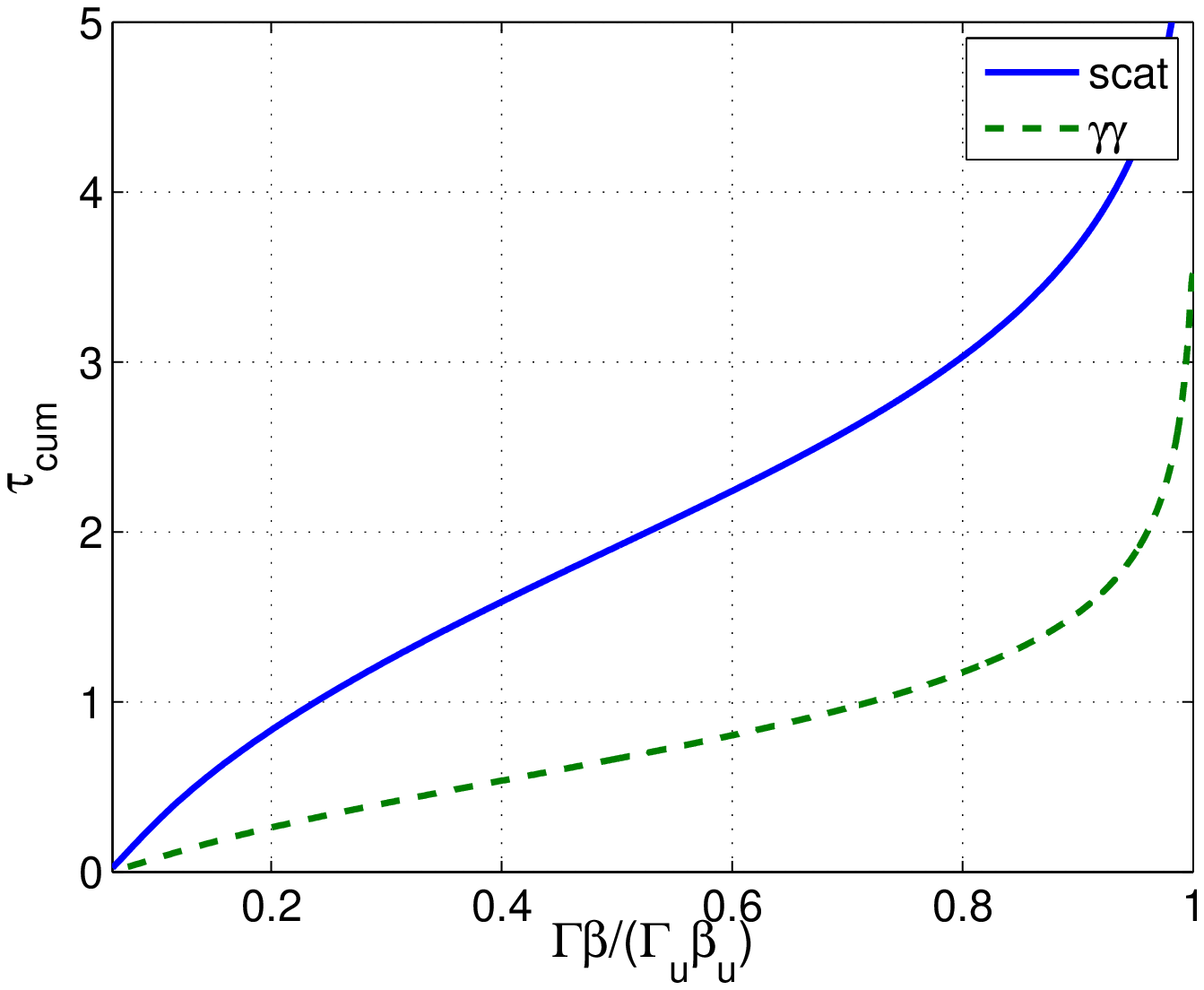}\caption{Cumulative optical depth of US going photons with $ \hat \nu_{sh}=1 $ leaving the subshock  vs.  $ \Gamma\beta/(\Gamma_u\beta_u) $, due to Compton scattering and photon-photon pair production, $ \Gamma_u=10 $.  }
\label{fig:tau_us_all_nu1Gu_10}
\end{minipage}
\hspace{0.5cm}
\begin{minipage}[t]{0.5\linewidth}
\centering
\includegraphics[scale=0.5]{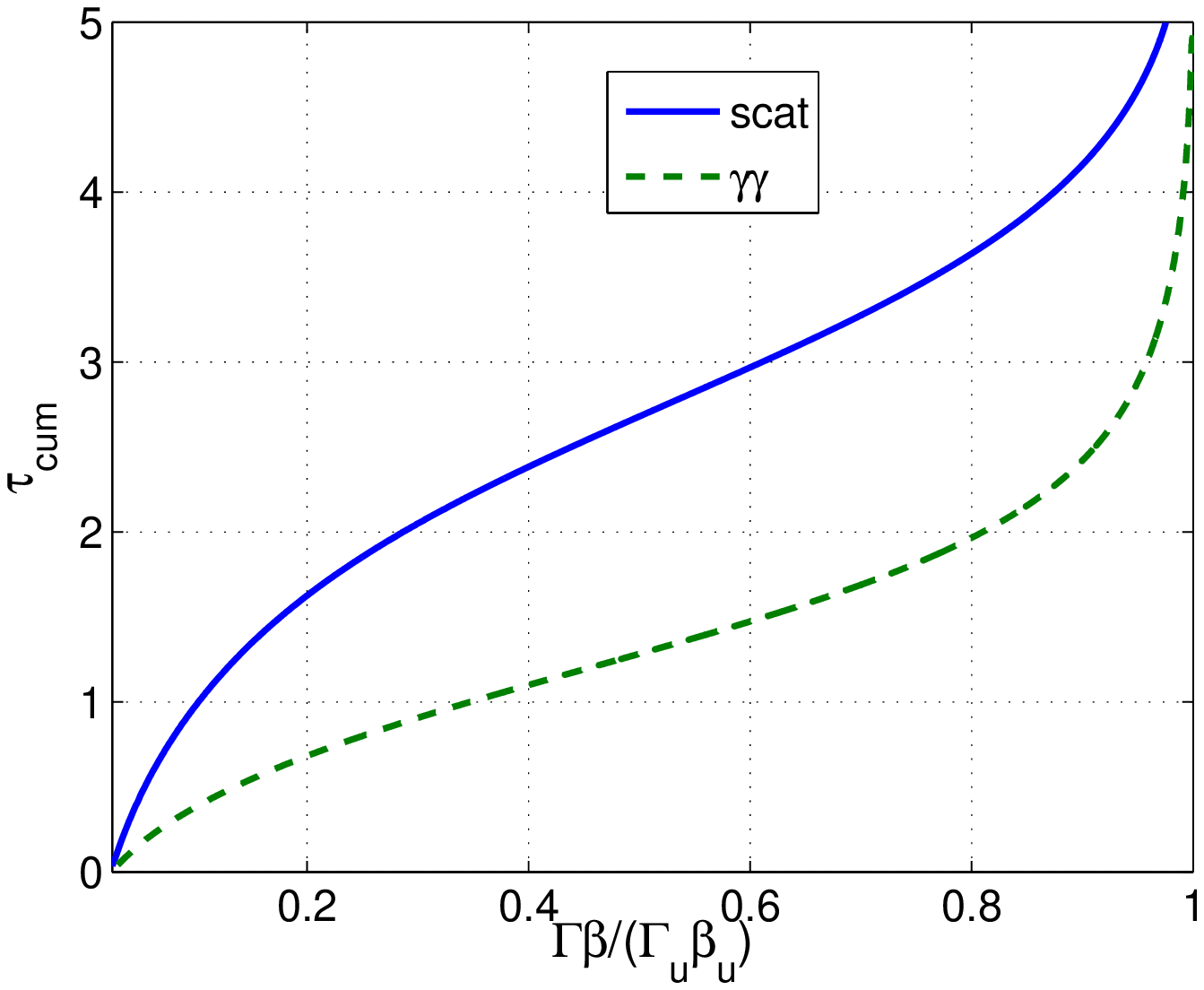}\caption{Cumulative optical depth of US going photons with $ \hat \nu_{sh}=1.1 $ leaving the subshock  vs.  $ \Gamma\beta/(\Gamma_u\beta_u) $, due to Compton scattering and photon-photon pair production, $ \Gamma_u=30 $. }
\label{fig:tau_us_all_nu1Gu_30}
\end{minipage}
\end{figure*}

Figures \ref{fig:tau_us_all_nu1Gu_10} and \ref{fig:tau_us_all_nu1Gu_30} show the cumulative optical depths for US going photons with $ \hat \nu_{sh}\approx 1 $ leaving the subshock, vs. the relativistic velocity $ \Gamma\beta $ of the flow in the transition region. These photons constitute the majority of the photon flux leaving the immediate DS in the US direction. It can be seen that most of the shock profile, up to $ \Gamma\sim 0.9\Gamma_u $, has a total optical of $ \sim 5 $ for these photons, most of it due to Compton scattering, and order unity  optical depth due to photon-photon pair production.

\begin{figure*}[ht]
\begin{minipage}[t]{0.5\linewidth}
\centering
\includegraphics[scale=0.5]{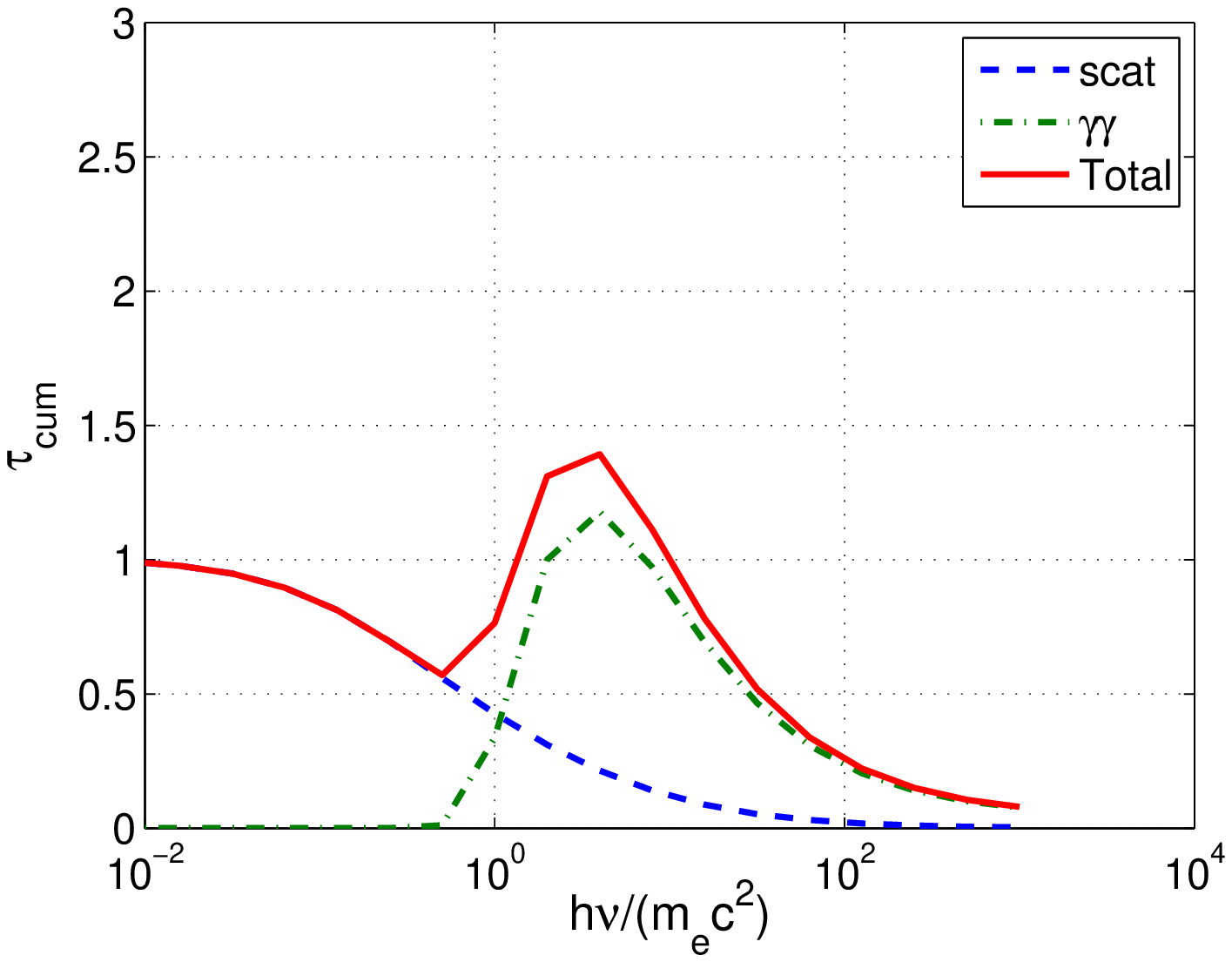}\caption{Cumulative optical depth of DS going photons   from the point $ \Gamma=3 $ to the subshock  vs.  $ \hat \nu_{sh}$, due to Compton scattering and photon-photon pair production, $ \Gamma_u=10 $.}
\label{fig:tau_ds_G3_10}
\end{minipage}
\hspace{0.5cm}
\begin{minipage}[t]{0.5\linewidth}
\centering
\includegraphics[scale=0.5]{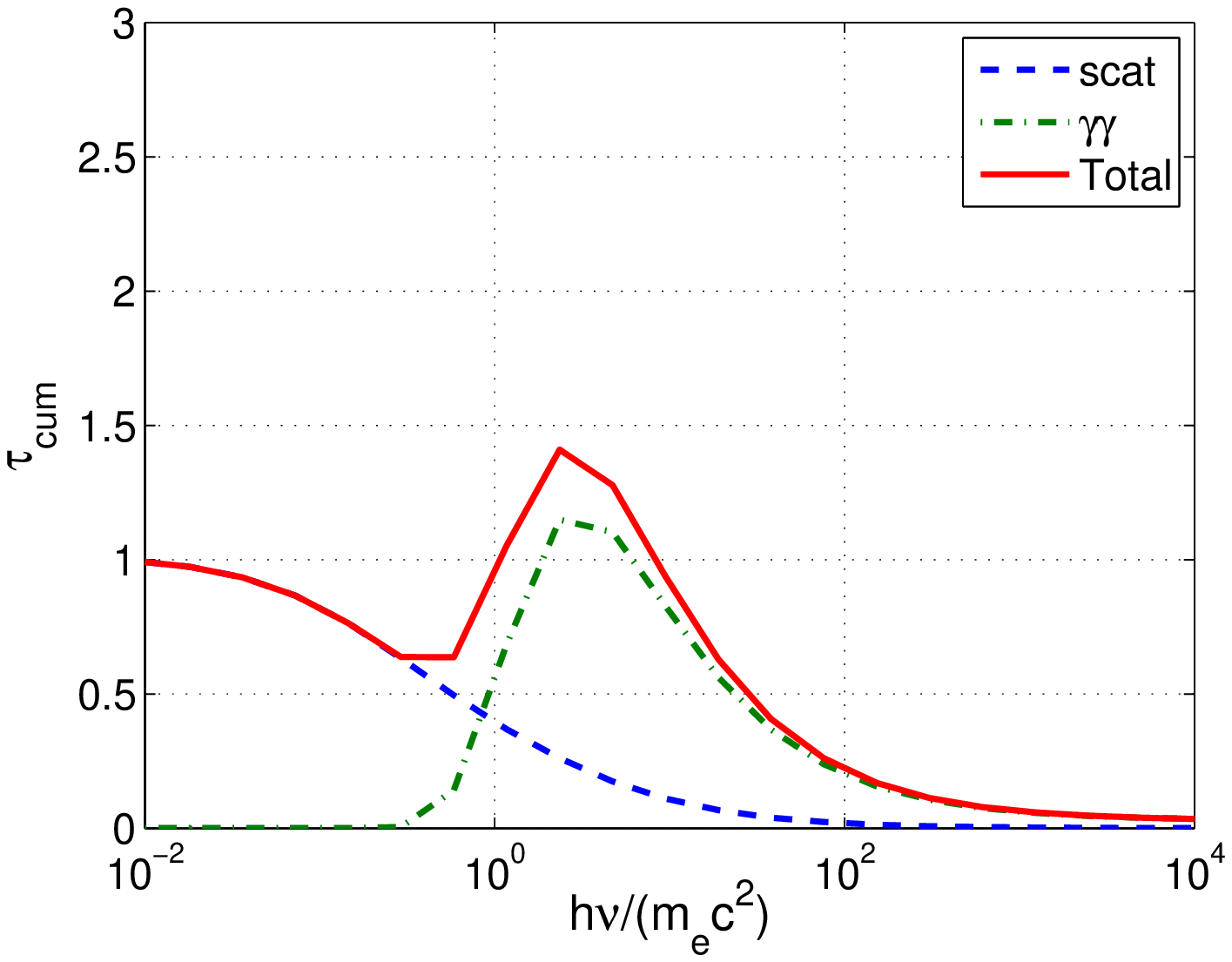}\caption{Cumulative optical depth of DS going photons   from the point $ \Gamma=3 $ to the subshock  vs.  $ \hat \nu_{sh}$, due to Compton scattering and photon-photon pair production, $ \Gamma_u=30 $. }
\label{fig:tau_ds_G3_30}
\end{minipage}
\end{figure*}

Figures \ref{fig:tau_ds_G3_10} and \ref{fig:tau_ds_G3_30} show the cumulative optical depths for DS going photons, starting from the point $ \Gamma=3 $ in the transition and reaching the subshock, as a function of $ \hat \nu_{sh} $. Comparing the results for $ \Gamma_u=30 $ and $ \Gamma_u=10 $ we find that the optical depth due to both scattering and photon-photon pair production are very similar for both values of $ \Gamma_u $, suggesting a common structure and a common upstream going photon spectrum in this region.

\begin{figure*}[ht]
\begin{minipage}[t]{0.5\linewidth}
\centering
\includegraphics[scale=0.5]{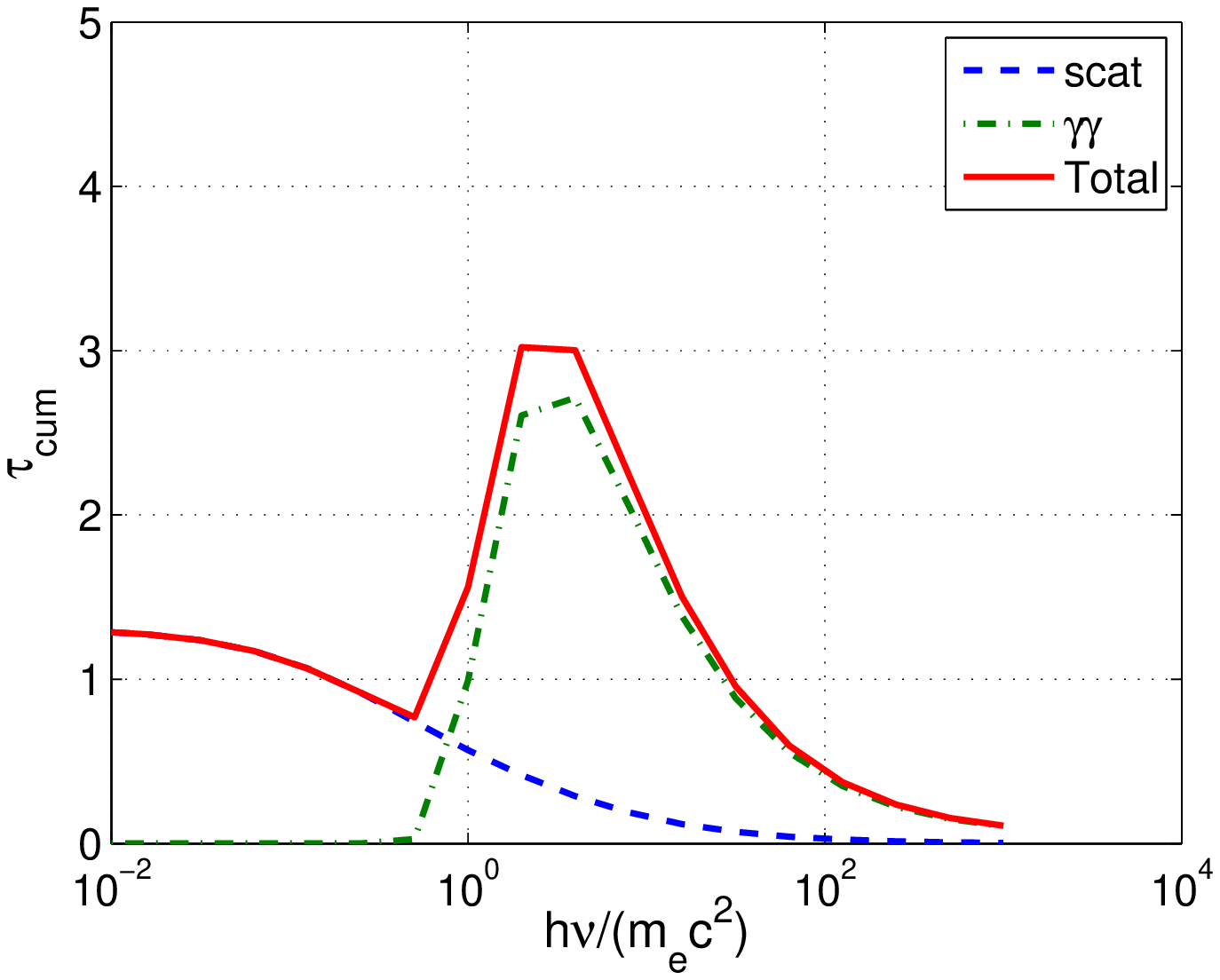}\caption{Cumulative optical depth of DS going photons   from the point $ \Gamma=\Gamma_u/2 $ to the subshock  vs.  $ \hat \nu_{sh}$, due to Compton scattering and photon-photon pair production, $ \Gamma_u=10 $.  }
\label{fig:tau_ds_halfGu_10}
\end{minipage}
\hspace{0.5cm}
\begin{minipage}[t]{0.5\linewidth}
\centering
\includegraphics[scale=0.5]{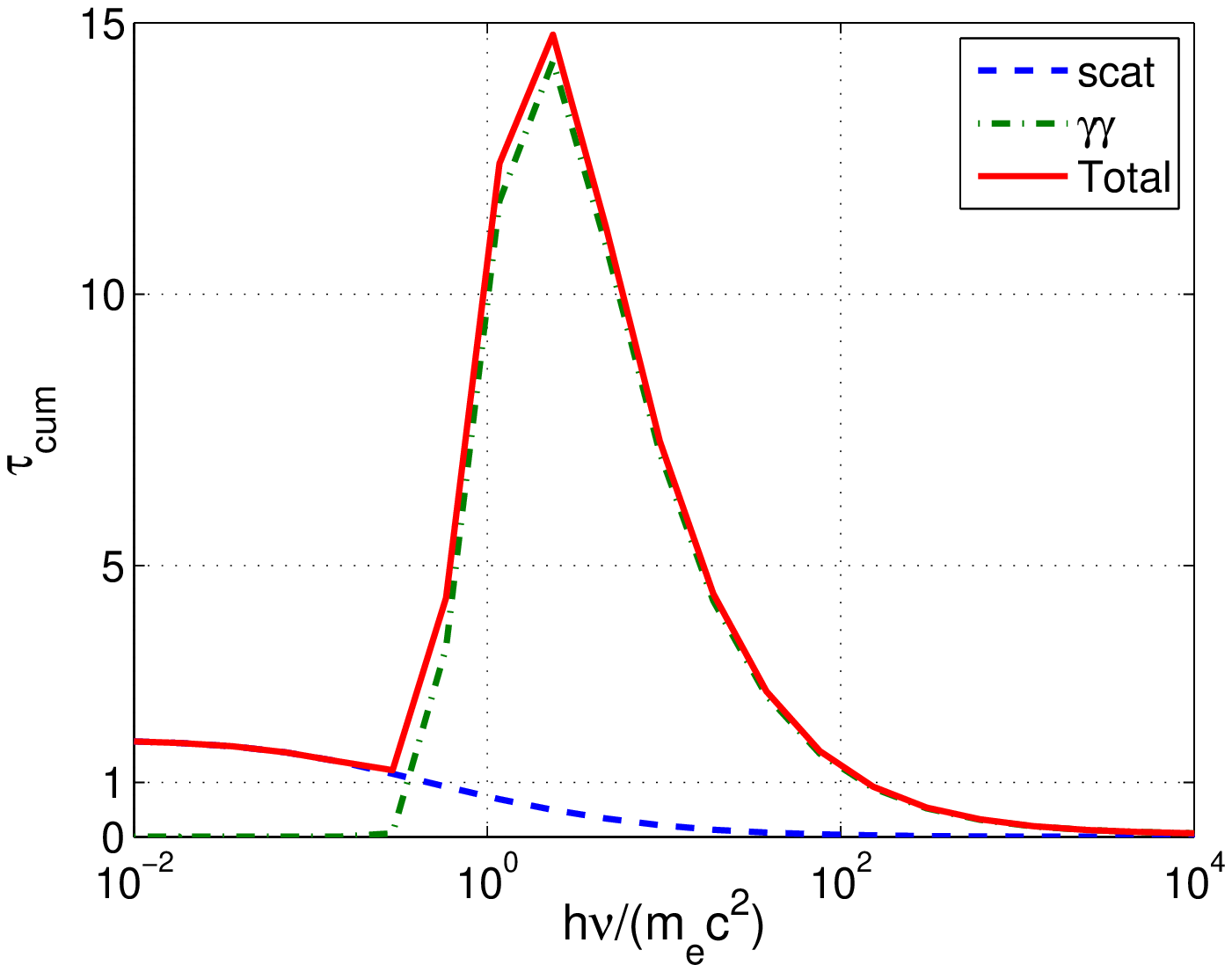}\caption{Cumulative optical depth of DS going photons   from the point $ \Gamma=\Gamma_u/2 $ to the subshock  vs.  $ \hat \nu_{sh}$, due to Compton scattering and photon-photon pair production, $ \Gamma_u=30 $.  }
\label{fig:tau_ds_halfGu_30}
\end{minipage}
\end{figure*}

Figures \ref{fig:tau_ds_halfGu_10} and \ref{fig:tau_ds_halfGu_30} show the cumulative optical depths for DS going photons, starting from the point $ \Gamma=\Gamma_u/2 $ in the transition and reaching the subshock, as a function of $ \hat \nu_{sh} $. As was shown earlier, the shock frame radiation in the transition region is dominated by photons with energy $ \sim \Gamma_u^2 m_ec^2 $ propagating towards the DS. The figures illustrate that the optical depth for these photons to reach the immediate DS is less than unity. On the other hand, photons with energies around the pair production threshold in the shock frame, $ 0.1 \lesssim \hat \nu_{sh}\lesssim 10 $, will suffer a strong attenuation due to pair production.

\subsection{Numerical convergence}\label{sec:NumConv}

\subsubsection{Resolution}

The solution of the equations is obtained using iterations, as described in \sref{sec:iter_scheme}. Iterations are continued until the changes in integral quantities ($ T $, $ \Gamma $, $ x_+ $, $ P_{rad} $ etc.) are less than $ \sim 1\% $ between successive iterations. 
The resolution used for the solutions presented in the preceding sub-sections is given in table \ref{tab:nom_res}.

\begin{table}
\caption{The resolution used for the calculations of the shock profiles.}
\centering\begin{tabular}{|c|c|c|c|c|c|c|c|}
\hline
$\Gamma_u$ & $\delta \tau_*$ & US $\tau_*$ & DS $ \tau_*$ & $\nu_{i+1}/\nu_i$ & $ \hat \nu_{min} $ & $ \hat \nu_{max} $  &  $ N_\mu $ \tabularnewline
\hline
\hline
6 & 0.1 & 200 & 3.5 & 2 & $ 10^{-9} $&$  10^3 $ & 8 \tabularnewline
\hline
10 & 0.1 & 500 & 5 & 2 & $ 10^{-9} $&$  10^3 $ & 13 \tabularnewline
\hline
20 & 0.1 & 1000 & 7 & 2 & $ 10^{-9} $&$  10^4 $ &  18 \tabularnewline
\hline
30 & 0.2 & 2000 & 7 & 2 & $ 10^{-9} $&$ 2\times 10^4 $ & 18 \tabularnewline
\hline
\end{tabular}
\label{tab:nom_res}
\end{table}

We found that the solutions are modified by $ \sim 1\% $ when the resolution in $\tau_*$ is increased from $ \delta \tau_*=0.2 $  to $  \delta \tau_*=0.1  $, and therefore concluded 
that solutions obtained with either resolution are satisfactory. The convergence of
the solutions with respect to the resolution in $ \nu_{sh} $ and $
\mu_{sh} $  was tested using solutions with lower and higher
resolutions for  $ \Gamma_u=10 $. 
We used several  properties of
the solution to quantify the convergence. The solution properties we checked
were:
\begin{itemize}
\item $ T_{jump} $ - the temperature immediately behind the subshock;
\item The maximal $ x_+ $ value;
\item $ P_{sh,jump} $ - the value of the radiation pressure in the shock frame at the subshock;
\item $ -\tau_*(\Gamma\beta=5) $, the normalized optical depth upstream of the subshock at which the Lorentz factor drops by $ \sim$half;
\item $ -\tau_{*,nl} $ - the normalized optical depth upstream of the subshock at which the US evolution becomes nonlinear (see detailed explanation in \sref{sec:far-US});
\end{itemize}
The value of $\tau_{*,nl} $ is very sensitive to small
changes in resolution, since it is set by the exponential decay of the number
of photons arriving from the immediate DS. However, its exact value does not affect
significantly the structure of the deceleration region. We use it here merely as a stringent test of numerical
convergence. 

\begin{figure}\centering
\includegraphics[scale=0.45]{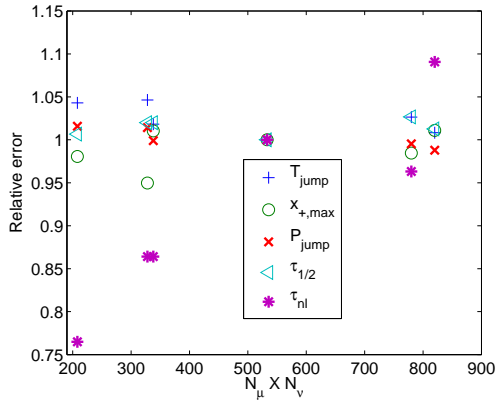}\caption{A summary of numerical convergence tests. Results obtained with different resolutions in $\nu$ and $\mu$ are shown for $ \Gamma_u=10 $, as function of $ N_\nu\times N_\mu $, the product of the number of discrete values chosen for $\mu$ and for $\nu$. The $ y $ axis shows the values of $ T_{jump} $, $ x_{+,max} $, $ P_{sh,jump} $, $ -\tau_*(\Gamma\beta=5) $  and $ -\tau_{*,nl} $ (see text for definitions), divided by their values obtained using the  resolution given in table \ref{tab:nom_res}. { The different sets of $ \{ N_\mu,N_\nu \} $ shown are: $ \{8,26  \} $,  $ \{ 8,41 \} $, 
$ \{13,16  \} $, $ \{ 20,41 \} $ and  $ \{ 13,60 \} $, while the reference is $ \{ 13,41 \} $.} }
\label{fig:res_check}
\end{figure}

The changes in the values of the test parameters as a function of resolution are given
in fig. \ref{fig:res_check}. The results were obtained using lower
and higher resolutions in $ \nu $ and $ \mu $, and are presented as a function of $ N_\nu\times N_\mu $, the product of the number of discrete values chosen for $\mu$ and for $\nu$.
The numerical error around the nominal resolution used in our calculations is few percent, except for the most sensitive parameter, $ \tau_{*,nl} $, for which the numerical error is around $ 10\% $.

\subsubsection{Changes in the length of the DS}\label{sec:ds_length}
\begin{figure*}[ht]
\begin{minipage}[t]{0.5\linewidth}
\centering
\includegraphics[scale=0.5]{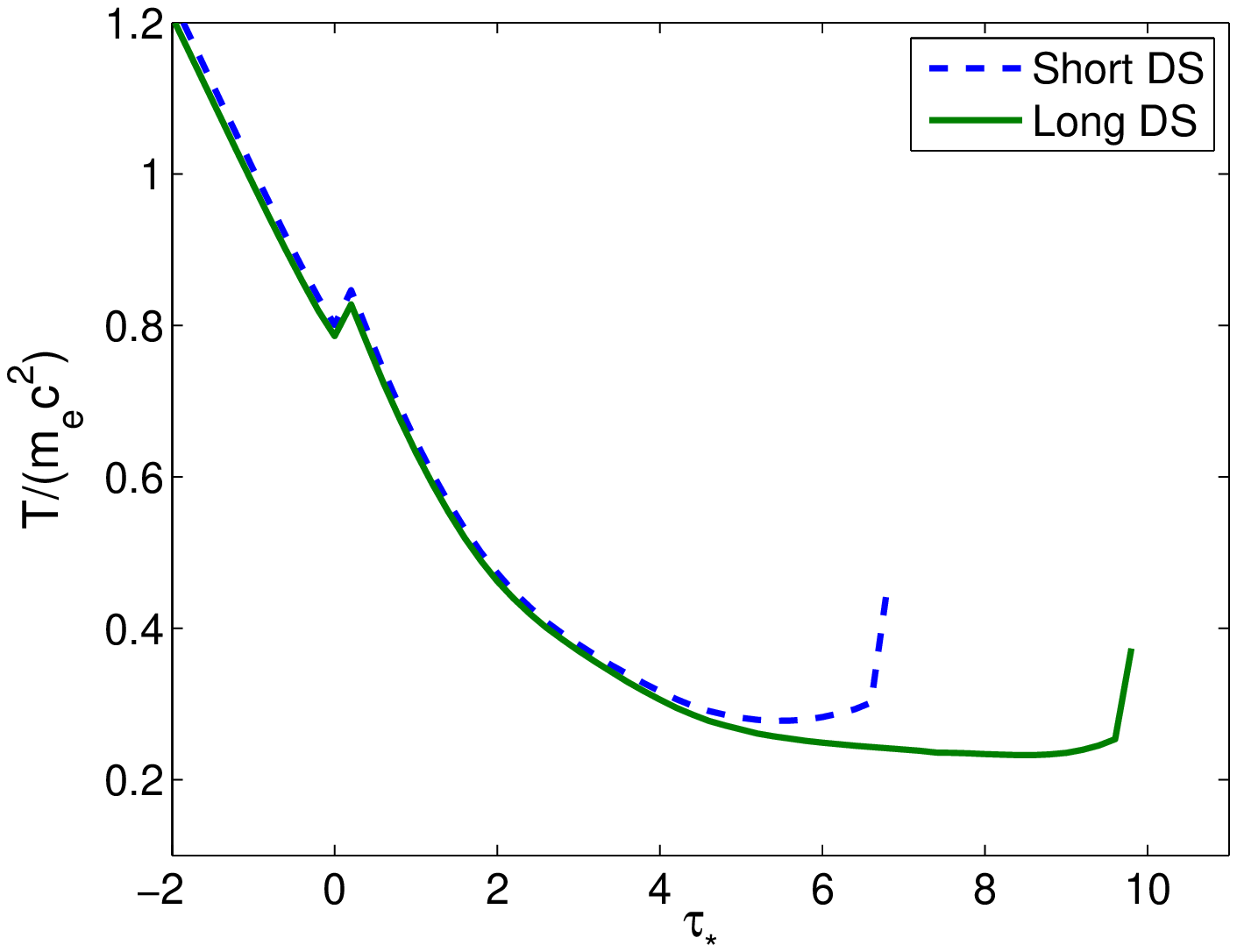}\caption{$ \hat T $ vs. $ \tau_* $ for $ \Gamma_u=30 $, obatined using a DS optical depth of $\tau_*=7 $ (dashed line) and $\tau_*=10 $ (solid line). }
\label{fig:Long_DS_T}
\end{minipage}
\hspace{0.5cm}
\begin{minipage}[t]{0.5\linewidth}
\centering
\includegraphics[scale=0.5]{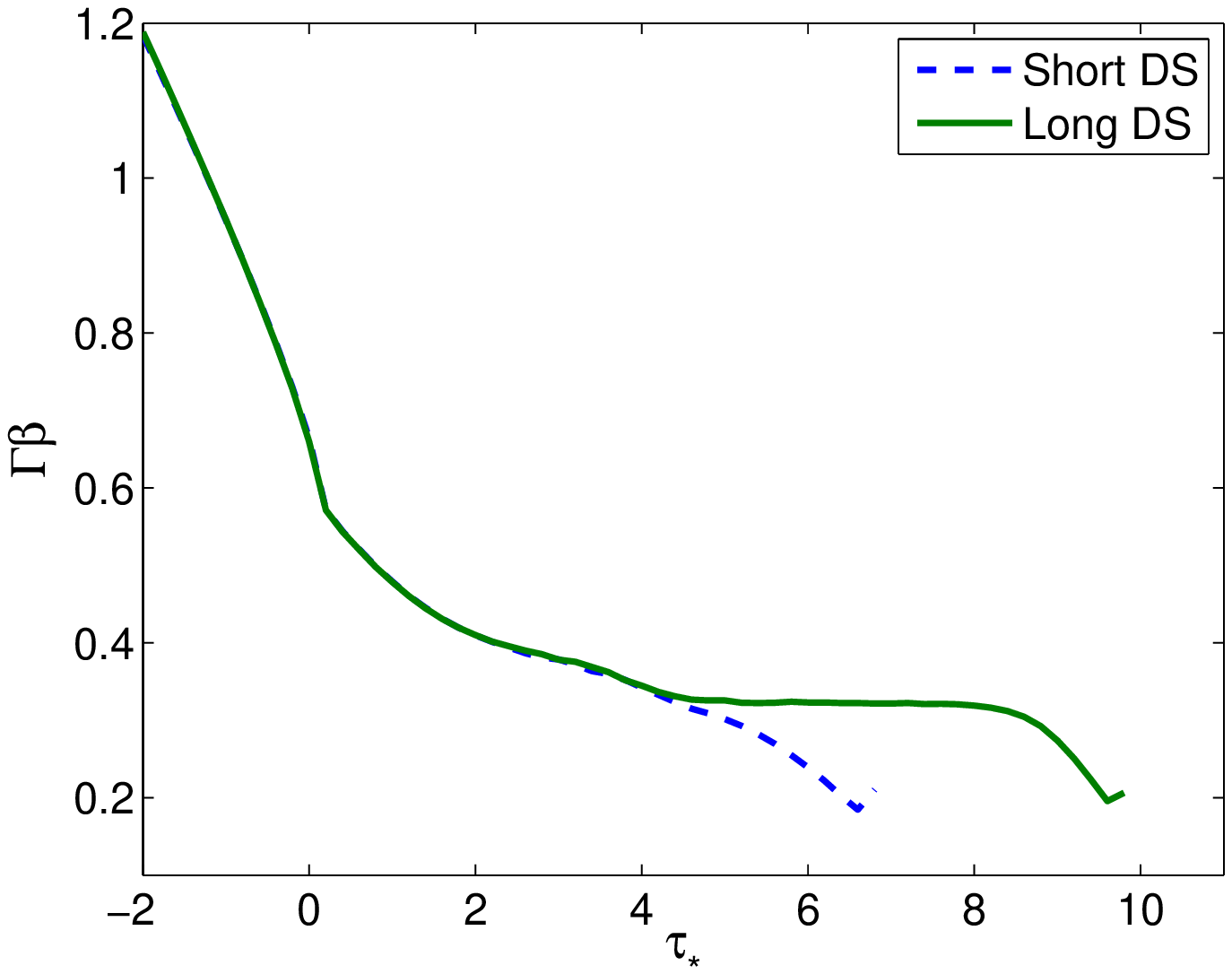}\caption{ $ \Gamma\beta $ vs. $ \tau_* $ for $ \Gamma_u=30 $, , obatined using a DS optical depth of $\tau_*=7 $ (dashed line) and $\tau_*=10 $ (solid line).  }
\label{fig:Long_DS_g}
\end{minipage}
\end{figure*}

In order to verify that the boundary conditions imposed on the DS
edge of the shock do not have a  significant effect on the final
results, around the subshock and in the shock transition region, we
compare the results shown above to the results obtained with a solution including a longer DS
region behind the subshock. We are limited in extending the DS because of
numerical problems, caused by the proximity to a second sonic point.
For this reason we extend only the DS of the calculation for $
\Gamma_u=30 $, from $ \tau_*=7 $ in the calculations presented above to $
\tau_*=10 $. The changes in integral
quantities resulting from this modification of the DS region length are of order of a percent. 
The only quantity which changes by a larger amount, $ \sim 3\% $, is $
\tau_{*,nl} $. The temperature and velocity profiles obtained in the two
calculations are compared in figs. \ref{fig:Long_DS_T} and
\ref{fig:Long_DS_g}.

\section{Simplified analytic modelling of RRMS structure}
\label{sec:analytic}

The key to understanding the structure of RRMS lies in the understanding of the behavior in the immediate DS. In our qualitative analysis of the immediate DS of RRMS, \S~\ref{sec:immDS_subson}, we have argued that the immediate DS photon-electron-positron plasma should be close to Compton pair equilibrium (CPE). This enabled us to demonstrate that the temperature in the immediate DS is expected to be $T_s\sim 0.4m_e c^2$, and that the immediate DS should be sub-sonic. These results are consistent with the numerical results presented in \S~\ref{sec:numerical}. We first discuss in some detail in \S~\ref{sec:simplified-im-ds} the accuracy of the CPE approximation for the description of the immediate DS.

Once the immediate DS is understood, a simple estimate of the spectrum of photons emanating from this region in the US direction leads to an understanding of the transition (deceleration) region, and of the asymptotic (far) US. These are discussed in \S~\ref{sec:simple-transition} and \S~\ref{sec:far-US}.

The flow downstream of the immediate DS is smooth and NR. As the plasma flows away from the shock transition, it slowly produces the photon density needed for thermal equilibrium, eventually reaching the asymptotic DS thermal equilibrium conditions. This "thermalization" phase is discussed in \S~\ref{sec:far-DS}. Finally, we discuss in \S~\ref{sec:HE-beam} the high energy photon "beam" propagating from the transition region into the DS, and comment on the behavior in the $\Gamma_u\rightarrow\infty$ limit in \S~\ref{sec:infin_gamma}.

\subsection{Immediate DS}\label{sec:simplified-im-ds}

\begin{figure*}
\includegraphics[scale=0.45]{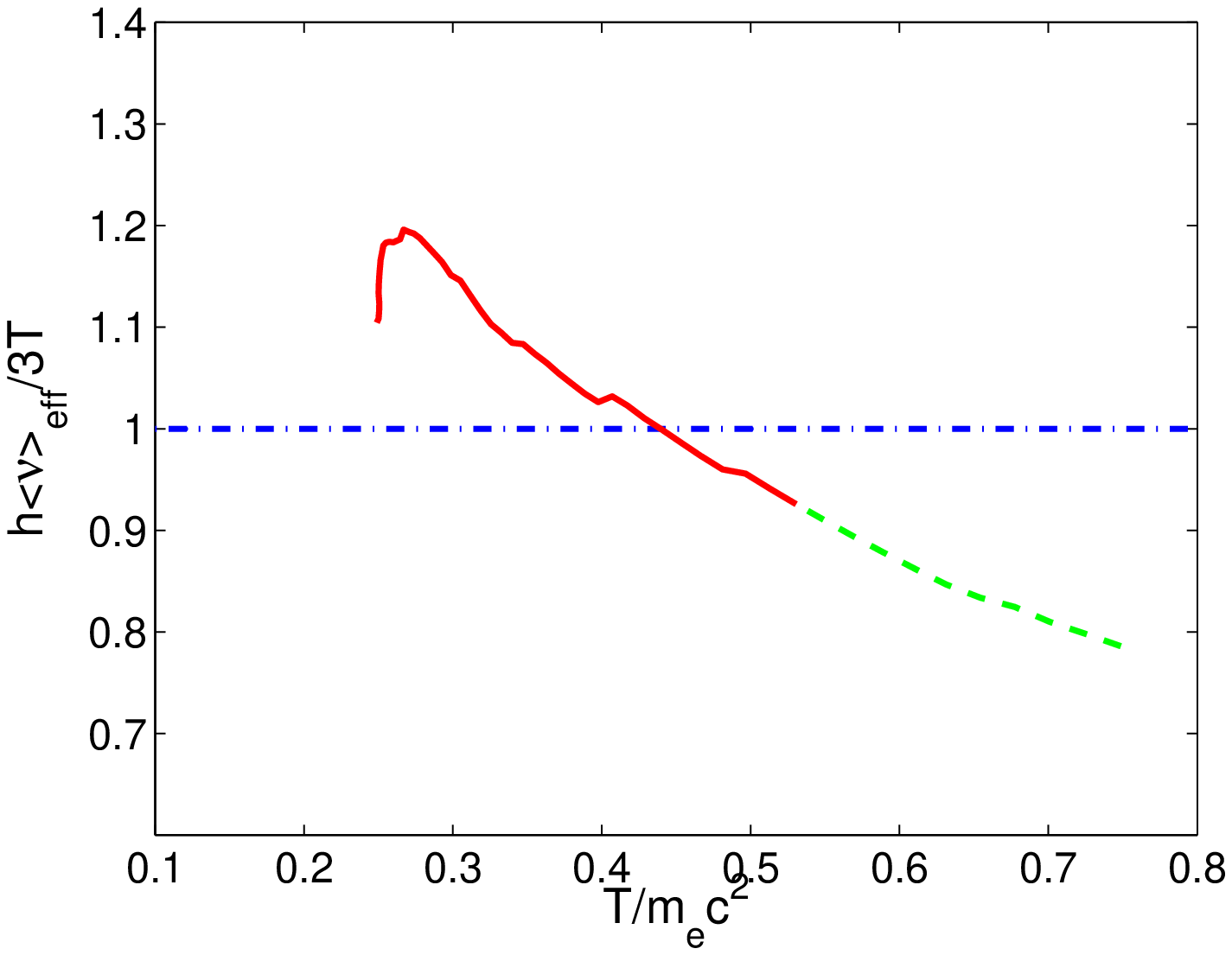}
\includegraphics[scale=0.45]{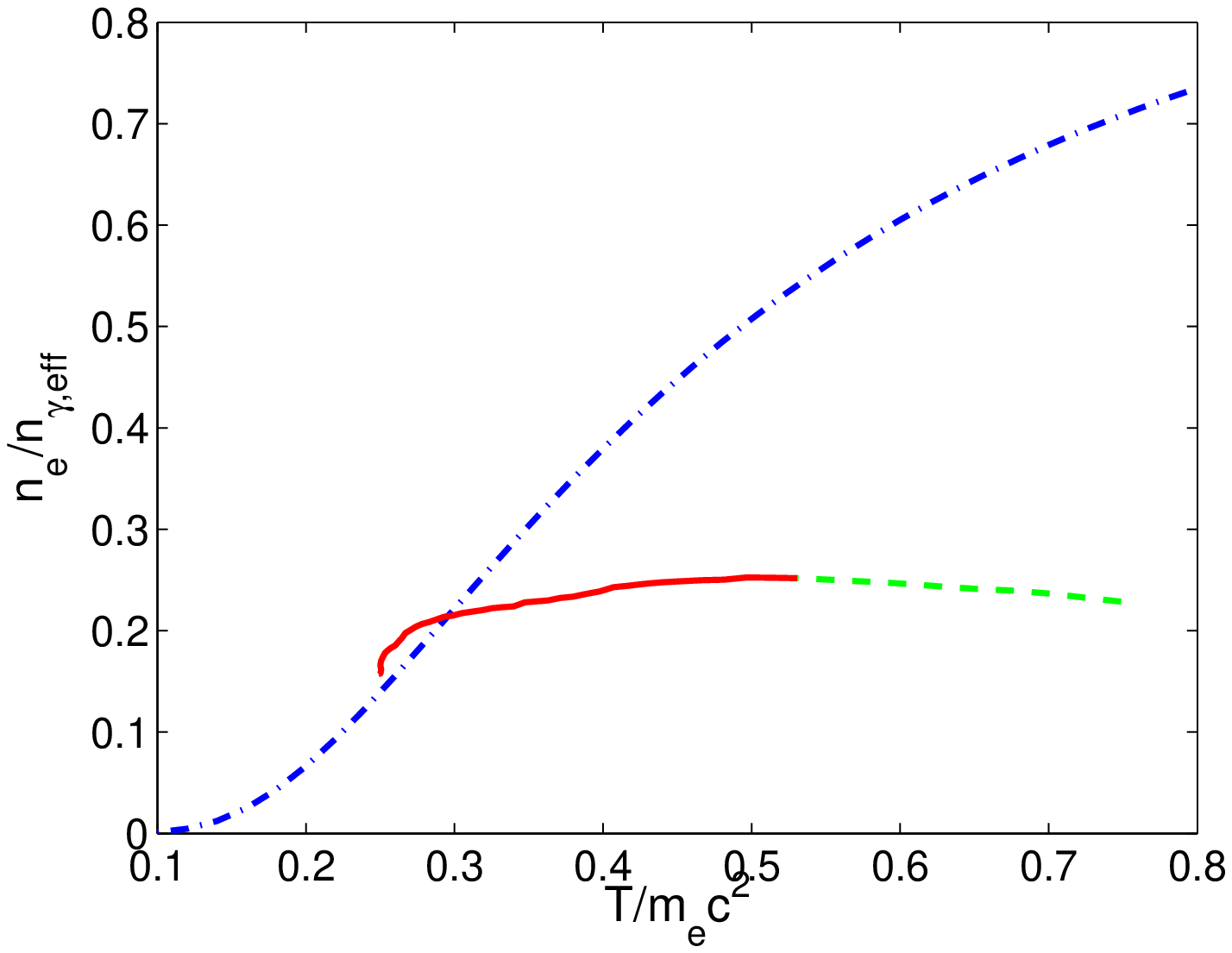}
\\
\includegraphics[scale=0.45]{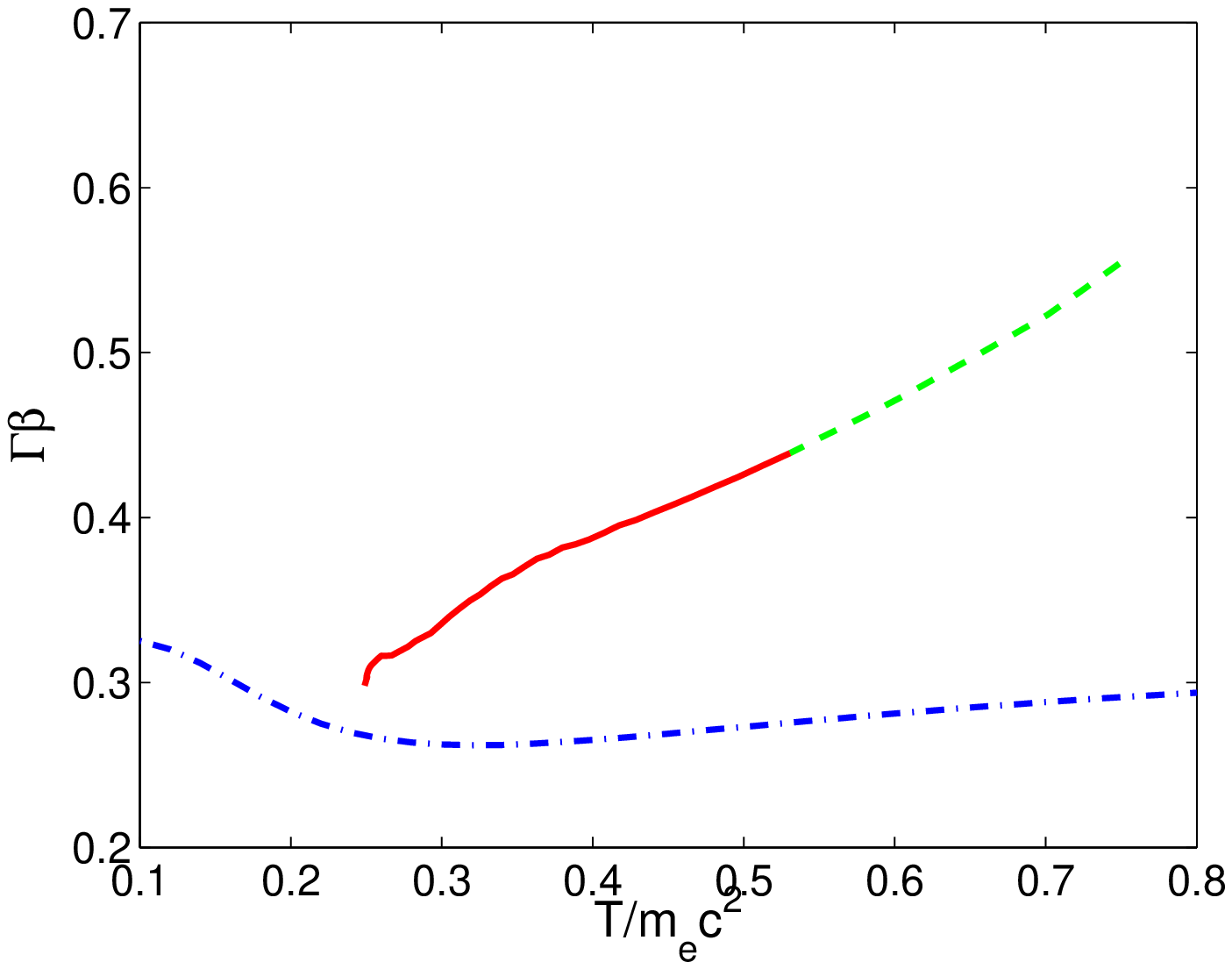}
\includegraphics[scale=0.45]{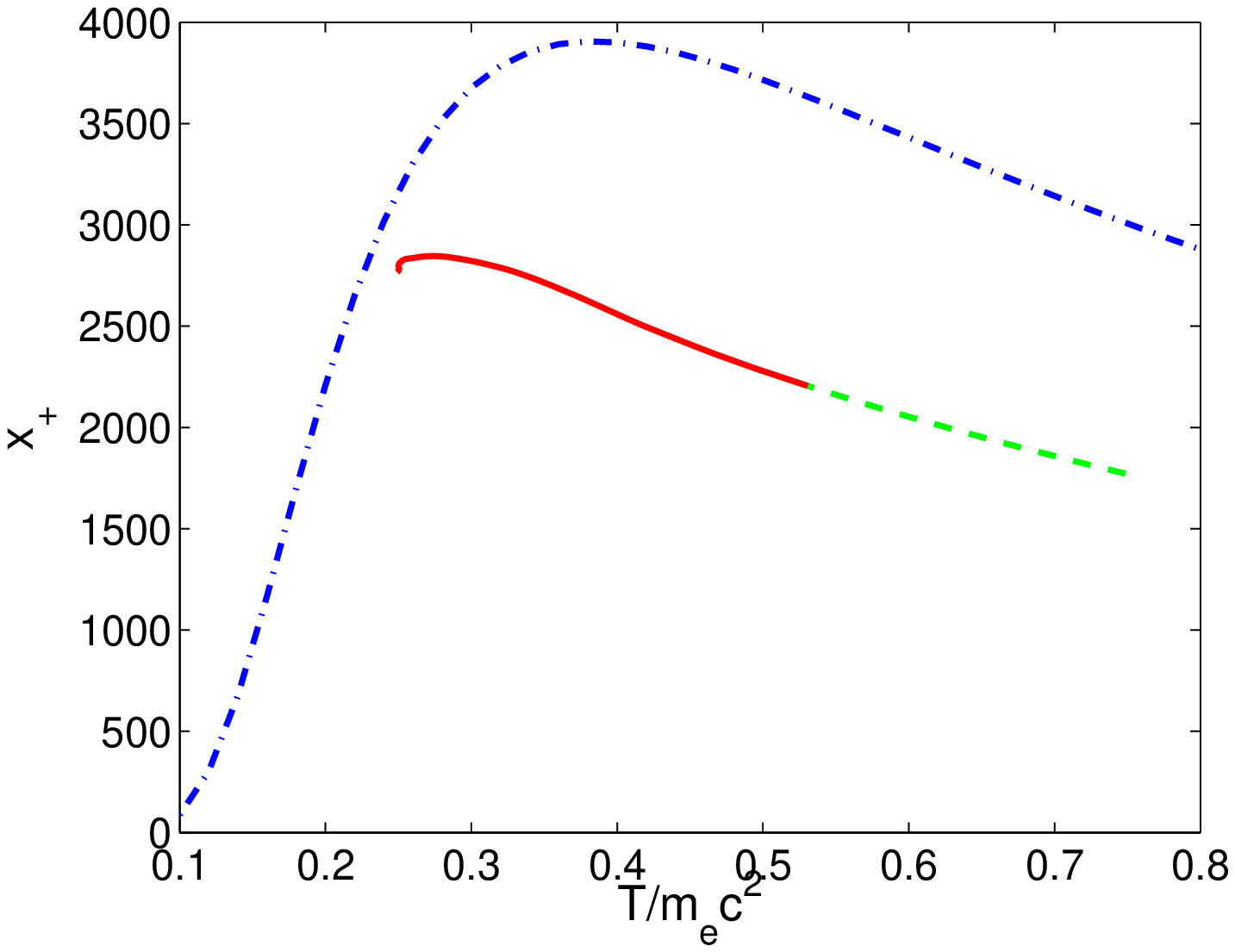}\\
\centering
\includegraphics[scale=0.45]{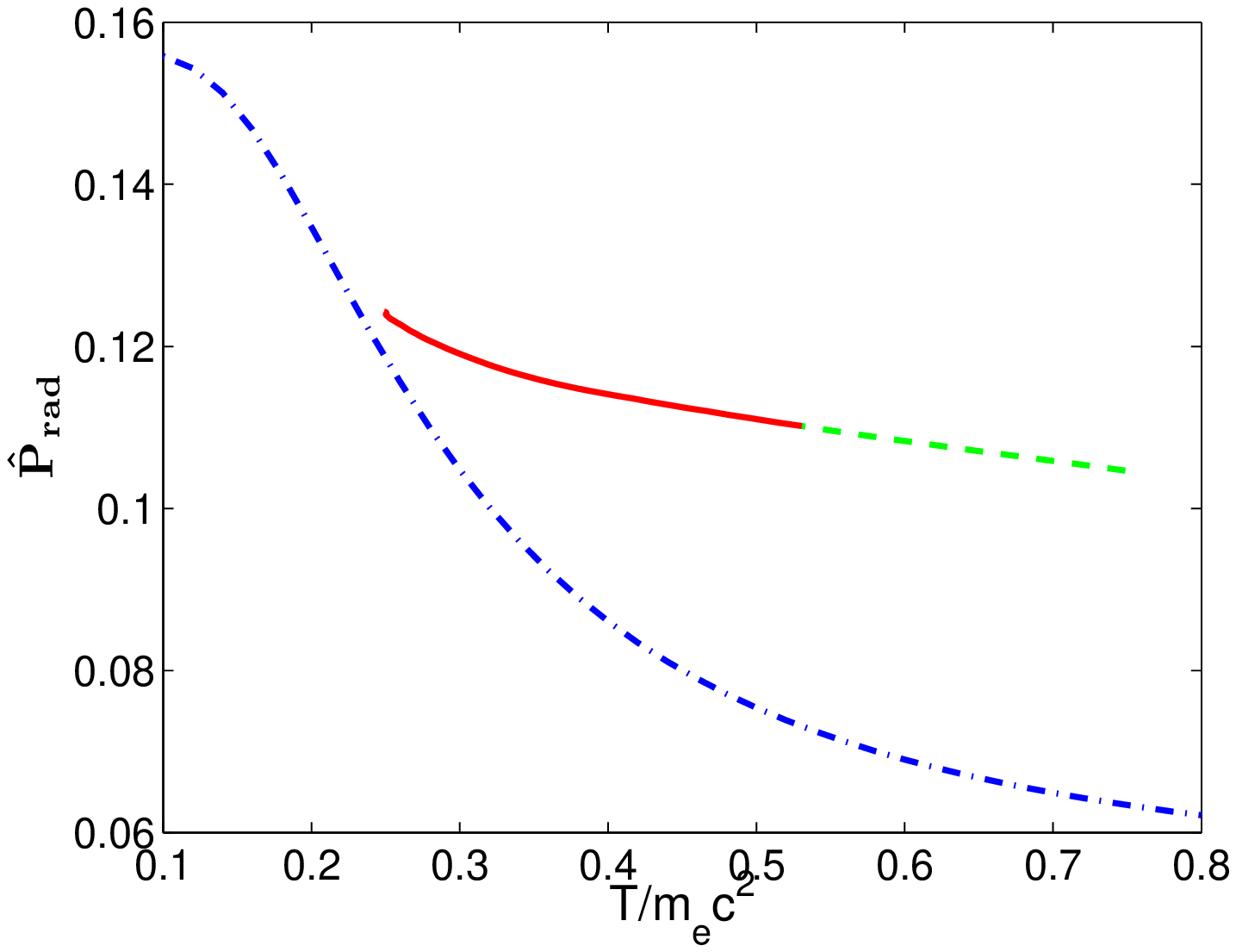}
\caption{Comparison of the numerical solution in the immediate downstream with results obtained assuming CPE, for the case $\Gamma_u=20 $. { The blue dash-dotted curves show the values obtained assuming CPE, while the green dashed and red solid lines show the numerical solution behind  the sub shock. Green dashed lines show the results in the first Thomson mean free path behind the sub shock and the red solid lines show the results deeper into the downstream.} } \label{fig:Gu20_CE}
\end{figure*}

Let us examine the accuracy of the CPE approximation, relating the temperature, the number of positrons and the spectrum of the photons.   
In the top panel of figure \ref{fig:Gu20_CE} we compare the average photon energy $\langle\hat{\nu}\rangle_\text{eff}$ and the $n_l/n_{\gamma,\text{eff}}$ ratio obtained in the immediate DS of the $\Gamma_{u}=20$ solution, with those expected at CPE, $\langle\hat{\nu}\rangle_{\text{eff}}=3\hat{T}$, and  
\begin{equation}\label{eq:lepton-frac-exact}
\frac{n_{l}}{n_{\gamma,\text{eff}}}|_{\text{eq}}=\int_0^{\infty}dxx^2e^{-\sqrt{x^2+\hat T^{-2}}}=\frac{K_2(\hat T^{-1})}{\hat T^2},
\end{equation}
where $K_2$ is the order 2 second kind modified Bessel function. 
{ The effective number of photons and the average energy per photon were calculated using the spectrum around the maximum of $  I_\nu $ in the rest frame of the plasma, $ \nu_\text{peak} $. Specifically, the numerical values shown for $  n_{\gamma,\text{eff}} $  and $\langle\hat{\nu}\rangle_\text{eff}$ are the number of photons in the energy range $\left[  \nu_{\text{peak}}/10, ~10\nu_{\text{peak}} \right]$ and their average energy, respectively.}
The values of $\langle\hat{\nu}\rangle_\text{eff}$ and of $n_l/n_{\gamma,\text{eff}}$ are shown as functions of $\hat{T}$ in the vicinity of the sub-shock.
Immediately downstream of the sub-shock these values are far from those expected for CPE, and they approach the CPE values away from the sub-shock. The figures show a systematic deviation from CPE. This is expected, since the high energy DS photon beam, discussed in detail in \ref{sec:HE-beam}, carries a significant fraction of the energy and is very weakly coupled to the plasma, due to suppression of the cross sections at high photon energy.

The lower panels of fig.~\ref{fig:Gu20_CE} compare the values of $\Gamma$, $P_{rad,sh}$, and $x_+$ obtained in the numerical solution, with those obtained under the CPE approximation (note, that under the CPE approximation the conservation eqs., eqs. \eqref{eq:en_flux}, \eqref{eq:mom_flux}, and \eqref{eq:particle_cons}, allow one to determine  $\Gamma$, $P_{rad,sh}$, and $x_+$  as a function of $T$). Here too, the solution deviates from the CPE predictions immediately downstream of the sub-shock, and approaches the CPE prediction away from it. We conclude that the CPE approximation yields estimates of the global flow variables ($\Gamma$, $P_{rad,sh}$, and $x_+$) which are accurate in the immediate DS to within tens of percent.

\subsection{The transition region}
\label{sec:simple-transition}

The transition or deceleration region is the region in which the energy and momentum flux of the US plasma are transferred to the radiation and to the $ e^+e^- $ pairs. The behavior in the ${ \Gamma_u \gg \Gamma \gg 1 }$ regime may be understood using the following arguments.

\begin{figure}[ht]
\centering
\includegraphics[scale=0.45]{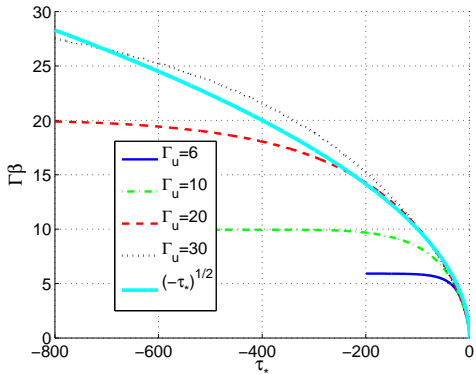}\caption{The relativistic velocity $ \Gamma\beta $ for different values of $ \Gamma_u $ and the simplified anlytical result for the structure $ \Gamma\beta \sim \sqrt{-\tau_*} $ (bold line) vs. $ \tau_* $.}
\label{fig:struct_g_simplified}
\end{figure}

\begin{enumerate}
\item \emph{The photons decelerating the plasma originate in the immediate DS and have a shock frame energy of $ \sim m_ec^2 $ and a rest frame energy of $ \sim \Gamma m_ec^2 $}. Since the immediate DS temperature is $T_s\sim 0.4m_e c^2$ (see \sref{sec:simplified-im-ds}), the characteristic shock frame energy of these photons is $ h\nu \sim 3 T_s \sim m_ec^2 $. Across the transition region, these photons dominate the energy density in the rest frame of the plasma, where their energy is $ \sim \Gamma m_ec^2 $ (see e.g. figure \ref{fig:spec_30_mid_rest_mus}).

\item \emph {The upstream going photons decelerate the plasma by Compton scattering, and by pair production interactions with photons, that are generated  within the transition region either by Bremsstrahlung emission or by inverse Compton scattering (upstream going photons that are back-scattered by the downstream flow).} The three processes similarly contribute to the deceleration, as explained in point \ref{item:trans-width} below.

\item \emph{$ T\sim \Gamma m_ec^2 $.} Both Compton scattering and photon-photon pair production generate electrons/positrons with characteristic energy $ \sim \Gamma m_ec^2 $, driving the plasma temperature to $ \sim \Gamma m_ec^2$.

\item \emph{Pairs produced in the deceleration region drift with the plasma all the way to $ \Gamma\sim 1 $ without annihilating}, due to the high temperatures that reduce the annihilation cross section ($ \propto \log 2\hat T /\hat T$) and to the $ \propto \Gamma^{-2}$ suppression of the collision rate.

\item {\it The plasma rest frame energy density is dominated by pairs rather than protons $ (n_e+n_+)(m_ec^2+3T)> n_pm_pc^2 $.} Once $ \Gamma\ll \Gamma_u $, most of the energy flux is carried by radiation and pairs. 
The pairs carry a significant fraction of the energy flux (see figure \ref{fig:struct_Fpos_F}), hence their energy in the rest frame dominates over the protons rest mass.

\item \label{item:one_interact_change} \emph{A significant deceleration of a fluid element, $ \Gamma \rightarrow \Gamma/2 $, requires that the number of Compton / pair production interactions occurring within it be similar to the number of leptons within it.}
A change of factor 2 in $ \Gamma  $ corresponds, in the plasma rest frame, to an acceleration to velocity $ \beta'=0.6 $ towards the US. This requires a momentum transfer of $ \sim 2  T/c $  to each lepton (recall, that for $\Gamma\ll\Gamma_u$ the plasma energy density is dominated by pairs). This is similar to the momentum transfer by Compton scattering or pair production interaction of a typical US going photon, for a plasma rest frame temperature of $ \sim 2\Gamma m_ec^2  $.  

\item \label{item:opt_depth_us_going} \emph{The optical depth for typical US going photons between $ \Gamma \rightarrow \Gamma/2 $ is $\Delta\tau \sim 1$. }
The number flux of US going photons is similar to the sum of number fluxes of typical DS going photons and pairs (pairs are downstream going). The  similarity between the number densities and the fact that DS going pairs undergo $ \sim 1 $ interaction between $ \Gamma \rightarrow \Gamma/2 $ implies that US going photons roughly interact once as well.

\item \label{item:trans-width} \emph{The Thompson optical depth in the range $\Gamma\ra \Gamma/2$, is roughly $ \Delta\tau_* \sim \Gamma^2 $, wether the deceleration is due to Compton scattering or due to pair creation on Bremsstrahlung generated photons. This implies $\tau_*(\Gamma)\sim \Gamma^{2}$.}

In the range $\Gamma\ra \Gamma/2$  there is $ \sim 1 $ interaction per lepton crossing (see point \ref{item:one_interact_change}). 
The Thomson optical depth required for a single Compton scattering is $ \delta\tau_{*,scat}\sim \Gamma\hat T\sim \Gamma^2 $ due to the KN correction to the cross section. Similarly, the Thomson optical depth required for a single pair production on a "returning" (downstream scattered) photon is $ \delta\tau_{*,ret}\sim \Gamma^2 \hat n_{\gamma,ret}/n_l \sim \Gamma^2 $, where $ n_{\gamma,ret}\sim n_l $ is the number density of returning photons.

Bremsstrahlung generated photons with energy $h\nu \sim m_ec^2/\Gamma$ have a large optical depth for pair creation
on the US going typical photons, since they do not suffer a suppression to the cross section. The number of these photons, produced up to a given point in the transition region  is given by
\begin{equation}
n_{\gamma,ff}= Q_{\gamma,ff}\frac {\delta z_{sh} }{c},
\end{equation}
where  $ Q_{\gamma,ff} $ is the production rate of photons that are able to upscatter to $ h\nu\sim m_ec^2/\Gamma $ in the rest frame of the plasma and $ \delta z_{sh} $ is the shock frame distance over which $ \Gamma $ changes significantly. The Thomson optical depth required for producing enough photons to decelerate the plasma, $ n_{\gamma,ff}\sim n_l, $ is thus
\begin{equation}
\delta \tau_{*,ff} \sim \frac{\Gamma^2}{\alpha_e \tilde g(\hat T,\hat \nu) \Lambda_{US}},
\end{equation}
where $10 \lesssim \tilde g(T,\nu)\lesssim 20 $ is the the Gaunt factor at high temperatures and low frequencies $ h\nu/ (m_ec^2)\sim \Gamma ^{-1}$ and $ \Lambda_{US}\sim 5 $ is a logarithmic correction accounting for photons that are produced at low energies and upscatter to the required energy by the available $\sim1$ number of Compton scatterings on the thermal electrons. 

Two conclusions can be drawn. First, the Thomson cross section needed to decelerate a Lorentz factor $ \Gamma $ is $ \tau_*\sim \Gamma^2 $. Second, all three processes discussed in this point are comparable. 
Simply taking  $\Gamma^2 = -\tau_* $ (where $ \tau_* $ is measured from the subshock) results in a qualitatively good fit to the numerical results, as can be seen in fig. \ref{fig:struct_g_simplified}. It is evident that the deceleration, when approaching the subshock, has a universal structure for different $ \Gamma_u $ values.
\end{enumerate}

The following additional properties are implied by the above considerations.
\begin{itemize}
\item {\it $ x_+ \sim (\Gamma_u/\Gamma^2)\times (m_p/m_e)/8 $ when $ \Gamma\ll \Gamma_u $.} This follows from conservation of momentum flux, and the significance of pairs in the flux. 
\item {\it Most of the  photons resulting from Compton scattering will propagate to the immediate DS without undergoing further interactions.} 
The optical depth for scattering of photons originating from scattering into the DS direction is negligible since they have shock frame energy of $ \sim \Gamma^2 m_ec^2 $ and suffer a $ \sim \Gamma^{-2} $ attenuation in interaction rate.{ The optical depth for pair production is of order unity. This can be seen by the fact that the cross section and target photons for pair production are similar to the Inverse Compton cross section and target photons of the $ e^+ $ and $ e^- $ in the deceleration region. In fact, the total optical depth   is  $\lesssim 1$ , as  can be seen in figure \ref{fig:tau_ds_halfGu_30}. }
\end{itemize}

\subsection{Far US}
\label{sec:far-US}

\begin{figure}\centering
\includegraphics[scale=0.65]{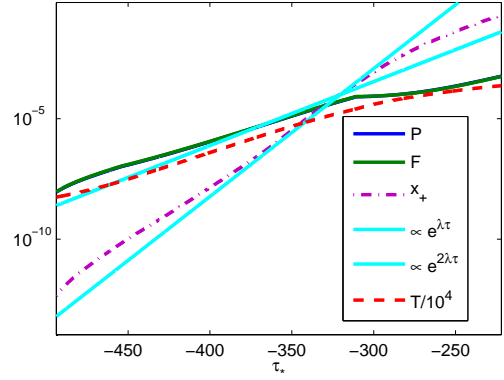}\caption{$ \Gamma_u=10 $ far US exponential growth of integral quantities. The Cyan lines show $ e^{\lambda\tau_*} $ and $ e^{2\lambda\tau_*} $, which are the simplified model exponentials (see eq.~\ref{eq:lambda_as}) expected for $ P $, $ F $ and $ T $ ($ \lambda $) and $ x_+ $ ($ 2\lambda $). }
\label{fig:US-exponents}
\end{figure}

As was shown in \ref{subsec:spectrum}, the radiation, as seen in the rest frame of the far US plasma, is strongly dominated by a beamed [$ \mu_{rest} \approx -1+ 1/ \Gamma_u^2 $ ],  radiation field with photons of typical energy of several times $ \Gamma_um_ec^2 $. To understand the main physical properties of this region, it is useful to approximate the radiation field as a delta function in energy and direction, going in the US direction. The asymptotic solution for such a radiation field can be easily found to be an exponential growth of the parameters $ P_{rad,sh} $,  $ F_{rad,sh} $, $ \Gamma_u\beta_u-\Gamma\beta $ and $ T $, with the same exponent, $ \lambda_{as} $, and an exponential growth of $ x_+ $ with an exponent $ 2\lambda_{as} $ [see \cite{sagiv06}]. An approximate value for $ \lambda_{as} $ is given by
\begin{equation}\label{eq:lambda_as}
\lambda_{as}\approx 0.28 \frac {1}{\Gamma_u\beta_u n_u \sigma_T}\left( \frac{\sigma_c}{\sigma_T} \right) ^{-1},
\end{equation}
where $ \sigma_c $ is the total cross section for the photons in the rest frame of the plasma. The results of the numerical calculations are shown in figure \ref{fig:US-exponents} to agree with the expected exponential growth for the case $ \Gamma_u=10 $, where the rest frame dominant frequency is $ h\nu_{rest}\approx 36 m_ec^2 $. 

The solution deviates from exponential growth when the temperature approaches $ m_ec^2 $ as, for example, the Compton cross section changes significantly.  For $ \Gamma_u=10 $ the transition from linear to non-linear evolution occurs at $ \tau_{*,nl}\approx -320 $, while for $ \Gamma_u=20 $ it occurs at $ \tau_{*,nl} \approx -890 $.  $ \tau_{*,nl} $ grows with $ \Gamma_u-1 $ (energy per proton) in a manner faster than linear.

\subsection{Far DS}
\label{sec:far-DS}

This region is characterized by an almost constant velocity and a slow growth in photon number that lowers the temperature. It can be divided into two regions: $ T \gtrsim 50 \keV $, where $ x_+>1 $ and electron-positron annihilation takes place, and $T \lesssim 50$~keV, where the number of positrons is small and they play no significant role.

Let us first consider the $ x_+>1 $ region. The low temperature limit of eq. \eqref{eq:lepton-frac-exact} yields
\begin{equation}\label{eq:lowTnenga}
  \frac {n_{l}}{n_{\gamma,\text{eff}}}(\hat{T} \ll 1)=
2 \sqrt{\frac{\pi}{8}}\frac{e^{-\inv{\hat {T}}}}{\hat{T}^{3/2}}.
\end{equation}
Using arguments similar to those used for the immediate DS estimates, we can write an equation for the evolution of photon number
\begin{equation}\label{eq:farDS-prod1}
\frac{1}{n_l(\tau_*)}\frac{dn_{\gamma,\text{eff}}(\tau_*)}{d\tau_*}\approx \frac{g_{ff}(\hat T)\Lambda_{eff}(\hat T)}{\beta_d \sqrt{\hat T}},
\end{equation}
and assuming that most of the energy flux is already in the radiation we can approximate $ n_{\gamma,\text{eff}} \hat T=\mathrm{Const} $. Using this assumption with Eqs. \eqref{eq:lowTnenga} and \eqref{eq:farDS-prod1} we obtain
\begin{equation}\label{eq:T-change-xpos}
\frac{d\hat T}{d\tau_*}=-2\sqrt{\frac{\pi}{8}}\frac{g_{ff}(\hat T)\Lambda_{eff}(\hat T)}{\beta_d }e^{-1/\hat T}.
\end{equation}
The flow reaches $ n_{e^+}\approx n_p $ ($ x_+\approx 1 $) when $ \hat T\approx 0.06 $, with a weak dependence on $ \Gamma_u $. From Eq. \eqref{eq:T-change-xpos} we see that the length scale is set by the lowest temperatures in this range, and reasonable parameters yield $ \sim 10^5 $ optical depths required for the positrons to annihilate.

In the the $ x_+<1 $ region, a gradual increase in photon number lowers the temperature until thermal equilibrium is reached. The length scale for this process is 
\begin{equation}
L_T=\beta_d c\frac{n_{\gamma,\mathrm{eq}}}{Q_{eff,d}},
\end{equation}
where $ n_{\gamma,\mathrm{eq}}\approx a_{BB} T_d^3/2.8 $ is the thermal equilibrium photon density and
\newline
$ Q_{eff,d} \approx g_{eff,d} \Lambda_{eff,d} n_d^2 \sigma_T c / \sqrt{\hat T _d} $ is the photon generation rate in the DS.
An estimate for $ L_T $ yields
\begin{equation}
\frac{L_T}{(\sigma_T n_d)^{-1}}\approx  3 \times 10^{6} \left(3\beta_d \right)\left( \frac{g_{eff,d} \Lambda_{eff,d}}{10}\right)^{-1}  \Gamma_{u,2}^{3/4}n_{u,15}^{-1/8}.
\end{equation}
In this region, the temperature drops as a power law $ T(\tau_*)\propto (\tau_*-\tau_0)^{-2} $, as was shown in \citet{Katz09}.

\subsection{The high energy photon component beamed in the DS direction}\label{sec:HE-beam}
As can be seen in Figs. \ref{fig:spec_10_DS_sh} and \ref{fig:spec_30_DS_sh},  the immediate DS has a high energy photon component beamed in the DS direction. 
{ We use below the simplified analysis presented in \sref{sec:simple-transition} to derive the characteristics of the spectrum of this beam. 

The photons in this beam originated from the immediate DS, propagated into the transition region and then were Compton scattered once before returning to the DS. Photons that were scattered at a point with Lorentz factor $ \Gamma $  return to the DS with an energy boosted to $ \sim \Gamma^2m_ec^2 $ and within a beaming angles $ \theta\sim \Gamma^{-1} $.}

Denote the shock frame intensity of US going photons with typical energies $ I_0(-\tau_*) $.
Conclusion \ref{item:opt_depth_us_going} in \sref{sec:simple-transition}, leads to the equation
 \begin{equation}
 I_0(\Gamma)=\varepsilon_\Gamma I_0(\Gamma/2)
\end{equation}
where $ \varepsilon_\Gamma\sim 1/3 $ is related to the exact total optical depth for typical US going photons from $ \Gamma/2 $ to $ \Gamma $. Assuming that the fraction of photons that scatter is constant with $ \Gamma $, the resulting intensity emitted at $ \Gamma $, $ I_B(\Gamma) $ will be
\begin{equation}
I_B(\Gamma)\approx 4\varepsilon_\Gamma I_B(\Gamma/2),
\end{equation}
since the photons gain a factor  of $ \sim \Gamma^2 $ to their energy when scattered at $ \Gamma $. The scattered photons are beamed into a cone with an opening angle $ \Gamma^{-1} $ in the DS direction. Since the losses of the scattered photons on the way to the immediate DS are less than a factor of 2 and depend weakly on the angle and energy of the photon, we find that the spectrum of the high energy beam $ I_B $ can be approximately described as
\begin{equation}\label{eq:beam-simple}
\hat \nu_{sh}I_B(\hat{\nu}_{sh},\theta_{sh})\propto \hat{\nu}_{sh}^{\alpha_1}\Theta(\theta_{sh}^{-1}-\hat{\nu}^{1/2}_{sh})\Theta (\hat{\nu}_{max}-\hat{\nu}_{sh}),
\end{equation}
where $ \hat{\nu}_{max} \approx \Gamma_u^2 $ and $ \alpha_1 \approx \log_2 (4\varepsilon_\Gamma)/2 $ is close to zero, and is equal to zero when $ 4\varepsilon_\Gamma=1 $.

\begin{figure*}[ht]
\begin{minipage}[t]{0.5\linewidth}
\centering
\includegraphics[scale=0.45]{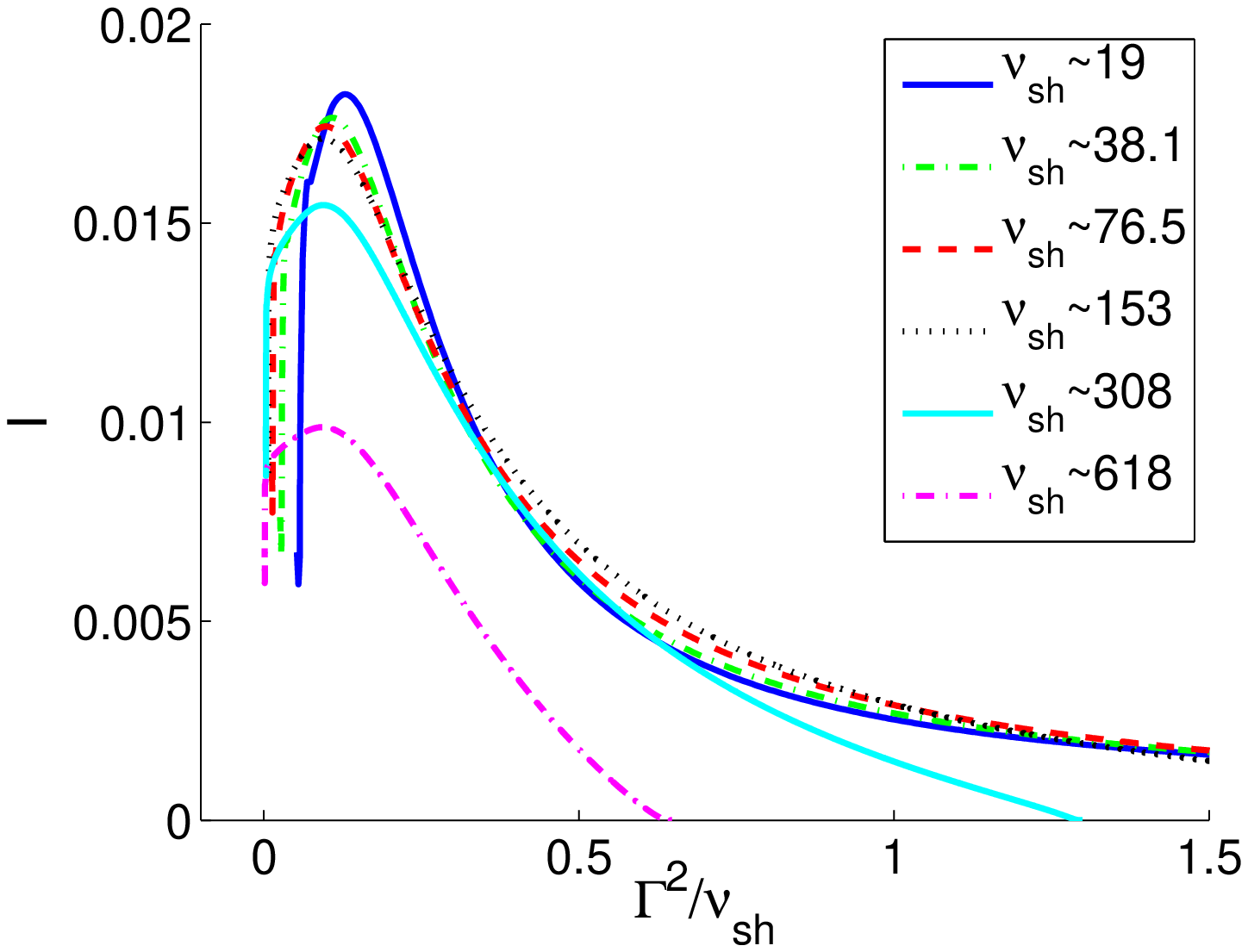}\caption{$ \hat I_{sh,\hat \nu_{sh}} $ directed towards the DS vs. $ \Gamma^2/\nu_{sh} $, for different high photon frequencies $ \hat \nu_{sh} $, $ \Gamma_u=20 $.}
\label{fig:ses_spectrum_1}
\end{minipage}
\hspace{0.5cm}
\begin{minipage}[t]{0.5\linewidth}
\centering
\includegraphics[scale=0.45]{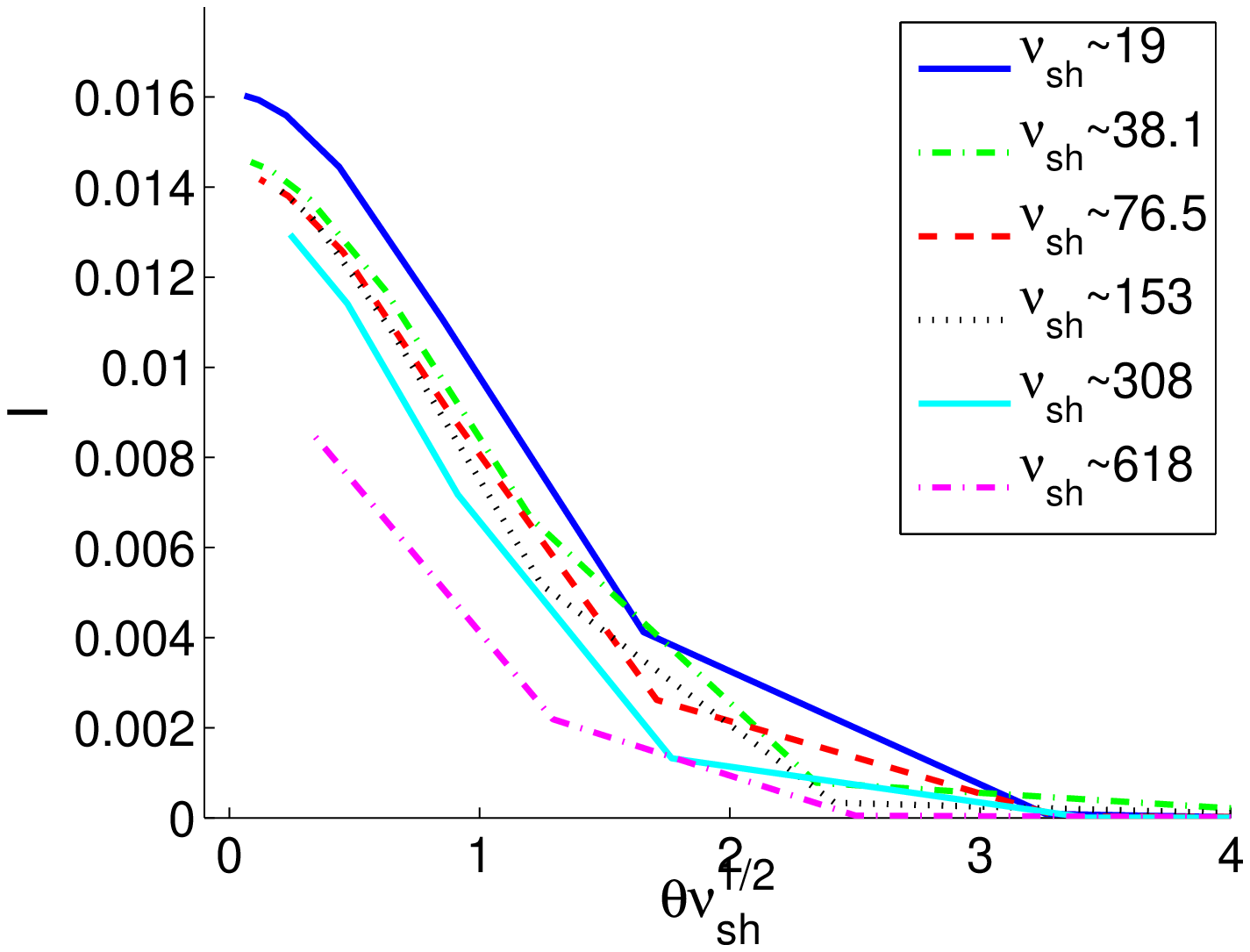}\caption{$ \hat I_{sh,\hat \nu_{sh}} $ at $ \tau_*=0 $, directed towards the DS vs. $ \theta_{sh} \hat \nu_{sh}^{1/2} $, for different angles with respect to the $ z $ axis in the shock frame  $\theta_{sh}$, $ \Gamma_u=20 $.}
\label{fig:ses_spectrum_2}
\end{minipage}
\end{figure*}

We next verify that this analysis complies with the numerical results (results shown for $ \Gamma_u=20 $ calculation). Fig. \ref{fig:ses_spectrum_1} shows the shock frame intensity of a beam with $ \theta_{sh}\approx 10^{-2} $ with different $ \nu_{sh} $ along the shock,  vs. $ \Gamma^2/\nu_{sh} $. We see that the intensity is mostly contributed by the part in the flow in which $ \Gamma^2\approx 200 \hat{\nu}_{sh} $, as the physical picture requires. Fig. \ref{fig:ses_spectrum_2} shows the  shock frame intensity immediately after the subshock,  at different $ \hat{\nu}_{sh}  $, as a function of $ \theta_{sh}\hat{\nu}_{sh}^{1/2} $. We see that the structure of the beam is such that the different energies are beamed according to Eq.  \eqref{eq:beam-simple}.

\subsection{The $  \Gamma_u \rightarrow \infty $ limit}
\label{sec:infin_gamma}

{ Based on the results for $ \Gamma_u  \le 30 $ and the analysis above, it appears that for $ \Gamma_u \rightarrow \infty $, $ T(\tau_*) $ and $ \Gamma(\tau_*) $ approach asymptotic profiles in the regime where $ \Gamma \ll \Gamma_u $ and $ {x_+ \gg 1} $.
In particular, $ T\sim m_ec^2 $ in the immediate DS and $ \Gamma\sim \tau_*^2 $ in the transition region.} However, we have also seen that the high energy beam becomes more dominant as $ \Gamma_u $ grows. The structure of the shock, particularly the immediate DS, may be different if the high energy beam becomes the dominant carrier of momentum and energy of the radiation. Unfortunately, a full calculation of very high $ \Gamma_u $ shocks is beyond our current numerical capabilities, and requires further investigation.

\section{NR RMS revisited}\label{sec:nonrel}

In this section we briefly describe a preliminary application of the code to NR shocks. 
Our numerical scheme was designed and optimized for
the solution of the relativistic problem, and is not efficient and
easy to use for NR problems. The main difficulties are 1. Solving the momentum and energy conservation equations for the velocity and temperature of the plasma is problematic due to the negligible contributions of the thermal energy and pressure. 2. Radiation field convergence requires a large number of iterations, roughly one iteration per single Compton scattering, implying  $ \propto \beta_d^{-2} $ iterations.

A different scheme for finding the plasma temperature and velocity and a different boundary condition in the far DS were used for solving the NR problem:
\begin{itemize}
\item The temperature was set to the local CE value calculated from the radiation field and the velocity was found by solving the momentum conservation only. This approximation is justified in the case where
$n_{\gamma}/n_{e}\gg1$, where $n_{\gamma}$ is the number density of
photons, which holds in the transition region (when the energy
density of the radiation is a fair fraction of the flow) and the
downstream of a NR RMS. Convergence required that the temperature be set to a value that is slightly smaller than the actual CE value. 
\item The following downstream boundary condition was used. 
The radiation field in the upstream direction $I_{\nu_{sh}}^{fl}(\mu_{sh}<0,\tau_{*}=\max(\tau))$,
was set to represent the radiation field at a chosen point in the downstream. This was done by assuming a Wien spectrum with a temperature lower than $ T_s $ [see Eq. \eqref{eq:btOfT}],  and an intensity that satisfies the equilibrium at the DS velocity as expected in the DS well behind the velocity transition. 
\end{itemize}
The radiation transport is solved similarly to the relativistic case.

Figures a preliminary solution for a shock with
upstream energy per proton $\varepsilon=50\MeV$ and a very low
density ($n_{u}=10^{6}\cm^{-3}$), which ensures that bremsstrahlung
absorption remains unimportant until after the velocity has already
reached its downstream value. In the calculation shown here,
absorption is everywhere unimportant, since it does not reach the
downstream temperature.
We stress that the resulting solution contains a limited region of optical depth $ \sim \beta_d^{-1} $ behind the velocity transition, hence the temperature profile may not correctly represent the actual solution. 

\begin{figure}[h]
\center
\includegraphics[scale=0.5]{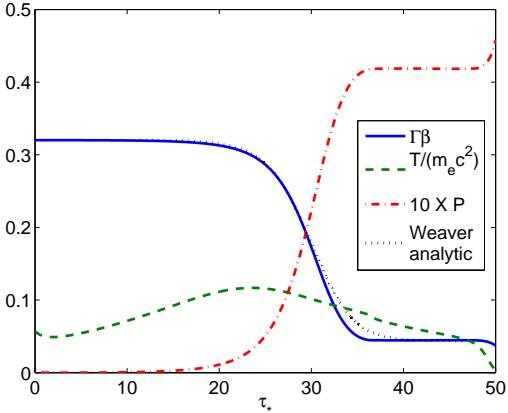}\caption{The shock structure for $\varepsilon=50\MeV$.  The dotted black line is the analytic solution for $ \Gamma\beta $ obtained by Weaver 1976 [equation (5.10) there], with average Compton cross section $ \bar \sigma_C=0.56 \sigma_T $.}
\label{fig:NR_struct}
\end{figure}

Figure \ref{fig:NR_struct} shows the structure ($ \Gamma\beta $, $ \hat T $ and $ \hat P $) of a shock with $ \varepsilon=50\MeV $ as a function of  $ \tau_* $. The dotted black line is the analytic solution for $ \Gamma\beta $ obtained by Weaver 1976 [equation (5.10) there], using $ \bar \sigma_C=0.56 \sigma_T $ for the average Compton cross section, suitable for $ h\nu\approx 0.5 m_ec^2 $ typical photon energy in the transition region. Weaver's solution deviates from the numerical solution near the immediate DS. This is due to the lower average photon frequency there, compared to that in the transition region, which leads to an increase in $ \bar \sigma_C $ in the (more accurate) numerical calculation. 

\begin{figure*}[h]
\includegraphics[scale=0.5]{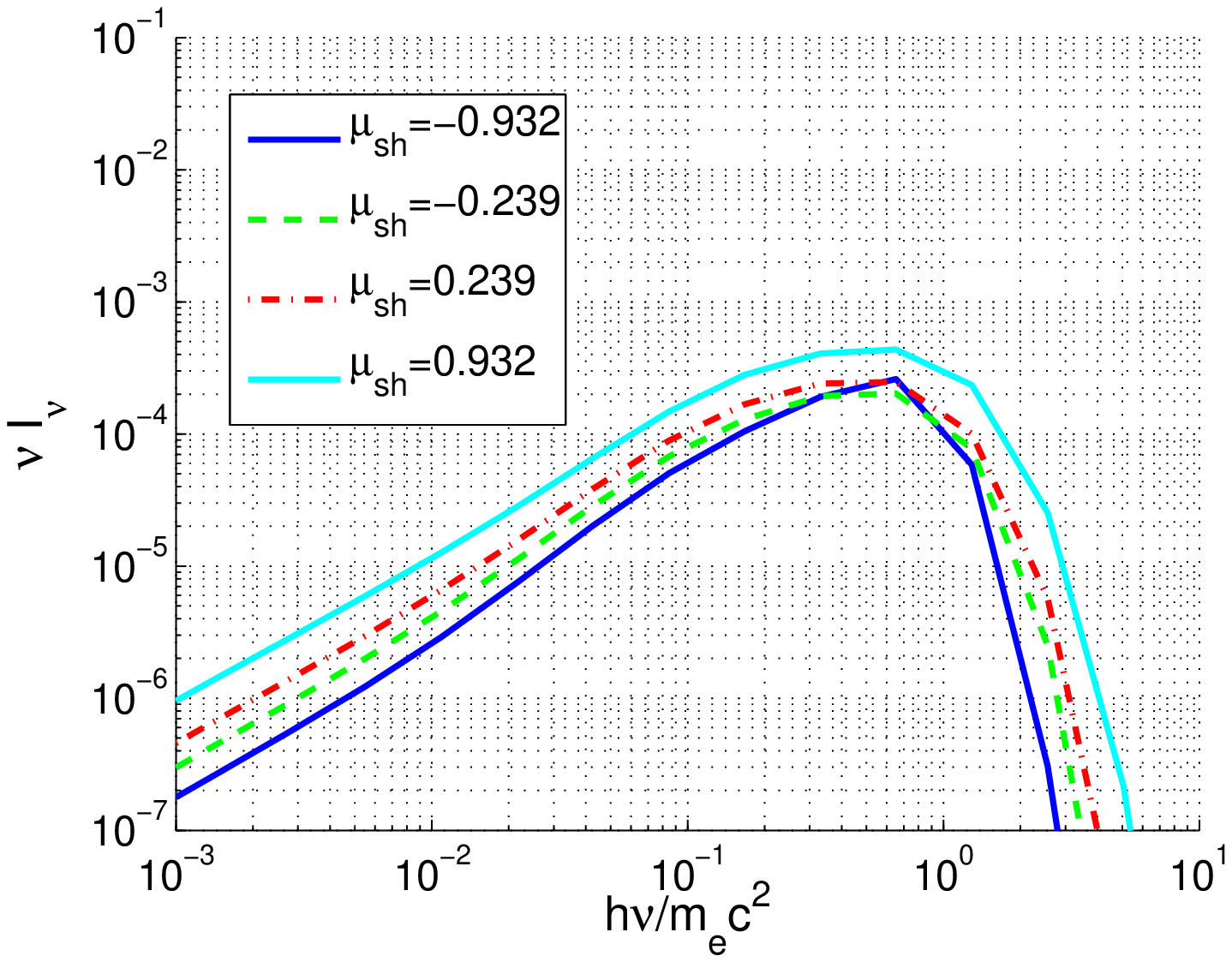}\includegraphics[scale=0.5]{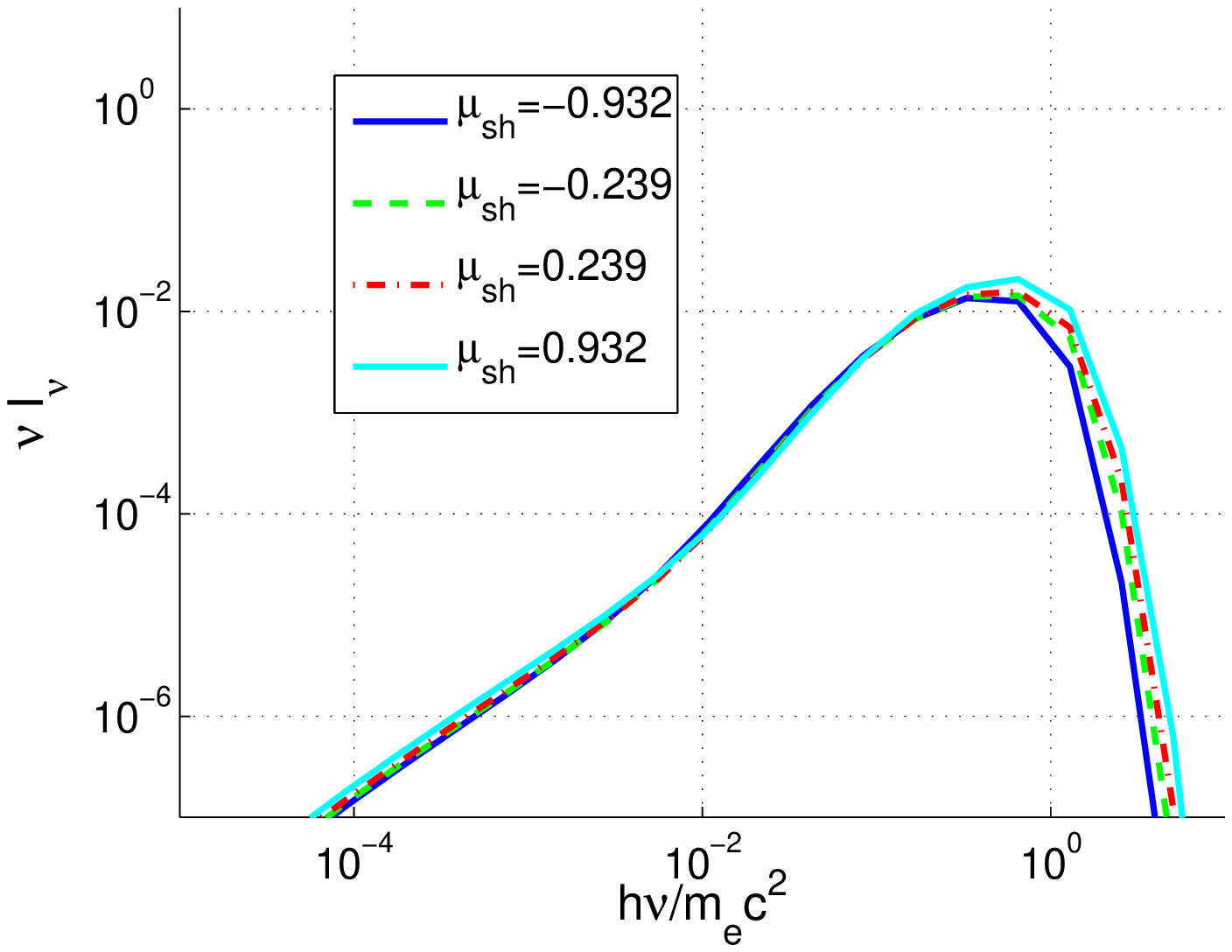} \\
\centering
\includegraphics[scale=0.5]{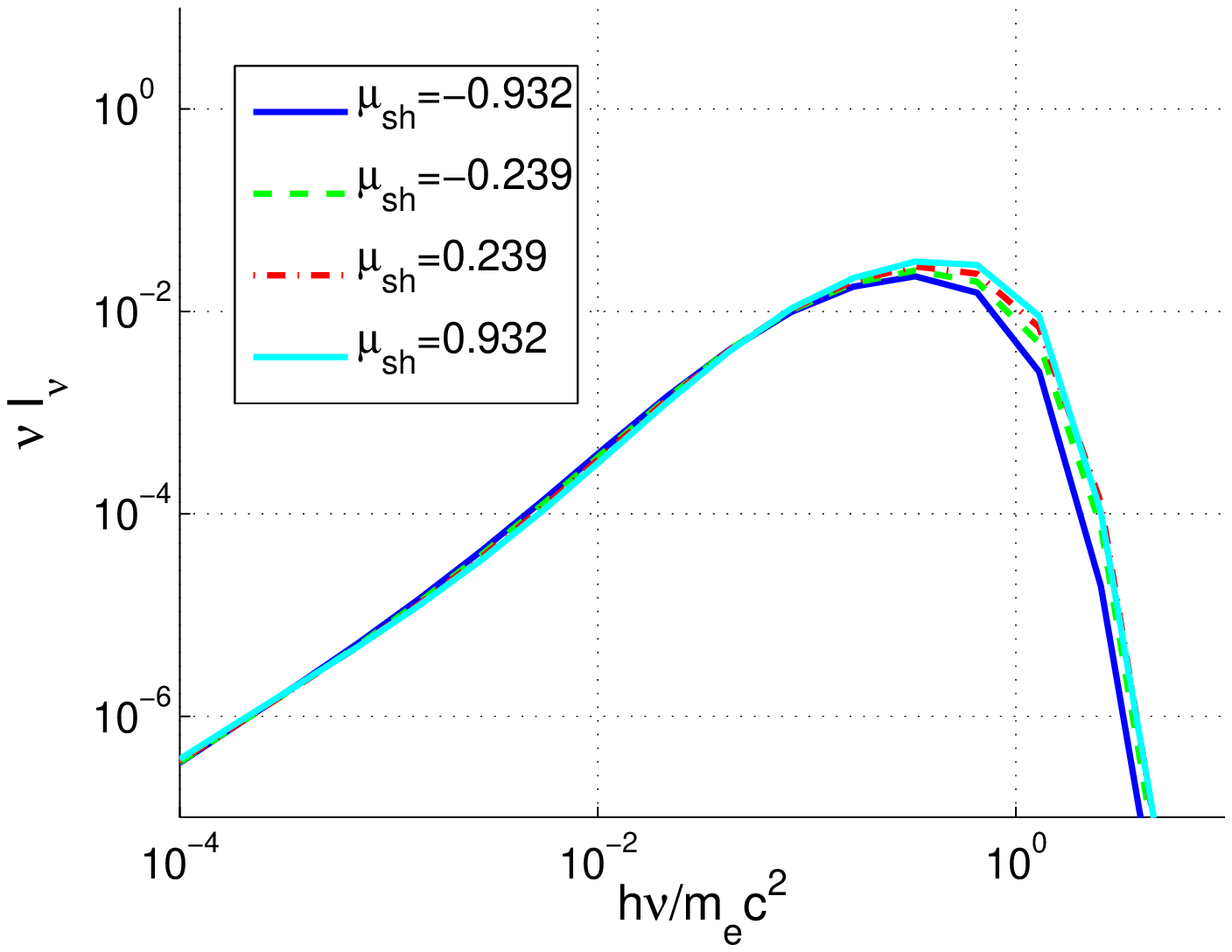}
\caption{Spectra of the radiation in the shock frame along the shock profile
for $\varepsilon=50\mathrm{MeV}$. Upper left: far upstream ($\beta=0.99\beta_u$),
upper right: inside the velocity transition ($\beta=0.5\beta_u$) and
lower: In the immediate downstream ($\tau_{*}=37)$.}
\label{fig:NR_spec}
\end{figure*}

Examining the radiation spectra obtained in our numerical
calculations, fig. \ref{fig:NR_spec}, we are able to verify the validity of two of
Weaver's assumptions. First, it is clear that the spectrum at each
point along the shock is dominated by photons within
a narrow energy range. Second, the anisotropy of the radiation is of
order $ \beta $, which is the expected anisotropy due to diffusion
of the radiation.

To conclude, the preliminary solution found using our
numerical scheme is consistent with Weaver's results. In addition,
the detailed spectra support the validity of Weaver's approximations
regarding the radiation spectrum.  The fact that the results for NR shocks
are in agreement with previous work supports the validity of the
numerical scheme.

\section{Discussion}\label{sec:discussion_RMS}

We have  calculated and analyzed  the structure of relativistic radiation mediated shocks (RRMS). A qualitative discussion of the shock physics was presented in \S~\ref{sec:phys}, including analytic estimates of the deceleration and thermalization length scales of non-relativistic (NR) RMS (equations \eqref{eq:ShockWidth}, \eqref{eq:LT}; see figure~\ref{fig:NR_scheme} for a schematic shock structure description) and of the immediate DS temperatures of both NR RMS [eq.~\eqref{eq:btOfT}] and RRMS [eq.~(\ref{eq:Ts_rel})]. We have also shown (in \sref{sec:immDS_subson}) that the immediate DS of RRMS is expected to be subsonic, and concluded that the structure of RRMS must include two sonic points.

In section \sref{sec:formulation} we derived a  dimensionless form of the equations describing the conservation and transport equations determining the structure of the shock, and described in detail the radiative processes included in our treatment and the approximations we used. In section \sref{sec:num_meth} we presented a novel iteration scheme for numerically solving the equations, and demonstrated its validity by applying it to several test cases. In section \sref{sec:numerical} we have presented numerical solutions for the profiles and radiation spectra of RRMS, for upstream Lorentz factors $ \Gamma_u $ in the range of $ 6 $ to $30$. The main results obtained are described below.
\\ \noindent {\bf[1] Structure and radiation spectrum.} In \sref{subsec:struct} we showed that the structure of RRMS can be divided into four regions, from upstream (US) to downstream (DS): The far US, the transition region, the immediate DS and the far DS. The far US is characterized by a velocity close to the US velocity and a radiation energy-momentum flux much smaller than that of the US plasma. The transition region is where the velocity ($ \Gamma\beta $) changes significantly, approaching $ \Gamma\beta\sim 1 $, while the momentum and energy fluxes are transferred to the $ e^+e^- $ pairs and to the radiation. In both regions, the radiation spectrum (shown in \sref{subsec:spectrum}) is dominated in the plasma rest frame by US going photons with energy of a few times $ \Gamma m_ec^2 $. In the shock frame the radiation is dominated by DS going photons, beamed into a cone with opening angle $ \sim \Gamma^{-1} $, and a typical energy $ \Gamma^2 m_ec^2$. In the far US the temperature grows exponentially with $\tau_*$ towards the downstream (fig.~\ref{fig:US-exponents}), { until it reaches $\sim m_e c^2$ } ($\tau_*$ is the Thomson optical depth for photons moving towards the upstream). The temperature then continues to grow at a slower rate until it reaches $ T/(m_ec^2)\sim \Gamma $ in the transition region, and then decreases, approximately following the deceleration, $ T/(m_ec^2)\sim \Gamma $ (fig.~\ref{fig:struct_T}).
 
The transition region ends at a subshock, possibly mediated by plasma instabilities, with a velocity
jump of $ \delta(\Gamma\beta)\sim 0.1 $ and a slight increase in 
temperature, to $ 0.4<T/(m_ec^2)<0.9 $ for $6<\Gamma_u<30$ (see Figs. \ref{fig:struct_g_DS.eps} and
\ref{fig:struct_T_DS}). The immediate DS, following the subshock, is
characterized by a small change of velocity, approaching the DS
value within $ \sim 2 $ Thomson optical depths, and a temperature that
decreases on a scale of a few Thomson optical depths to $
T/(m_ec^2)\sim 0.25$. The ratio of positron density to proton
density in the immediate DS reaches a maximum of $ \sim 140 \Gamma_u
$ (see Fig. \ref{fig:struct_x_DS}), approximately when the
temperature crosses $  T/(m_ec^2)\sim 0.3$, and then decreases. The
radiation spectrum in the immediate DS is dominated by a relatively
isotropic component with $ h\nu\sim 3 T $, but a fraction of $ 10\%
-20 \% $ of the energy flux is carried by a high energy photon tail,
strongly beamed towards the DS, with a cutoff at $ \sim \Gamma_u^2
m_ec^2 $ and a nearly flat spectrum, $ \nu F_\nu \propto \nu^0 $
(see Figs. \ref{fig:spec_30_DS_sh} to
\ref{fig:int_spec_ds_10_20_30}).
\\ \noindent {\bf[2] Optical depths due to Compton scattering and pair production:} In \sref{sec:results_opt_depth} we showed that the optical depth of typical photons ($ h\nu\sim m_ec^2 $) leaving the subshock in the US direction is a few. The optical depth is provided by both Compton scattering and pair production, the latter having a somewhat smaller contribution (see Figs. \ref{fig:tau_us_all_nu1Gu_10} and \ref{fig:tau_us_all_nu1Gu_30}). Photons with much smaller energies are scattered close to the immediate DS and do not reach the transition region (see Figs. \ref{fig:tau_us_halfGu_10} and \ref{fig:tau_us_halfGu_30}).
Typical DS going photons from the transition region (with shock frame energy $ \sim \Gamma^2 m_ec^2 $) undergo very few interactions on the way to the immediate DS (see Figs. \ref{fig:tau_ds_G3_10} to \ref{fig:tau_ds_halfGu_30}).
\\ \noindent {\bf[3] The importance of $ e^+e^- $ pairs.} In figure \ref{fig:struct_Fpos_F} we show that the pairs produced along the shock transition and in the immediate DS play an important role in decelerating the US plasma. The energy flux removed from the protons is dominated by pairs over radiation during most of the transition, and the ratio between the two becomes larger as $ \Gamma_u $ grows. The pair energy flux is dominated by thermal energy flux since the transition region temperatures are relativistic ($ T>m_ec^2 $).

We find several characteristics of the structure of RRMS, which are qualitatively different from those of NR RMS.
\\ \noindent 1. The Thomson
optical depth of the transition region is much larger than unity, is dominated by pairs, and
grows with $ \Gamma_u $ in a manner faster than linear. However, the
actual (KN corrected) optical depth (including pair production) for a typical photon crossing the shock is of order of a few. 
\\ \noindent 2. The temperatures of the pair plasma within the transition region are
relativistic, $ T>m_ec^2 $. 
\\ \noindent 3. The relativistic shock structure includes a sonic
point crossing, in which the flow changes from supersonic to subsonic. We find that this sonic point must be a sub-shock mediated by processes not included in our calculation, which operate on a scale much shorter than the radiation
mean free path [e.g. plasma instabilities, see eq.~(\ref{eq:t_pl})].
\\ \noindent 4. $ e^+e^- $ pairs carry most of the energy and
momentum flux in the transition region
\\ \noindent 5. In RRMS a fair fraction of
the energy density in the immediate DS is carried by a nonthermal
tail of high energy photons, where in the DS of NR RMS the radiation is in CE with the plasma.

We developed in \sref{sec:analytic} an analytical understanding of the key features of the shock structure and radiation spectrum. Several points should be highlighted.
\\ \noindent {\bf[1]  Immediate DS.} The key to understanding the structure and radiation spectrum of RRMS is the understanding of the immediate DS. The immediate DS of RRMS is close to CPE (see Fig. \ref{fig:Gu20_CE}), which, due to the fast increase of the number of pairs with temperature, sets the temperature to a large fraction of $ m_ec^2 $ \citep{Katz09}. { The large amount of positrons and the high temperature imply a relativistic speed of sound in matter $ \beta_{ss}\sim 1/\sqrt 3 $, and combined with the low velocity in this region that quickly approaches its DS value $ \beta_d \le 1/3 $, a subsonic regime is inevitable.}
The immediate DS acts as the supplier of photons directed towards the US, which decelerate the incoming plasma through Compton scattering and pair production.
\\ \noindent {\bf[2]  Deceleration region.} For $ \Gamma \ll \Gamma_u $, we find $ \Gamma(-\tau_*)\approx \sqrt{-\tau_*} $ (see fig. \ref{fig:struct_g_simplified}), where the subshock is located at $ \tau_*=0 $. This behavior is due mainly to the KN scaling of the cross sections, and to the fact that the optical depth for US going photons is of order few. This approximation follows closely the numerical results up to $ \Gamma\approx \Gamma_u/2 $. 
\\ \noindent {\bf[3]  High energy photon beam.} { The immediate DS has a high energy photon component narrowly beamed in the DS direction, with a nearly flat power-law like spectrum, $\nu I_\nu\propto\nu^0$ and an energy cutoff at $ \sim \Gamma_u^2 m_ec^2 $.  The photons in this beam originated from the immediate DS, propagated into the transition region and then were Compton scattered once, before returning to the DS.  
Photons that were scattered at a point with Lorentz factor $ \Gamma $  return to the DS with an energy boosted to $ \sim \Gamma^2m_ec^2 $ and within a beaming angle $ \theta\sim \Gamma^{-1} $.}
 An approximate description of the resulting spectral and azimuthal structure of the beam is given in Eq. \eqref{eq:beam-simple}. The total optical depth for these photons to reach the immediate DS is  small, and they carry $ 10\% - 20\% $ of the energy flux in the immediate DS.  The beam is only stopped far into the DS, producing pairs on low energy photons.
\\ \noindent {\bf[4]  Far US.} In the far US, $ P_{rad,sh} $,  $ F_{rad,sh} $, $ \Gamma_u\beta_u-\Gamma\beta $ and $ T $, all grow exponentially with $\tau_*$ with the same exponent, $ \lambda_{as} $ given in eq. \eqref{eq:lambda_as}, while $ x_+ $ grows exponentially with an exponent $ 2\lambda_{as}$ (see fig. \ref{fig:US-exponents}).
\\ \noindent {\bf[5] Thermalization length scale.} The thermalization length is much longer than the shock transition, both in terms of Thomson optical depth and in real distance. Thermal equilibrium is reached $ \sim 10^6 $  Thomson optical depths into the DS.
 
Finally we showed for completeness in \sref{sec:nonrel} the preliminary results of a detailed calculation of the structure of a NR RMS including full radiation transport. The results are consistent with previously published ones, and support the validity of the numerical methods we use and of the diffusion approximation used for solving the problem in earlier work.

\acknowledgements This research was partially supported by Minerva, ISF and AEC grants.

\appendix

\section{A. Notations frequently  used in the paper}\label{app:notations}

\subsection{Subscripts, superscripts and miscellanea}

as~: Asymptotic upstream behavior\\
US~: Upstream\\
DS~: Downstream\\
CE~: Compton equilibrium\\
CPE~: Compton-Pair equilibrium\\ 
d~: Asymptotic downstream (postshock) value\\
dec~: Deceleration\\
e~: Electron value\\
p~: Proton value\\
NR~: Non-relativistic\\
pl~: Plasma value\\
rad~: Radiation field value\\
sh~: Shock frame value\\
rest~: Plasma rest frame value\\ 
u~: Asymptotic upstream (preshock) value\\
$\gamma$~: Photon value\\
+~: Positron value\\
$\scriptscriptstyle{\wedge}$ (hat)~: Normalized units \\

\subsection{Symbols}

$a_{bb} =$~:
Radiation constant\,\\
$\sigma_c$~: Compton scattering cross section \,\\
$F_{rad} \units{ergs\; cm^{-2}s^{-1}}$~: Radiation energy flux \,\\
$h \units{ergs\; s}$~: Planck's constant\\\
$I(\Omega,\nu) \units{ergs\; cm^{-2}\,s^{-1}str^{-1}\,Hz^{-1}}$~:
Specific intensity of radiation field \,\\
$\eta(\Omega,\nu) \units{ergs\; cm^{-3}\,s^{-1}str^{-1}\,Hz^{-1}}$~:
Emissivity coefficient \,\\
$\ell \units{cm}$~: Photon mean free path \,\\
$n_e, n_+, n_i, n_{\gamma,\text{eff}} \units{cm^{-3}}$~: Number density of electrons, positrons, ions (protons) and typical photons\\
$n_u \units{cm^{-3}}$~: Upstream proton (and electron) number density\\
$P \units{ergs\; cm^{-3}}$~: Pressure\,\\
$P_{rad} \units{ergs\; cm^{-3}}$~: Radiation pressure \,\\
$Q_+ \units{cm^{-3}s^{-1}}$~: Net rate of positron production \,\\
$T \units{erg}$~: Electrons \& positron temperature\\
$\hat{T} \equiv T/ m_ec^2$\\
$T^{0z},\, T^{zz} \units{ergs\; cm^{-3}}$~: Components of
stress-energy tensor (energy and momentum fluxes, \\
$\phantom{aaaaaa}$ respectively)
\,\\
$x_+ = n_+/n_i$~: Positron fraction \,\\
$z \units{cm}$~: Length along flow direction\\
$\chi(\Omega,\nu) \units{cm^{-1}}$~: absorption coefficient \,\\
$\alpha_{e}$~: Fine structure constant\\
$\beta \equiv \sqrt{1-\Gamma^{-2}}$~: Flow velocity (units of
$c$)\\
$\Gamma$~: Flow Lorentz factor\\
$\Gamma_u$~: Upstream flow Lorentz factor\\
$\gamma_{e,th}$~: Lorentz factor associated with random motion of
$e^+$ and $e^-$ \,\\
$\delta = 1-\Gamma\beta/\Gamma_u\beta_u$~: Asymptotic deceleration
parameter \,\\
$\epsilon_{sc},\, \units{ergs}$~:
Radiation emission cutoff energy due screening \,\\
$\zeta$~: Riemann's zeta function\\
$\eta \equiv \exp(-\gamma_E) = 0.5616$ (where $\gamma_E \simeq
0.5772$ is Euler's constant) \,\\
$\lambda^{(ff)}$~: Correction factor for
bremsstrahlung emission \,\\
$\lambda_D \units{cm}$~: Debye length [$\equiv \sqrt{T/4\pi e^2
(n_e+n_+)}$] \,\\
$\mu$~: Cosine of angle relative to positive $z-$axis (flow direction)\\
$\nu \units{Hz}$~: Photon frequency\\
$\hat{\nu} \equiv h\nu/m_ec^2$\\
$\sigma_c \units{cm^{2}}$~: Total Compton scattering cross
section \,\\
$\sigma_{\gamma\gamma} \units{cm^2}$~: Cross section for
$\gamma\gamma \ra \epm$ pair production
\,\\
$\sigma_T \units{cm^2}$~: Thomson cross section ($8\pi r_0^2/3$)\\
$\tau_\star$~: Thomson optical depth for upstream-going photons, given by $ \tau_*\equiv \int \Gamma(1+\beta)(n_e+n_+)\sigma_Tdz_{sh}  $. \\
$\phantom{aaa}$ $ \tau_*=0 $ at the subshock and grows towards the downstream. \,\\

\section{B. Compton scattering approximation}\label{app:compton}

In order to reduce (significantly) the computing time, we use an approximate Compton Scattering Kernel (CSK) that represents the physically important features of the exact CSK. We make the approximation that the scattering is isotropic in the plasma frame, and write the differential cross section as
\begin{equation}
\frac{d\sigma_s}{d\nu d\Omega}\left( \nu , \Omega  \rightarrow  \nu' , \Omega'  \right)=
\frac{1}{4\pi} \sigma_c(\nu , T)
f_d\left( \nu, T,\nu' \right).
\end{equation}
Here, $\sigma_c$ is the total cross section given in eq. \eqref{eq:tot_scat_cs}, and $ f_d $ is the spectral redistribution function of the photons. We require scattering to conserve photon number and require $f_d$ to satisfy
\begin{equation}
\int _0^\infty f_d\left(\hat  \nu, \hat T,\hat \nu' \right)d\hat \nu'=1,
\end{equation} 
and
\begin{equation}
 \int _0^\infty f_d\left( \hat \nu, \hat T,\hat \nu' \right)\hat \nu' d\hat \nu'=\hat \nu_0(\hat \nu,\hat T),
\end{equation}
where $ \nu_0 $ is the average frequency of scattered photons. The approximations used for $ \nu_0 $ and $ f_d $ are given below. We use different approximations for low (NR) temperatures and for high (relativistic) temperatures, with a transition temperature $\hat{T}_{m}=0.25$. We use a smooth interpolation between the two temperature regimes (over a $ \sim 10 \% $ interval in $ \hat T $). 

\subsection{Low T ($ \hat T<0.25 $)}

\paragraph{Average energy shift} We chose $\hat{\nu}_0$ to produce the correct average energy shift for $ \hat \nu\ll 4\hat T $ and for $ \hat \nu\gg 4\hat T $, and no energy shift for NR Compton equilibrium, $ \hat \nu= 4\hat T $. We use
\begin{equation}
\frac{\hat{\nu}_0}{\hat{\nu}}=\min\left[\left(1+\frac{4\hat{T}(4\hat{T}+1)-\hat{\nu}(\hat{\nu}+1)}{(1+a_\nu \hat{\nu})^{3}}\right),\:\frac{4\hat{T}}{\hat{\nu}}\right]  
\end{equation}
for $ \hat{\nu}<4\hat{T} $, and 
\begin{equation}
\frac{\hat{\nu}_0}{\hat{\nu}}=\frac{1}{1+\log\left(\frac{\hat{\nu}+1}{4\hat{T}+1}\right)} 
\end{equation}
for $ \hat{\nu}>4\hat{T} $. $a_\nu( \hat T)$ is determined by requiring that for a Wien spectrum, the energy gain of photons with energy less than $4\hat{T}$,
\begin{equation}
P_{gain}(\hat{T},a)\propto\intop_{0}^{4\hat{T}}d\hat{\nu}\hat{\nu}^{2}e^{-\hat{\nu}/\hat{T}}\sigma_c(\hat{\nu},\hat{T})\left(\hat{\nu}_0\left(\hat{\nu},\hat T,a_\nu\right)-\hat{\nu}\right),
\end{equation}
be equal to the energy loss of higher energy photons,
\begin{equation}
P_{loss}(\hat{T})\propto
\intop_{4\hat{T}}^{\infty}d\hat{\nu}\hat{\nu}^{2}e^{-\hat{\nu}/\hat{T}}\sigma_c(\hat{\nu},\hat{T})
\left(\hat{\nu}_0\left(\nu,T\right)-\hat{\nu}\right).
\end{equation}
We use a 4-th order polynomial for $a(\log\hat{T})$,
\begin{align}
a_\nu(\hat{T})=-0.003763\log(\hat{T})^{4}-0.0231\log(\hat{T})^{3} -0.01922\log(\hat{T})^{2}-0.129\log(\hat{T})+3.139,
\end{align}
which is accurate to better than a percent, and set $a_\nu(\hat{T}<0.01)=a_\nu(\hat{T}=0.01)$. 

\paragraph{Photon redistribution}
We choose a photon re-distribution function that follows the
shape of a thermal spectrum with a target temperature $ \hat T_{tar}=\hat \nu_0(\nu,\hat T)/4 $,
\begin{equation}
f_d(\hat \nu,\hat T, \hat \nu')=A\hat{\nu}'^{3}e^{-\hat{\nu}'/\hat{T}_{tar}},
\end{equation}
where
\begin{equation}
A=\left(\intop_{0}^{\infty}\hat{\nu}'^{3}e^{-\hat{\nu}'/\hat{T}_{tar}}d\hat{\nu}'\right)^{-1}=\frac{1}{6\hat{T}_{tar}^{4}}.
\end{equation} 

\subsection{High T ($ \hat T>0.25 $)}

\paragraph{High $\nu$ - Klein Nishina corrections}
For $\hat{\nu}>1/(4\hat{T)}$, in the Klein-Nishina regime, we choose 
\begin{equation}
f_d(\hat \nu, \hat T, \hat \nu ') =\frac{\hat{\nu}'^{2}e^{-\hat{\nu}'/\hat{T}}\sigma(\hat \nu',\hat T)}{a_{d}(\hat T)\hat{T}^{3}\sigma_{T}}.
\end{equation}
The value of $a_{d}$ is chosen so that the integral over $f_d(\hat \nu, \hat T, \hat \nu ')d\hat\nu'$
 is  1. This form ensures that a Wien spectrum with a relativistic temperature is unchanged by scattering of electrons with the same temperature. We use a 4th order polynomial approxiomation,
\begin{align}
a_{d}(\hat{T})=-0.004611\log(\hat{T})^{4}+0.007197\log(\hat{T})^{3} +0.09079\log(\hat{T})^{2}-0.3166\log(\hat{T})+0.3146
\end{align}
and set $a(\hat{T})=a(5)(\hat{T}/5)^{-1.7}$ for $\hat{T}>5$. This approximation is accurate to better than a percent for temperatures below $ m_ec^2 $, and to better than 10\% everywhere.

\paragraph{Low $\nu$ - Inverse Compton} In order that a power law spectrum of the form $I_{\nu}\propto\nu^{2}$ retains its form after scattering, and in order to reproduce the ultra relativistic limit of the energy boost, $16\hat{T}^{2}$, we choose
\begin{equation}
f_d(\hat \nu, \hat T, \hat \nu ')\propto\sqrt{\nu'}e^{-\sqrt{\frac{\nu'}{\frac{4}{3}\hat{T}^{2}\nu}}}
\Theta(\nu'-\nu).
\end{equation} 
We use a cutoff at $8\hat{T}$ to avoid overproducing photons at high frequencies, and normalize accordingly.

\section{C. Plasma speed of sound}\label{sec:sound_speed}
Below is a short derivation of a general formula for the speed of sound in a plasma of electrons, protons and $ e^+e^- $ pairs, neglecting the thermal pressure of the protons (valid for $ T\ll m_pc^2 $).
Here $n_l$ is number density of leptons (electrons+positrons), and $n_{p}=n_l/(2x_++1)$. For covenience we use $ \hat p \equiv p/m_ec^2 $, $ \hat e \equiv e/m_ec^2 $.
Let $s$ be the entropy per lepton. Rewriting Eqs. \eqref{eq:epl} and \eqref{eq:ppl} we have
\begin{align}
 & \hat p=n_l \hat T,\\
 & \hat e=n_l\left(1+\tilde{f}(\hat T)+\frac{m_{p}}{2x_++1}\right),\end{align}
where we define
\begin{equation}
 \tilde{f}(\hat T)=\frac 32 f(\hat T)\hat T,
\end{equation} 
where $ f(\hat T) $ was defined in Eq. \eqref{eq:epl}.
The derivative with respect to $ \hat T$ at
constant $s$ is denoted by $'$. Constant entropy per lepton implies
\begin{equation}
 \left(\frac{\hat e}{n_l}\right)'=-\hat p\left(\frac{1}{n_l}\right)'.
\end{equation}
We therefore have
\begin{equation}
\tilde{f}'=\frac{\hat Tn_l'}{n_l},
\end{equation}
and
 \begin{align}
 & \hat p'=n_l\tilde{f}'+n_l\\
 & \hat e'=n_l\tilde{f}'+\left( 1+\tilde{f}+\frac{m_{p}/m_e}{2x_++1}\right) \frac{n_l\tilde{f}'}{\hat T}.
 \end{align}
 The speed of sound is finally given by
 \begin{equation}\label{eq:css}
\beta_{ss} =c_{ss}/c=\sqrt{\frac{\hat p'}{\hat e'}}=\sqrt{\hat T}\sqrt{\frac{1+1/\tilde{f}'}{1+\tilde{f}+\frac{m_{p}/m_e}{2x_++1}+\hat T}}.
\end{equation}
This equation can be easily verified to obey the asymptotic NR and ultra relativistic limits.

\section{D. Frame transformations}\label{sec:trtans-def}
Below are some useful transformation rules relating the values of $\nu$, $I$, $\eta$ and $\chi$ measured in the shock and plasma rest frames:
\begin{equation}
\mu_{sh}=\frac{\mu+\beta}{1+\beta\mu},\quad
\mu=\frac{\mu_{sh}-\beta}{1-\beta\mu_{sh}},
\end{equation}
\begin{equation}
\nu_{sh}=\nu\Gamma(1+\beta\mu),\quad
\nu=\nu_{sh}\Gamma(1-\beta\mu_{sh}),
\end{equation}
\begin{equation}
\frac{I(\mu,\nu)}{I_{sh}(\mu_{sh},\nu_{sh})}=\left(\frac{\nu}{\nu_{sh}}
\right)^3,
\end{equation}
\begin{equation}
\frac{\eta(\mu,\nu)}{\eta_{sh}(\mu_{sh},\nu_{sh})}=\left(\frac{\nu}{\nu_{sh}}
\right)^2,
\end{equation}
\begin{equation}
\frac{\chi(\mu,\nu)}{\chi_{sh}(\mu_{sh},\nu_{sh})}=\left(\frac{\nu}{\nu_{sh}}
\right)^{-1}.
\end{equation}

\input{bib1.tex}
\end{document}

%% file: Definitions.tex
\newcommand{\DDir}{\relax{D\kern-.7em{/}}}

\newcommand{\inv}[1]{\frac{1}{#1}}

\newcommand{\ra}{\rightarrow}

\newcommand{\be}{\begin{equation}}
\newcommand{\ee}{\end{equation}}
\newcommand{\bea}{\begin{equation*}}
\newcommand{\eea}{\end{equation*}}

\newcommand{\abs}[1]{\left\vert#1\right\vert}

\newcommand{\nin}{\relax{\in\kern-.8em{/}}}

\newcommand{\al}{\alpha}
\newcommand{\bt}{\beta}

\newcommand{\Lm}{\Lambda}

\newcommand{\sig}{\sigma}

\newcommand{\vep}{\varepsilon}

\newcommand{\cm}{\mbox{ cm}}

\newcommand{\keV}{\mbox{ keV}}
\newcommand{\MeV}{\mbox{ MeV}}

\newcommand{\gr}{\mbox{g}}
\newcommand{\sref}{\S~\ref}

%% file: bib1.tex
\bibliographystyle{apj}